\documentclass[fleqn,usenatbib]{mnras}

\usepackage{newtxtext,newtxmath}

\usepackage[T1]{fontenc}

\usepackage{tablefootnote}
\usepackage{multirow}
\newcommand{\angstrom}{\text{\normalfont\AA}}
\newcommand{\rasex}[3]{#1$^\mathrm{h}$\,#2$^\mathrm{m}$\,#3$^\mathrm{s}$}
\newcommand{\decsex}[3]{+#1$^\circ$\,#2$^\prime$\,#3$^{\prime\prime}$}

\DeclareRobustCommand{\VAN}[3]{#2}
\let\VANthebibliography\thebibliography
\def\thebibliography{\DeclareRobustCommand{\VAN}[3]{##3}\VANthebibliography}

\usepackage{booktabs}

\usepackage{graphicx}	
\usepackage{amsmath}	

\defcitealias{best2023lofar}{B23}
\defcitealias{das2024lofar}{D24}
\defcitealias{drake2024lofar}{Dr24}
\defcitealias{cidfernandes2010alternative}{C10}
\defcitealias{kauffmann2003host}{Ka03}
\defcitealias{kewley2001theoretical}{Ke01}

\title[Spectroscopic Classifications]{The LOFAR Two-metre Sky Survey Deep Fields: new probabilistic spectroscopic classifications and the accretion rates of radio galaxies}

\author[M. I. Arnaudova]{M. I. Arnaudova$^{1,2}$\thanks{E-mail: m.i.arnaudova@gmail.com},
D. J. B. Smith$^{1}$,
M. J. Hardcastle$^{1}$,
P. N. Best$^{2}$,
S. Das$^{1}$,
S. Shenoy$^{1}$,
K. J. Duncan$^{2}$,\and
L. R. Holden$^{1}$,
R. Kondapally$^{2, 3, 4}$,
L. K. Morabito$^{3, 4}$, 
H. J. A. R$\ddot{\rm{o}}$ttgering$^{5}$\\
$^{1}$Centre for Astrophysics Research, University of Hertfordshire, Hatfield, AL10 9AB, UK\\
$^{2}$Institute for Astronomy, University of Edinburgh, Royal Observatory, Blackford Hill, Edinburgh, EH9 3HJ, UK\\
$^{3}$Centre for Extragalactic Astronomy, Department of Physics, Durham University, Durham DH1 3LE, UK \\
$^{4}$Institute for Computational Cosmology, Department of Physics, University of Durham, South Road, Durham DH1 3LE, UK\\
$^{5}$Leiden Observatory, Leiden University, PO Box 9513, NL-2300 RA Leiden, The Netherlands\\
}

\date{Accepted 2025 August 12. Received 2025 August 11; in original form 2025 June 17}

\pubyear{2025}

\begin{document}
\label{firstpage}
\pagerange{\pageref{firstpage}--\pageref{lastpage}}
\maketitle

\begin{abstract}
The faint radio-source population includes sources dominated both by star formation and active galactic nuclei (AGN), encoding the evolution of activity in the Universe. To investigate its nature, we probabilistically classified 4,471 radio sources at $z < 0.947$ using low-frequency radio data from the LoTSS Deep Fields alongside a multi-component model for nebular emission, sampled by spectra obtained with the Dark Energy Spectroscopic Instrument (DESI).
This was done by combining three tools: (i) the identification of a radio excess, (ii) the BPT diagram, and (iii) a modified Mass Excitation diagram, alongside Monte Carlo methods to estimate the probability that each source is either a star-forming galaxy (SFG), a radio-quiet AGN (RQ AGN), or a high-\slash low-excitation radio galaxy (HERG or LERG). This approach extends the probabilistic classification framework of previous works by nearly doubling the redshift range, such that we can now probabilistically classify sources over the latter half of cosmic history. Often regarded as the `gold standard' method, spectroscopic classifications allow us to evaluate the performance of other methods. Using a 90 per cent reliability threshold, we find reasonable overall agreement ($\sim77$ per cent) with state-of-the-art photometric classifications, but significant differences remain, including that we identify $2$–$5$ times more RQ AGN. Furthermore, our high-confidence spectroscopic classifications show that radiatively-efficient and inefficient AGN exhibit clearly distinct Eddington-scaled accretion rate distributions, contrary to recent findings in the literature. Overall, our results highlight the need for new and forthcoming spectroscopic campaigns targeting radio sources, on the pathway to the SKA.

\end{abstract}

\begin{keywords}
catalogues -- galaxies: active -- galaxies: evolution -- radio continuum: galaxies -- techniques: spectroscopic
\end{keywords}



\section{Introduction}

Understanding the co-evolution of galaxies and their supermassive black holes (SMBH; $M_{\rm{BH}}\sim10^{6-10}M_{\odot}$) is one of the main drivers of modern astronomy, fuelling the development of ground and space-based observatories. Over the past three decades, an avalanche of multi-wavelength data has shed new light on this intricate relation: the mass of the SMBH correlates with a number of host galaxy properties (e.g. \citealt{Ferrarese2000}; \citealt{Gebhardt2000}; \citealt{ferrarese2005supermassive}; \citealt{bernardi2007selection}; \citealt{graham2016galaxy}); the evolution of the cosmic star formation history (CSFH) mirrors that of the black hole accretion history (BHAH), where both reach a peak of activity at $z\sim2$, followed by a decline to the present day (e.g. \citealt{boyle1998cosmological}; \citealt{hopkins2007observational}; \citealt{madau2014cosmic}; \citealt{cocrane2023lofar}); the energy released by active galactic nuclei (AGN) influences star-formation by either triggering (`positive' AGN feedback; e.g. \citealt{kalfountzou2012star}; \citealt{silk2013unleashing}; \citealt{maiolino2017star}) or suppressing it (`negative' AGN feedback; e.g. \citealt{heckman2014coevolution}; \citealt{harrison2017impact}). However, the underlying physical processes that drive feedback remain poorly understood.

The radio band plays a crucial role in such studies, as it can provide a clear view into the dust-obscured activity in the Universe, where it is believed that most of the star formation and AGN activity take place (e.g. \citealt{madau2014cosmic}; \citealt{hickox2018obscured}; \citealt{Zavala2021evolution}). The primary source of extragalactic radio emission at low frequencies ($\lesssim$ few GHz) is synchrotron radiation, which is a result of relativistic particles having been accelerated by either supernovae, the end products of short-lived massive stars (e.g. \citealt{condon1992radio}), or by jets (e.g. \citealt{begelman1984theory}). The development of next-generation radio interferometers (e.g. LOFAR; \citealt{vanhaarlem2013lofar} \& MeerKAT; \citealt{jonas2016meerkat}), which continue to improve their sensitivity and angular resolution, has allowed us to create large statistical samples of star-forming galaxies (SFGs) and AGN out to high redshifts (e.g. \citealt{smolcic2017vla}; \citealt{novak2017}; \citealt{whittam2022mightee}; \citealt{kondapally2022lofar}; \citealt{cocrane2023lofar}). At present, the second data release of the LOFAR Two-metre Sky Survey (LoTSS DR2) wide area presents the largest radio catalogue containing 4,396,228 sources across an area of 5,700 deg$^{2}$ (see \citealt{shimwell2022lofar} for further details). This is further complemented by the LoTSS Deep Fields (\citealt{tasse2021lofar}; \citealt{sabater2021lofar}), which are 3–5 times more sensitive in radio flux, covering a few tens of square degrees.

Studying the interplay between SF and AGN activity also involves investigating the various types of AGN. Depending on the efficiency of matter accreted onto the SMBH, AGN can be classified into two fundamental modes (e.g. see reviews by  \citealt{heckman2014coevolution}; \citealt{hardcastle2020radio}). Radiatively-efficient AGN (generally believed to be accreting at high Eddington-scaled accretion rates, $L/L_{\rm{Edd}}\gtrsim0.01$) produce electromagnetic radiation by efficiently converting potential energy from cold, accreted gas. 
This emission ionizes the surrounding gas, resulting in broad permitted ($\mathrm{FWHM}>1000$\,kms$^{-1}$) and higher excitation forbidden lines compared to those produced by star-forming processes (e.g. \citealt{kauffmann2003host}; \citealt{ho2008nuclear}; \citealt{best2012on}; \citealt{hardcastle2020radio}). This type of AGN is further divided into a subset of sources which possess powerful twin radio jets, referred to as high-excitation radio galaxies (HERG), and those that lack (or are associated with weaker) radio jets (e.g., \citealt{jarvis2019prevalence}; \citealt{gurkan2019lotss}; \citealt{macfarlane2021radio}; \citealt{morabito2022identifying}), called `radio-quiet' AGN (RQ AGN). This distinction raises the question of whether the HERGs and the RQ AGN represent physically distinct populations or different evolutionary stages of a single one (e.g. \citealt{kellermann1989vla}; \citealt{ivezic2002optical}; \citealt{cirasuolo2003there,cirasuolo2003radio}; \citealt{balokovic2012disclosing}; \citealt{gurkan2019lotss}; \citealt{macfarlane2021radio}; \citealt{arnaudova2024}; \citealt{yue2024, yue2025}).
Radiatively-inefficient AGN (generally with $L/L_{\rm{Edd}}\lesssim0.01$) are also associated with the presence of two-sided, collimated jets of charged particles but, in contrast, they emit very little radiation at other wavelengths, which is believed to be due to advection-dominated accretion of hot gas (e.g. \citealt{Narayan1994}; \citealt{Narayan1995}). Such sources are also referred to as low-excitation radio galaxies (LERGs), as they do not exhibit strong forbidden lines, or any indications of AGN activity at other wavelengths (e.g. \citealt{best2005}; \citealt{hardcastle2007}; \citealt{best2012on}; \citealt{hardcastle2020radio}; \citealt{kondapally2022lofar}).

Considerable efforts have been made in differentiating between these types of sources in radio surveys (e.g. \citealt{best2012on}; \citealt{smolcic2017vla}; \citealt{sabater2019lotss}; \citealt{whittam2022mightee}; \citealt{best2023lofar}; \citealt{das2024lofar}; \citealt{drake2024lofar}). Recently, \citeauthor{best2023lofar} (\citeyear{best2023lofar}; hereafter B23) made use of the high-quality radio observations and the available multi-wavelength data spanning from the ultraviolet to far-infrared in the LoTSS Deep Fields (\citealt{kondapally2021lofar}) to classify $\sim80,000$ radio sources. Employing a combination of four SED fitting codes, they were able to categorise up to 95 per cent as either SFGs, RQ AGN, LERGs, and HERGs. However, using four SED-fitting codes is time-consuming and necessitates substantial efforts in maintaining consistency and accuracy across the different results. To address these issues, \citeauthor{das2024lofar} (\citeyear{das2024lofar}; hereafter D24) used the SED fitting code \texttt{\textsc{Prospector}} to revisit the classification of radio sources in the European Large Area Infrared Space Observatory Survey-North 1 (ELAIS-N1).
They showed that they can achieve comparable results with a single SED fitting code, which would not only expedite efforts in classifying larger statistical samples in future studies, but also simplify the classification process by reducing the complexity and resources required. Nonetheless, relying solely on photometry may not always produce reliable classifications, as it inevitably depends on photometric redshifts, which may not always be accurate, and on SED fitting, which relies on models and prior assumptions that may not be universally applicable. Furthermore, these works have prioritised providing a best-estimate classification for as many sources as possible irrespective of the confidence in the results. The `gold-standard' for obtaining reliable classification involves optical spectroscopy (e.g. \citealt{kauffmann2003host}; \citealt{kewley2006host}; \citealt{cidfernandes2010alternative}; \citealt{best2012on}; \citealt{drake2024lofar}). 

Fortunately, with the advent of new and forthcoming spectroscopic surveys, such as the WEAVE-LOFAR survey (\citealt{smith2016weave}), the Optical, Radio Continuum and \textsc{Hi} Deep Spectroscopic Survey (ORCHIDSS; \citealt{duncan2023}) and the Dark Energy Spectroscopic Instrument (DESI; \citealt{desi2016desiI, desi2016desiII}), it will become possible to obtain spectroscopic information of large statistical samples of faint radio sources, and thus evaluate the performance of photometric classifications needed in their absence. The WEAVE-LOFAR survey is designed to produce a catalogue containing complete spectroscopy of all radio sources in the LoTSS Deep Fields. However, a fraction of these sources have already been observed and are part of the early data release of DESI (DESI EDR; \citealt{desi2023early}). 

In this work, we perform independent spectral fitting on a sample of available DESI spectra in ELAIS-N1 and use it to expand upon the probabilistic spectroscopic classification from \citeauthor{drake2024lofar} (\citeyear{drake2024lofar}; hereafter Dr24), which investigates the radio source population in the shallower LoTSS wide area using spectroscopy from the Sloan Digital Sky Survey (SDSS). A new spectral fitting approach is necessary since radio sources often exhibit more complex emission line profiles due to AGN-driven winds and outflows (e.g., \citealt{molyneux2019extreme}; \citealt{girdhar2022quasar}; \citealt{Escott2025}), which may necessitate more than the single Gaussian component used in some current DESI value-added catalogues (e.g. \citealt{Zou2024large}). This will further allow us to effectively validate the performance of future WEAVE derived data products that will underpin significant aspects of the WEAVE-LOFAR science case. We therefore extract emission line information to classify the radio sources into four categories (SFGs, RQ AGN, LERGs, and HERGs) and compare our results to those of B23 and D24.

This paper is structured as follows. Section \ref{sec:ch4_data} describes the radio and spectroscopic data used, along with the sample selection process. Section \ref{sec:line_fitting} describes the fitting technique used to obtain emission lines fluxes, necessary for the source classifications. In section \ref{sec:class_scheme}, we outline the classification method, and in section \ref{sec:results} we validate our results. We then compare our classifications in section \ref{sec:compare_SED} with those of B23 and D24. Finally, section \ref{sec:ch4_summary} gives a summary of our main results. Throughout this work, we use air wavelengths and a flat $\Lambda$CDM cosmology with $\Omega_{\Lambda}$ = 0.7, $\Omega_{M}$ = 0.3 and $H_{0}$ = 70 km s$^{-1}$ Mpc$^{-1}$.

\section{Data}\label{sec:ch4_data}

\subsection{Radio Data}

The radio data used in this work are taken from the first data release of the LOFAR Two-metre Sky Survey Deep Fields (LoTSS Deep DR1; \citealt{tasse2021lofar}; \citealt{sabater2021lofar}). This dataset includes observations centred on 150\,MHz at 6 arcsec resolution of three well-studied extragalactic fields: Lockman Hole (\rasex{10}{47}{00}, \decsex{58}{05}{00}), Bo{\"o}tes (\rasex{14}{32}{00}, \decsex{34}{30}{00}), and ELAIS-N1 (\rasex{16}{11}{00}, \decsex{55}{00}{00}). Each field is included in the value-added catalogue by \cite{kondapally2021lofar}, where a combination of the likelihood ratio (LR) method (e.g. \citealt{sutherland1992likelihood}; \citealt{smith2011herschel}) and visual classification with the LOFAR Galaxy Zoo \citep{williams2019} was used to associate the radio sources with their multi-wavelength counterparts across $\sim$26\,deg$^{2}$. In particular, the LR method incorporates both colour and magnitude information and, thanks to the deep ancillary data available ranging from UV to far-IR, achieves identification rates of up to 97\% on its own. Sources not suitable for LR matching (e.g. those with complex/extended morphologies) were visually inspected, mitigating potential sources of confusion such as blended or ambiguous counterparts. Therefore, the optical cross-identifications are expected to result in low misidentification rates and unlikely to significantly impact our results. This catalogue has also been used to estimate photometric redshifts and stellar masses for each source (\citealt{duncan2021lofar}).

In this work, we focus only on the 31,610 radio sources in ELAIS-N1, since this field contains the most sensitive 150\,MHz data (a root mean square noise level of $\lesssim$20\,$\mu$Jy\,beam$^{-1}$ in the central region and below 30\,$\mu$Jy\,beam$^{-1}$ over 10\,deg$^{2}$) compared to the other LoTSS Deep fields, as well as having the deepest wide-field optical, near- and mid-infrared data, and being included in both B23 and D24 classifications\footnote{We do not use the second data release of the LoTSS Deep Fields \citep{shimwell2025} since optical cross-matching has not been performed, and we want to directly compare to the B23 and D24 classifications.}. 

\subsection{Spectroscopic Data}

The spectroscopic data used in this work are taken from the early data release of the Dark Energy Spectroscopic Instrument (DESI EDR; \citealt{desi2023early})\footnote{This sample is not extended by DESI's 1st data release.}, which is accessible through DESI’s public website\footnote{\url{https://data.desi.lbl.gov}}. The spectra were taken using the DESI instrument, mounted on the 4m Mayall Telescope at Kitt Peak National Observatory, which consists of 10 identical spectrographs. Each spectrograph is equipped with 500 fibres with a 1.5$\arcsec$ entrance diameter, and covers the wavelength range of $3600 - 9824\,\angstrom$ at a resolving power of $\lambda/\Delta\lambda\approx2000-5500$ (\citealt{guy2023spec}).

The data encompass three stages of survey validation, referred to as ``sv1'', ``sv2'' and ``sv3''. Each stage was conducted with a different targeting algorithm, and further subdivided into programs based on observing conditions. The ``bright'' program targeted sources for the Bright Galaxy Survey (BGS) and the Milky Way Survey (MWS), while the ``dark'' program focused on fainter objects such as Luminous Red Galaxies (LRG), Emission-Line Galaxies (ELG), and Quasi-Stellar Objects (QSO). In addition, the ``backup'' program was implemented under bad conditions, and the ``other'' program was dedicated for secondary targets (see \citealt{Myers2023target} for further details).  This resulted in a total of 1,636,256 unique objects included in the primary surveys, and an additional 137,148 objects as part of a series of secondary programs across 1,390 deg$^{2}$. As we are interested in maximising our spectroscopic sample to obtain the largest possible number of classified sources, we consider all surveys and programs.

\subsection{Sample Selection}\label{sec:sample}

\begin{figure}
    \centering
    \begin{minipage}{0.45\textwidth}
        \centering
        \includegraphics[width=\textwidth]{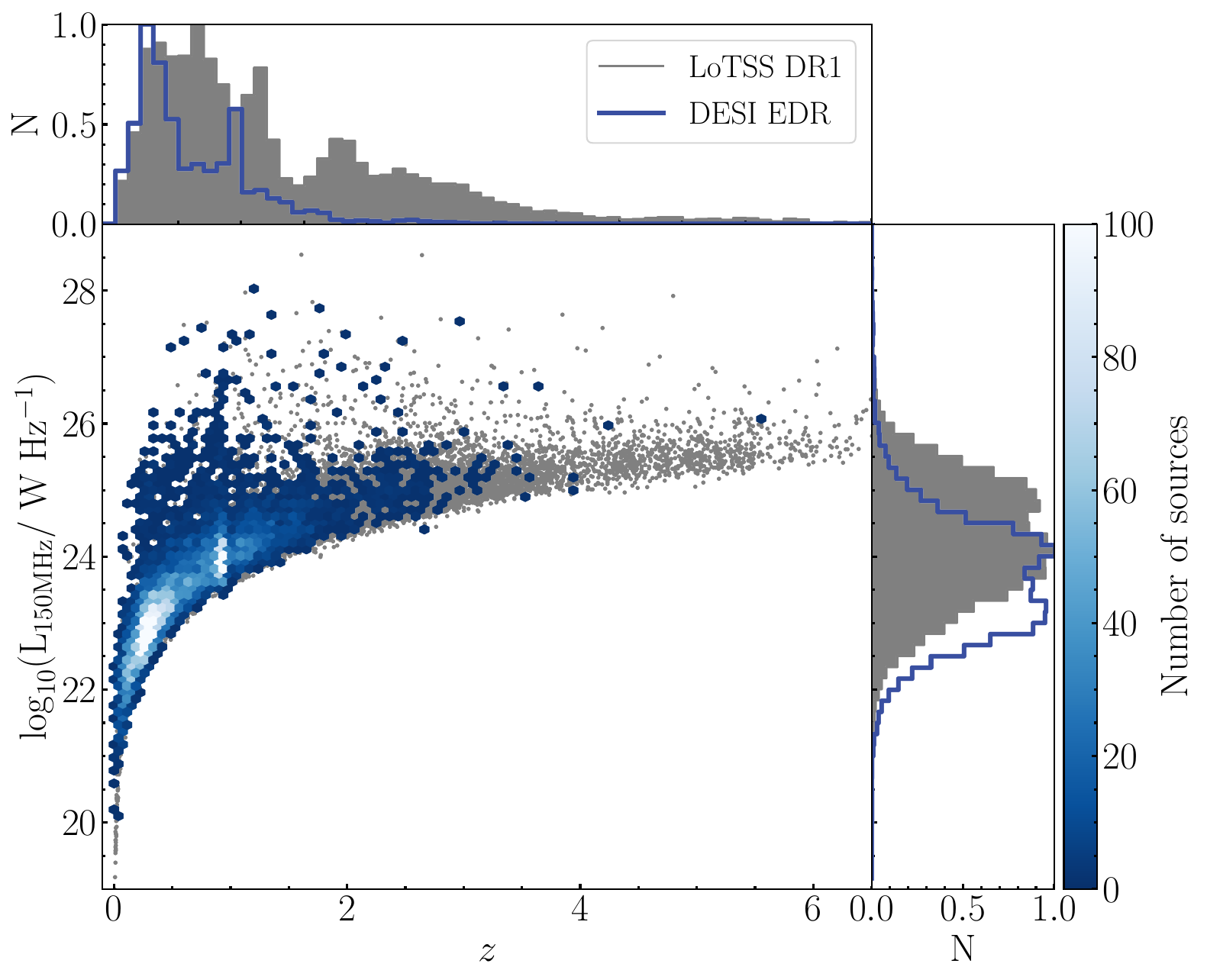}
        \label{fig:fig1}
    \end{minipage}%
    \hspace{0.05\textwidth}
    \begin{minipage}{0.45\textwidth}
        \centering
        \includegraphics[width=\textwidth]{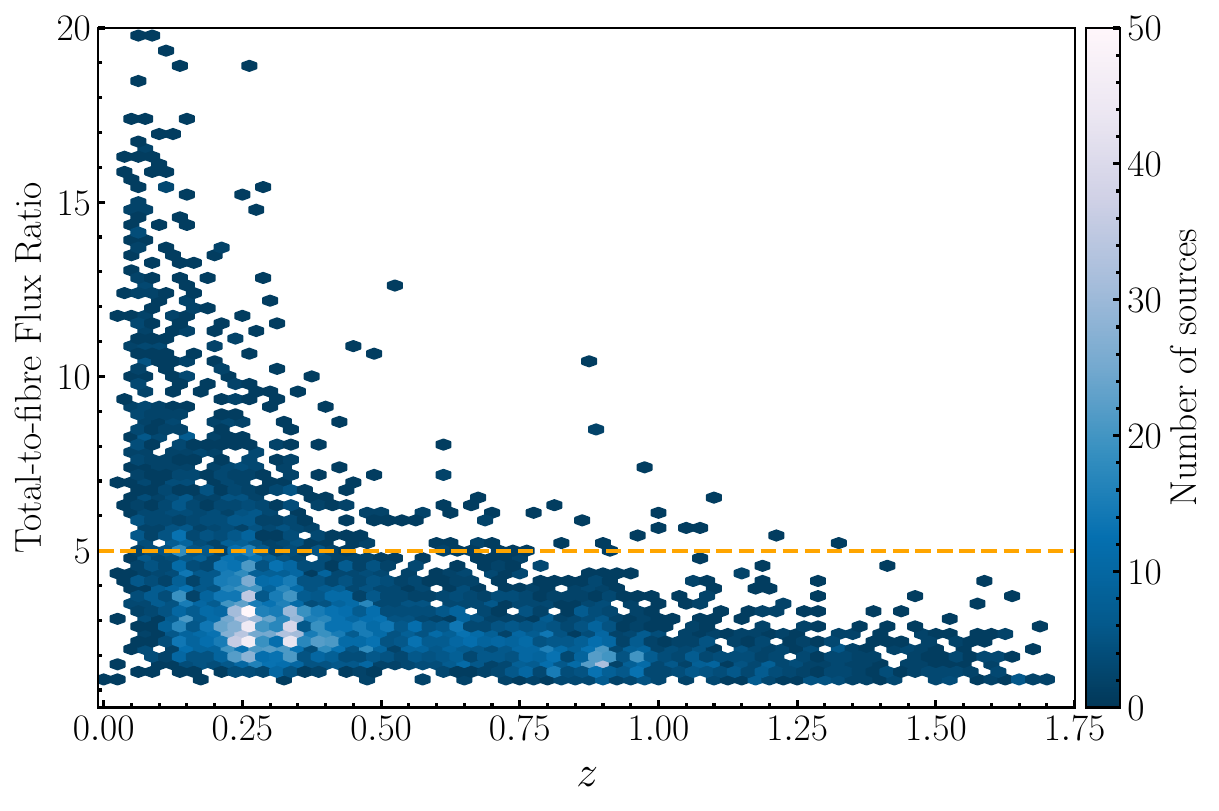}
        \label{fig:fig2}
    \end{minipage}
    \caption{The top panel presents the redshift-luminosity plane for all radio sources with a multi-wavelength counterpart in ELAIS-N1 (in grey), and the sources with spectroscopic information from DESI EDR (colour-coded to indicate their number as presented by the colour bar to the right), along with their respective luminosity and redshift distributions. The bottom panel shows the total-to-fibre $r$-band flux ratio distribution as a function of redshift, with the dashed line representing our threshold for excluding sources with the largest fibre aperture losses.}
    \label{fig:z_vs_L150}
\end{figure}

Starting with a sample of 31,610 radio sources located in the ELAIS-N1 field and possessing a corresponding multi-wavelength counterpart in the DR1 catalogue (\citealt{kondapally2021lofar}), we identify 8,248 with spectroscopic information available from DESI EDR. This was done by using a positional cross-match with a maximum search radius of 1 arcsec between the multiwavelength counterparts reported by \cite{kondapally2021lofar}, whose positions are derived from stacked $\chi^{2}$ S/N images combining optical to near-infrared bands, and sources reported as primary objects in DESI EDR (ZCAT\_PRIMARY==True) with fibres placed on objects (OBJTYPE==TNG). 
The redshift-radio luminosity ($L_{\mathrm{150MHz}}$)\footnote{To calculate the radio luminosity, we used the integrated 150\,MHz flux density ($S_{\mathrm{150MHz}}$) from the value added catalogue by \cite{kondapally2021lofar}, a radio spectral index of $\alpha=-0.7$ (assuming $S_{\nu} \propto \nu^{\alpha}$), and the spectroscopic redshifts as reported in DESI EDR.} plane of all radio sources in ELAIS-EN1, both with and without spectroscopic information from DESI EDR, is shown in the top panel of Figure \ref{fig:z_vs_L150}. We can see that DESI probes a different parameter space than the full radio source population (lower redshift, less radio bright). However, this parameter space coverage will change once WEAVE-LOFAR becomes operational and provides us with complete spectroscopic coverage. Although WEAVE-LOFAR will typically target sources of lower continuum SNR compared to DESI due to its selection on radio flux (but have higher SNR for the same sample of sources as a result of longer exposure time), most sources are expected to be rich in emission lines (\citealt{smith2016weave}). This will enable reliable measurements of redshifts and emission line fluxes, allowing the classification techniques developed in this work to be readily applied to WEAVE-LOFAR data.

Next, we remove sources with a non-zero `zwarn' flag from DESI EDR to ensure reliability of redshift estimates, and thus emission line identification\footnote{We note that `zwarn == 0' selects sources for which the difference in $\chi^2$ between the best and second-best redshift solutions ($\Delta\chi^2$) exceeds 9, which is the standard threshold used by DESI to indicate a reliable redshift. However, 98\% of our sources have $\Delta\chi^2 > 40$, satisfying the more stringent criterion used in the Bright Galaxy Survey.}.
This leaves us with a sample of 7,631 extragalactic sources (`spectype' $\neq$ STAR) sources, out of which 566 are spectroscopically defined as QSOs (i.e. Type 1 QSOs). We remove such objects from our sample, as the stellar population library from \cite{bruzual2003stellar}, which we use in this work, does not account for the various continuum components used for quasars\footnote{While it is possible that a small number of QSO-like sources remain in the sample despite not being classified as SPECTYPE = QSO, such sources are likely to result in `bad' spectral fits. These are typically excluded during the quality filtering applied as part of our spectral fitting procedure described in Section~\ref{sec:line_fitting}, and are therefore not expected to significantly affect our results.} (e.g. see \citealt{arnaudova2024} for details). 
We also account for fibre aperture losses, since we are using radio information which is representative of the entire source, while the optical spectra are limited to the DESI fibre aperture (1.5 arcsec diameter). To approximate this effect, we use the ratio of total-to-fibre $r$-band flux (both provided in the DESI EDR catalogue) as a function of redshift, as shown in the bottom panel of Figure \ref{fig:z_vs_L150}. We can see that at $z > 0.5$ most sources exhibit flux losses of up to a factor of 5, whereas a subset of low-redshift sources are prone to more extreme losses, making a simple scaling correction for these highly uncertain. We therefore restrict our analysis to sources with flux losses less than a factor of 5, reducing our sample to 6,185 sources. We also note that 1,295 sources are secondary targets which lack $r$-band fibre flux measurements in the DESI EDR. For these sources, we assign the median total-to-fibre flux ratio computed within redshift bins of width $\Delta z=0.05$. This approach is found to give comparable results when comparing spectroscopic and photometric SFR estimates (as discussed in Appendix~\ref{sec:SFR}), which is particularly important for the identification for radio-excess sources (see section \ref{sec:radio_excess}).

Finally, due to the changing detectability of optical emission lines with redshift, in the coming sections, we define our sample in two different redshift ranges:

\begin{itemize}
    \item $0<z<0.483$ (hereafter the low-$z$ sample), which includes 3,258 sources with detectable [\textsc{Nii}]\,$\lambda$6583 emission lines, given the upper limit of $\lambda=9824$\AA\ for the DESI spectrograph. This is necessary for the classification method described by Dr24, which uses a combination of  H$\alpha$ and $L_{150\mathrm{MHz}}$ to identify radio-loud AGN and the BPT-NII diagram (\citealt{baldwin1981}) for identifying radiatively-efficient AGN. 
    
    \item $0.483<z<0.947$ (hereafter the high-$z$ sample), which includes 1,851 sources with detectable [\textsc{Oiii}]\,$\lambda$5007 emission lines. Given that the redshift success rate of WEAVE-LOFAR is expected to approach 100 per cent up to $z=1$ (\citealt{smith2016weave}), we expand the Dr24 method to higher $z$ by making use of the H$\beta$ line and the Mass-Excitation (hereafter MEx; \citealt{Juneau2011, Juneau2014}) diagnostic (see section \ref{sec:class_scheme} for details).
\end{itemize}

\section{Spectral Fitting Method}\label{sec:line_fitting}
\begin{figure*}
    \centering
    \includegraphics[width=0.95\textwidth]{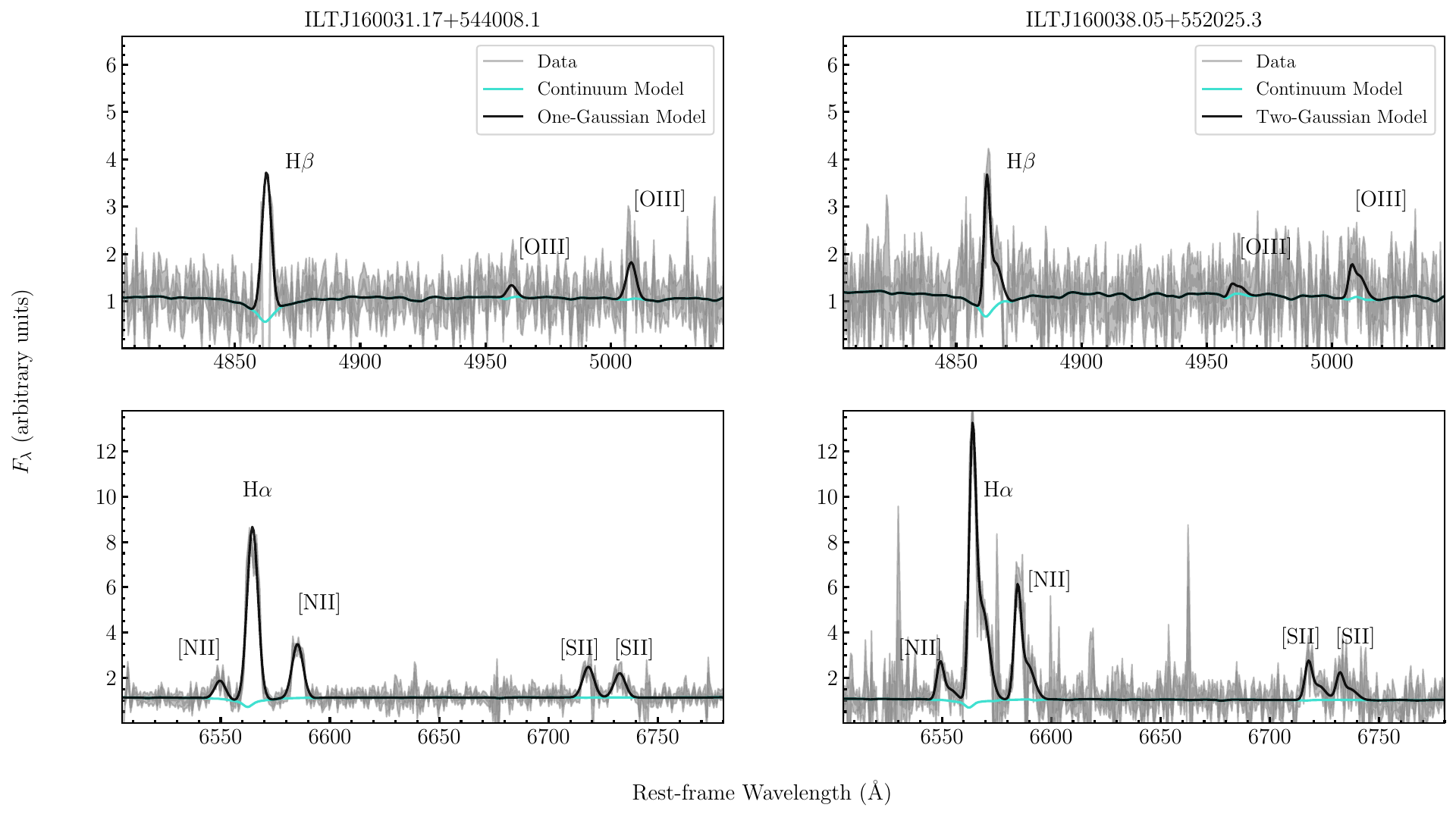}
    \caption[Example fits of DESI spectra]{Example fits of galaxy spectra using our spectral fitting method, where the data are represented in grey with a shaded region enclosing the 1$\sigma$ uncertainties, the continuum model is denoted in turquoise, and the continuum+emission line best fit model is overlaid in black. The top panels are centred on the H$\beta$ and [\textsc{Oiii}]\,$\lambda\lambda$4959, 5007 emission lines, whereas the bottom panels show the H$\alpha$, [\textsc{Nii}]\,$\lambda\lambda$6548, 6583 and [\textsc{Sii}]\,$\lambda\lambda$6716, 6732 lines. The left panels show a random galaxy from the sample, where a single Gaussian component has been determined to model the emission line profiles, whereas the right panels present another example of a galaxy spectrum, whose emission lines are well-fitted with two Gaussian components.}
    \label{fig:galaxy_fit_example}
\end{figure*}

To obtain emission line fluxes needed for the source classification scheme, we apply the spectral fitting approach described by \cite{arnaudova2024b}. To summarise, for each galaxy spectrum we use a Monte Carlo Markov Chain (MCMC) algorithm to model the continuum locally to each emission line complex and use multiple Gaussian components with a fixed line width and velocity offset for each emission line considered, after correcting all spectra for foreground extinction using the re-calibrated reddening data, $E(B-V)$, from \citet*{schlegel1998maps} with the Milky Way reddening curve from \cite{fitzpatrick1999correcting} for an extinction-to-reddening ratio of $R_{V}=3.1$. However, given the less complex velocity structure of galaxies in general compared to the shocked gas studied by \cite{arnaudova2024b}, we limit the use of Gaussian components to a maximum of two per individual emission line. 
This simultaneous fitting of emission lines with tied parameters benefits from using constraints from neighbouring lines not only to deblend emission lines, as in \citet{arnaudova2024b}, but also to mitigate the impact of skyline contamination or telluric absorption on individual line fluxes. We also allow negative amplitudes for individual Gaussian components to ensure that the posterior peak-flux distributions for faint emission lines are not truncated, and thus that our uncertainty estimates remain robust. In addition, we employ the stellar population library from \cite{bruzual2003stellar} to limit the impact of continuum features (e.g. Balmer absorption) on our emission line flux estimates. 

There are 39 templates included in this library computed using the \cite{chabrier2003} initial mass function (IMF), which correspond to three metallicities ($Z=0.4$, 1.0 and 2.5\,$Z_{\odot}$) and 13 star formation history models: 10 instantaneous-burst models with ages of 0.005, 0.025, 0.10, 0.29, 0.64, 0.90, 1.4, 2.5, 5 and 11\,Gyr\footnote{Whilst including the full range of templates is unphysical for the highest redshift sources in our sample (since the model stellar populations are older than the Universe), in practice this makes no difference to our analysis since our focus is on the continuum subtracted emission lines.}; a constant star formation model with an age of 6\,Gyr; and two models with exponentially declining star formation histories with a timescale of $\tau_{\rm{SFR}}=5$\,Gyr and 9\,Gyr and an age of 12\,Gyr. All templates are considered in the single Gaussian component fits, where they are shifted to the observed frame by using the spectroscopic redshifts from the DESI EDR catalogue and resampled onto the DESI wavelength grid using the \textsc{\texttt{SpecRes}} resampling algorithm \citep{carnall2017spectres} to match the data. Additionally, the dust extinction law from \cite{calzetti2000dust} with $R_{V}$ = 4.05 is applied to the continuum templates, where the V-band extinction ($A_{V,\star}$) is allowed to vary as a free parameter to account for internal dust attenuation\footnote{We note that this parameter is not expected to be physically meaningful, as the continuum is fitted only locally to each emission line complex. Its inclusion serves primarily to improve the quality of the emission line fits rather than to recover global dust properties.}.
The Bayesian Information Criterion (BIC) is initially applied to select the optimal template for each spectrum based on the lowest BIC value, after which the same template is re-fitted with a double Gaussian component. The BIC is then used again to assess whether a single or double Gaussian model provides a better fit to the emission lines. Thus, each galaxy is modelled a total of 40 times: once for each of the 39 continuum models with a single Gaussian component and once for the optimal template with a double Gaussian component. For the low-$z$ sample, two primary spectral windows are used: the rest-frame wavelength range $4265<\lambda<5045$\AA\ covering H$\gamma$, H$\beta$, and [\textsc{Oiii}]\,$\lambda\lambda$4959, 5007, and $6225<\lambda<6624$\AA\ containing [\textsc{Oi}] $\lambda$6364, H$\alpha$, [\textsc{Nii}]\,$\lambda\lambda$6548, 6583, where the continuum template normalisation is allowed to differ between the two different wavelength ranges considered for the fitting\footnote{Note that we treat this as a nuisance parameter, which is not used in the analysis.}. For the subset of sources where the [\textsc{Sii}]\,$\lambda\lambda$6716, 6732 doublet falls within the wavelength range sampled by DESI, we extend the fitting window to $\lambda = 6675$ \AA \ to include it in the fit, as we are interested in calculating the Excitation Index (e.g. \citealt{best2012on}; hereafter EI; see section \ref{sec:bpt_class}). For the high-$z$ sample, only the first spectral window is used. 

To remove the bad fits from our sample (where `bad fits' are defined as those for which the probability that the data are consistent with the model is less than 1 per cent), we model the histogram of reduced chi-squared values from each fit (either with one- or two-Gaussian components,  depending on the BIC) across both the low- and high-$z$ samples, using a $\chi^{2}$ distribution. Having determined the best-fit parameters of the $\chi^2$ distribution, we consider only the subset with best fitting $\chi^2$ less than the 99th percentile of the fitted distribution as having acceptable fit quality. This gives us a total of 4,471 galaxies with good fits (3,027 for the low-$z$ and 1,444 for the high-$z$ sample), out of which 794 required a two-Gaussian component model. From these fits, we calculate the total emission line flux per line species (where we sum the fluxes of the individual components if a two-Gaussian model is needed), and apply the aperture flux correction using the total to fibre $r$-band flux, as discussed in section \ref{sec:sample}. The uncertainties are evaluated by propagating the values extracted from the MCMC chains. This is important to ensure robust uncertainty estimates (e.g. derived by properly accounting for varying noise levels including those from sky lines and telluric absorption), which play a critical role in assigning classification probabilities in the spectroscopic classification (see section \ref{sec:full_approach}). The results of this procedure are found to be comparable to those produced using the \textsc{\texttt{FastSpecFit}} Spectral Synthesis and Emission-Line Catalog (\citealt{Moustakas2023fastspecfit}; Moustakas et al. \textit{in prep.}), provided as part of DESI EDR (see Appendix \ref{appendix:fastspec} for details). However, we adopt our own measurements due to our use of multiple Gaussian components and uncertainty estimates that we consider more appropriate for the needs of our classification framework.

Examples demonstrating the performance of the fitting method are shown in Figure \ref{fig:galaxy_fit_example}, where we have chosen random galaxies determined to be best modelled with one (left panels) and two  Gaussian components (right panels), centred on the brighter H$\beta$, [\textsc{Oiii}]\,$\lambda$5007, H$\alpha$, [\textsc{Nii}]\,$\lambda$6583 and [\textsc{Sii}]\,$\lambda\lambda$6716, 6732 lines.

\section{Spectroscopic Classification Scheme}\label{sec:class_scheme}

\begin{figure*}
    \centering
    \includegraphics[width=0.9\textwidth]{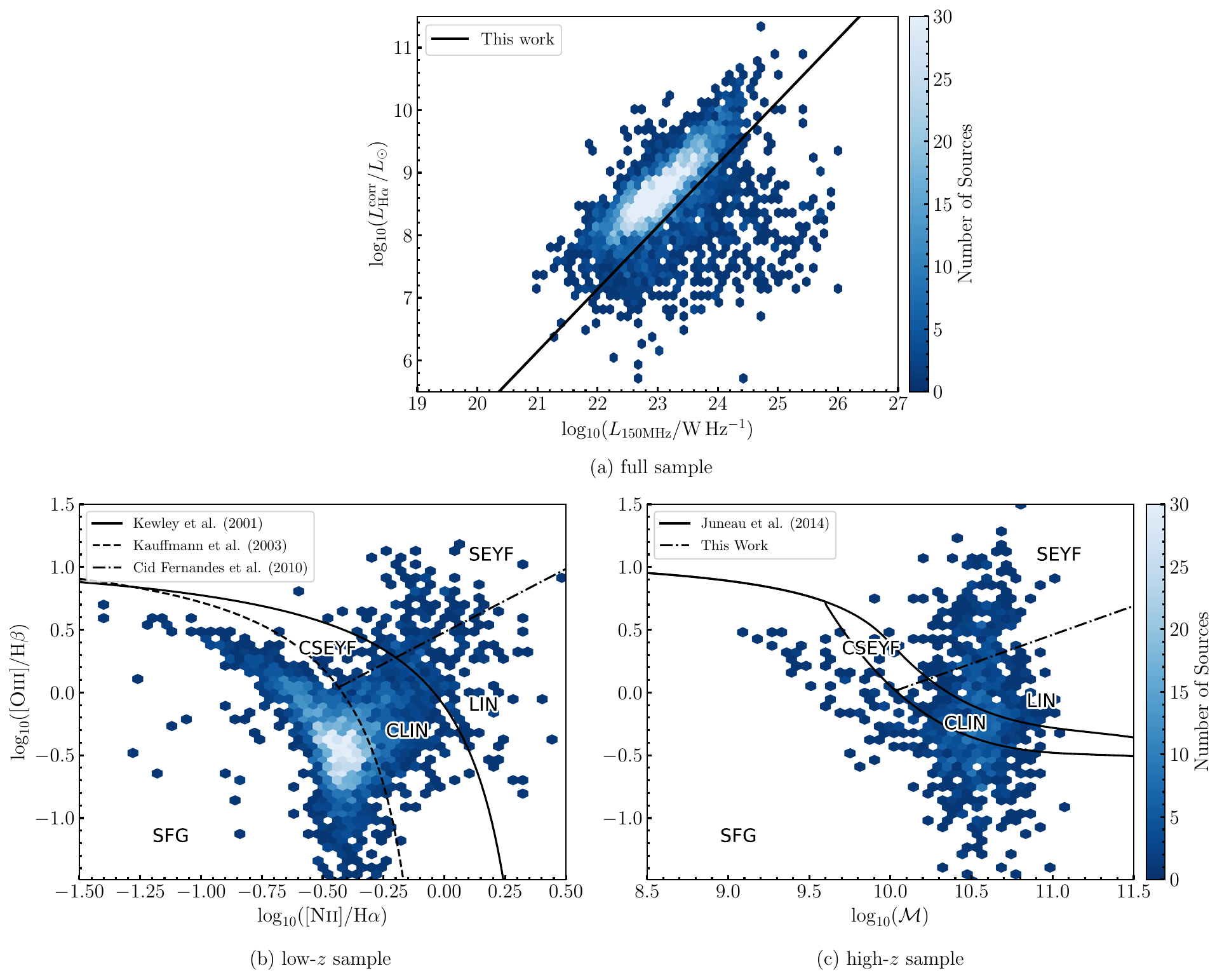}
    \caption[The spectroscopic classification scheme]{The spectroscopic classification scheme. Panel (a) presents the radio excess diagnostic, where the solid line is used in this work to select sources from both the low-$z$ and high-$z$ sample, which have a radio luminosity in excess of what can be explained by star-forming processes alone. Panel (b) shows the BPT diagram for the low-$z$ sample, which is used to distinguish between SFGs and radiatively-efficient AGN, as well as between their low- and high-ionisation categories, as defined by the \citet{kewley2001theoretical}, \citet{kauffmann2003host} and \citet{cidfernandes2010alternative} diagnostic lines. Similarly, panel (c) shows the MEx diagram for the high-$z$ sample, where the separation between SFG, low- and high-ionisation AGN are defined by the modified \cite{Juneau2014} demarcation lines, and the demarcation line based on the Excitation Index defined in this study. The colourbars present the number of sources in each diagnostic.}
    \label{fig:spec_class}
\end{figure*}

\begin{figure}
    \centering
    \includegraphics[width=1\columnwidth]{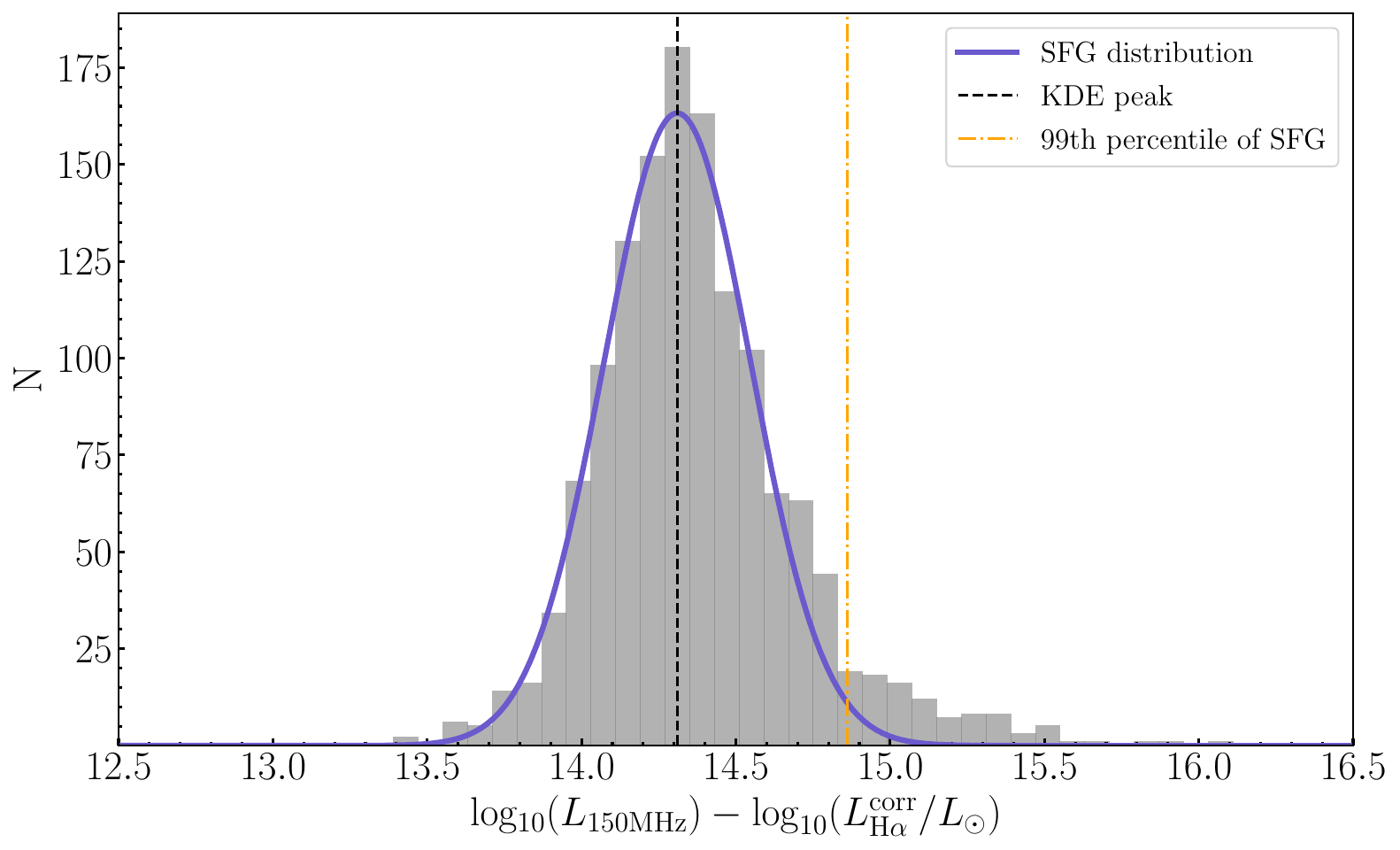}
    \caption{    
    The distribution of $\log L_{150\mathrm{MHz}}-\log L_{\mathrm{H}{\alpha}}^{\mathrm{corr}}$ for all sources with a $\geq5\sigma$ detection in H$\beta$, H$\alpha$ and $S_{\mathrm{150MHz}}$. The blue solid line represents a Gaussian distribution approximating the SFG population, derived by reflecting the left-hand side of the distribution about the KDE peak (black dashed line). The orange dotted line indicates the 99th percentile of the inferred SFG distribution, above which sources are classified as having a radio excess.}
    \label{fig:LHa_L150}
\end{figure}

Following Dr24, we use a combination of the radio excess and hardness of ionisation diagnostics to estimate the probability of each object belonging to the four physical classes: star-forming galaxies (SFG), radio-quiet AGN (RQ AGN), low-excitation (LERG), and high-excitation radio galaxies (HERG). The following sections discuss this classification method for both the low- and high-$z$ samples and outline the differences in our approach compared to Dr24.

\subsection{The radio excess diagnostic}\label{sec:radio_excess}
We use a radio excess (hereafter RX) diagnostic to examine the correlation between the extinction corrected H$\alpha$ ($L_{\rm{H}\alpha}^{\mathrm{corr}}$) and the 150\,MHz radio luminosity ($L_{150\mathrm{MHz}}$), which enables the identification of AGN with a radio excess compared to star-forming processes (i.e LERGs and HERGs). The extinction correction in $L_{\mathrm{H}\alpha}$ is obtained by calculating $A_{V}$ from the Balmer decrement in H$_{\alpha}$, assuming case B recombination (e.g. \citealt{Osterbrock2006}) and a reddening curve provided by \cite{calzetti2000dust} with $R_{V}=4.05$ (see \citealt{dominguez2013dust} for details).
For the 299 sources where $A_{V}$ is negative, we do not correct the H$_{\alpha}$ luminosity since these values are unphysical. As explained by \cite{best2012on}, this diagnostic relies on the expected correlation between the $L_{\rm{H}\alpha}^{\mathrm{corr}}$ and $L_{150\mathrm{MHz}}$ in SFGs, where both serve as direct star-formation rate (SFR) indicators (e.g. \citealt{kennicutt1998}; \citealt{gurkan2018lofar}; \citealt{smith2021lofar}, D24). Conversely, in radio-excess AGN, the $L_{150\mathrm{MHz}}$ is significantly higher compared to SFGs, leading to a noticeable deviation from this correlation. The relationship between the $L_{\rm{H}\alpha}^{\mathrm{corr}}$ and $L_{150\mathrm{MHz}}$ is presented in panel (a) of Figure \ref{fig:spec_class}, where we can see that the majority of sources lie on a tight locus, with some scatter in the lower-right quadrant, where radio-excess AGN are situated.

To determine the demarcation line to use for radio excess identification, we use our low-$z$ sample and apply a similar process to that described by Dr24. We are using a much deeper radio sample (with an rms$\sim$20--35\,$\mu$Jy as opposed to $\sim100\mu$Jy in the wide area), as well as a spectroscopic sample associated with different aperture sizes, for which we apply an aperture correction unlike Dr24. For these reasons, we cannot adopt the radio-excess demarcation line for H$\alpha$ used in their work, but derive our own instead. To do this, we examined the distribution of $\log L_{150\mathrm{MHz}}-\log L_{\mathrm{H}{\alpha}}^{\mathrm{corr}}$ shown in Figure \ref{fig:LHa_L150}. 

To identify the optimal demarcation line, we approximate the SFG population as a Gaussian distribution symmetric around the peak of $\log L_{150\mathrm{MHz}}-\log L_{\mathrm{H}{\alpha}}^{\mathrm{corr}}$ estimated from the kernel density estimate (KDE) and calculate the 99th percentile, such that objects falling to the right are considered to have a radio excess. Therefore, sources are classified as having a radio excess if:
\begin{equation}
\log_{10}( L_{150\mathrm{MHz}}/\mathrm{W Hz}^{-1}) > \log_{10}(L_{\mathrm{H}{\alpha}}^{\mathrm{corr}}/ L_{\odot})+14.86
\end{equation}\label{eqn:rexcess}

\noindent where 14.86 corresponds to the 99th percentile of the inferred SFG distribution.

To apply this diagnostic to our high-$z$ sample, where H$\alpha$ is not accessible, we use the H$\beta$ flux and the median likelihood V-band extinction measurements obtained with \texttt{\textsc{Prospector}} ($A_{V, \mathrm{SED}}$; S. Das, \textit{private communication}) to predict $L_{\mathrm{H}{\alpha}}^{\mathrm{corr}}$. Since $A_{V, \mathrm{SED}}$ indicates the dust attenuation of the stellar continuum (as opposed to the nebular emission lines discussed here), we study its relation to the Balmer derived $A_{V}$ from the low-$z$ sample. Focusing on sources with a 5$\sigma$ detection in both H$\beta$ and H$\alpha$, along with a reliable fit to the photometry from \texttt{\textsc{Prospector}}, we find the following linear relationship between the stellar and nebular dust attenuation: $A_{V, \mathrm{SED}}$~=~(0.52 $\pm$ 0.01)$A_{V}$. This result is in good agreement with previous studies (e.g. \citealt{Papovich2022}; \citealt{pirzkal2024}), which also find that the nebular gas is more extinguished by a factor of two compared to the stellar continuum. Thus, using this relationship and assuming case B recombination, we are able to calculate the H$\alpha$ luminosity for the high-$z$ sample, and apply the same radio excess criterion (equation \ref{eqn:rexcess}) to identify high-$z$ radio excess sources.

\begin{figure*}
    \centering
    \includegraphics[width=1\textwidth]{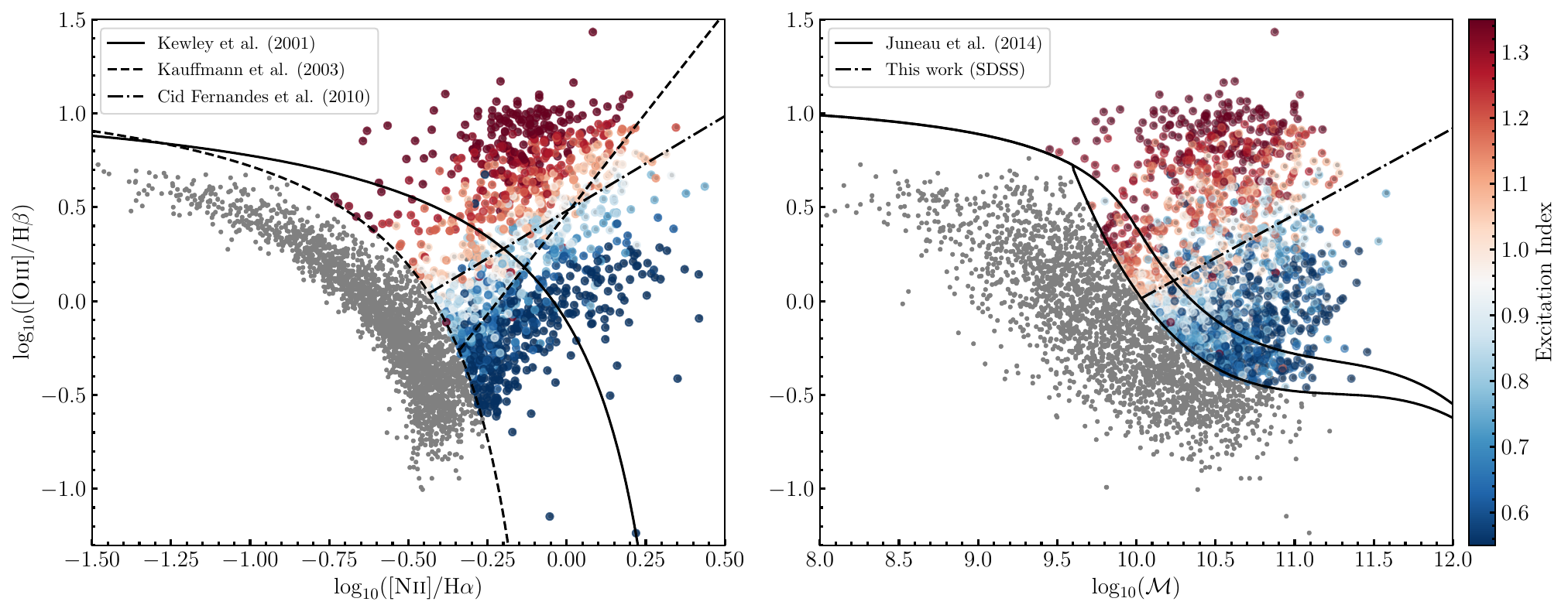}
    \caption{The left panel shows the BPT diagram for all SDSS sources in Dr24 (grey) and those with a $\geq3\sigma$ detection in H$\beta$, [\textsc{Oiii}]\,$\lambda$5007, [\textsc{Oi}]\,$\lambda$6364, H$\alpha$, [\textsc{Nii}]\,$\lambda$6583 and [\textsc{Sii}]\,$\lambda\lambda$6716, 6732 that lie above the Ka03 line, colour-coded based on the Excitation Index (EI). Similarly, the right panel shows the $\mathcal{M}$Ex diagram for all Dr24 sources (grey), and those above the lower curve from J14 with a 3$\sigma$ detection in the same emission lines, including the stellar mass. The colourbar presents the EI for both diagrams and is centred on $\mathrm{EI}=0.95$, the value separating low- and high-ionisation sources.}
    \label{fig:BPT_MEx}
\end{figure*}
\subsection{The BPT diagnostic diagram}\label{sec:bpt_class}

The BPT-NII (hereafter BPT; \citealt{baldwin1981}) diagram is a widely used tool for distinguishing between SFGs and radiatively-efficient AGN, which involves the use of the [\textsc{Oiii}]$\lambda$5007/H$\beta$ and the [\textsc{Nii}]$\lambda$6583/H$\alpha$ emission line flux ratios (e.g. \citealt{stasinska2006semi}; \citealt{best2012on}; \citealt{cidfernandes2010alternative}; \citealt{sabater2019lotss}; \citealt{kewley2019understanding} and references therein). 
Depending on the science case, different demarcation lines are employed. Some studies use the empirically-defined line by \citeauthor{kauffmann2003host} (\citeyear{kauffmann2003host}; hereafter Ka03), some use the theoretically-defined maximum starburst line by \citeauthor{kewley2001theoretical} (\citeyear{kewley2001theoretical}; hereafter Ke01), and some use both to define an additional `composite' class, where both star-formation and AGN activity significantly contribute to the emission lines. Following Dr24, we adopt the Ka03 line for our classification method but also include the composite region (further divided into LINER/Seyfert, see below) in our catalogue to allow other users the flexibility to define an alternative classification.

In addition to these, other demarcation lines are used to differentiate between sources with hard ionizing spectra, such as Seyferts, and those that lack high-ionization lines, like low-ionization nuclear regions (LINERs). Dr24 used the LINER/Seyfert separation proposed by Ka03. However, when using the EI, defined as: 
\begin{equation}
\begin{aligned}
\mathrm{EI} &\equiv \log_{10}([\textsc{Oiii}]\lambda5007/\mathrm{H}\beta) - \frac{1}{3}\big[\log_{10}([\textsc{Nii}]\lambda6583/\mathrm{H}\alpha)\\
&\hspace{0em} +\log_{10}([\textsc{Oi}]\lambda6364/\mathrm{H}\alpha)+\log_{10}([\textsc{Sii}]\lambda\lambda6716, 6732/\mathrm{H}\alpha)\big]
\end{aligned}
\end{equation}
\newline
\noindent where EI = 0.95 marks the boundary between low- and high-ionisation sources (e.g. \citealt{Buttiglione2010optical}; \citealt{best2012on}), and colour-coding the AGN class on the BPT diagram  according to this parameter for all sources with a $>3\sigma$ detection in the relevant lines, they found a better agreement with the demarcation line from \citeauthor{cidfernandes2010alternative} (\citeyear{cidfernandes2010alternative}; hereafter C10), as can be seen in Figure \ref{fig:BPT_MEx}; therefore, we adopt the C10 line. In this way, we can use the low-z scheme to classify each set of emission line fluxes for a given source into one of
five BPT classes: \texttt{BPT\_SFG},\texttt{ BPT\_CLIN}, \texttt{BPT\_SEYF}, \texttt{BPT\_LIN} and \texttt{BPT\_SEYF}, where LIN and SEYF refer to LINERs and Seyferts, respectively, and the prefix `C' denotes objects in the composite region, as can be seen in panel (b) of Figure \ref{fig:spec_class}. This is done in a probabilistic manner for all sources, irrespective of the SNR of the emission lines (see section \ref{sec:full_approach}).

\subsection{The Mass-Excitation diagnostic diagram}\label{sec:MEx_diagnostic}

\begin{figure}
    \centering
    \includegraphics[trim={0cm 0.3cm 0cm 0cm},clip,width=1\columnwidth]{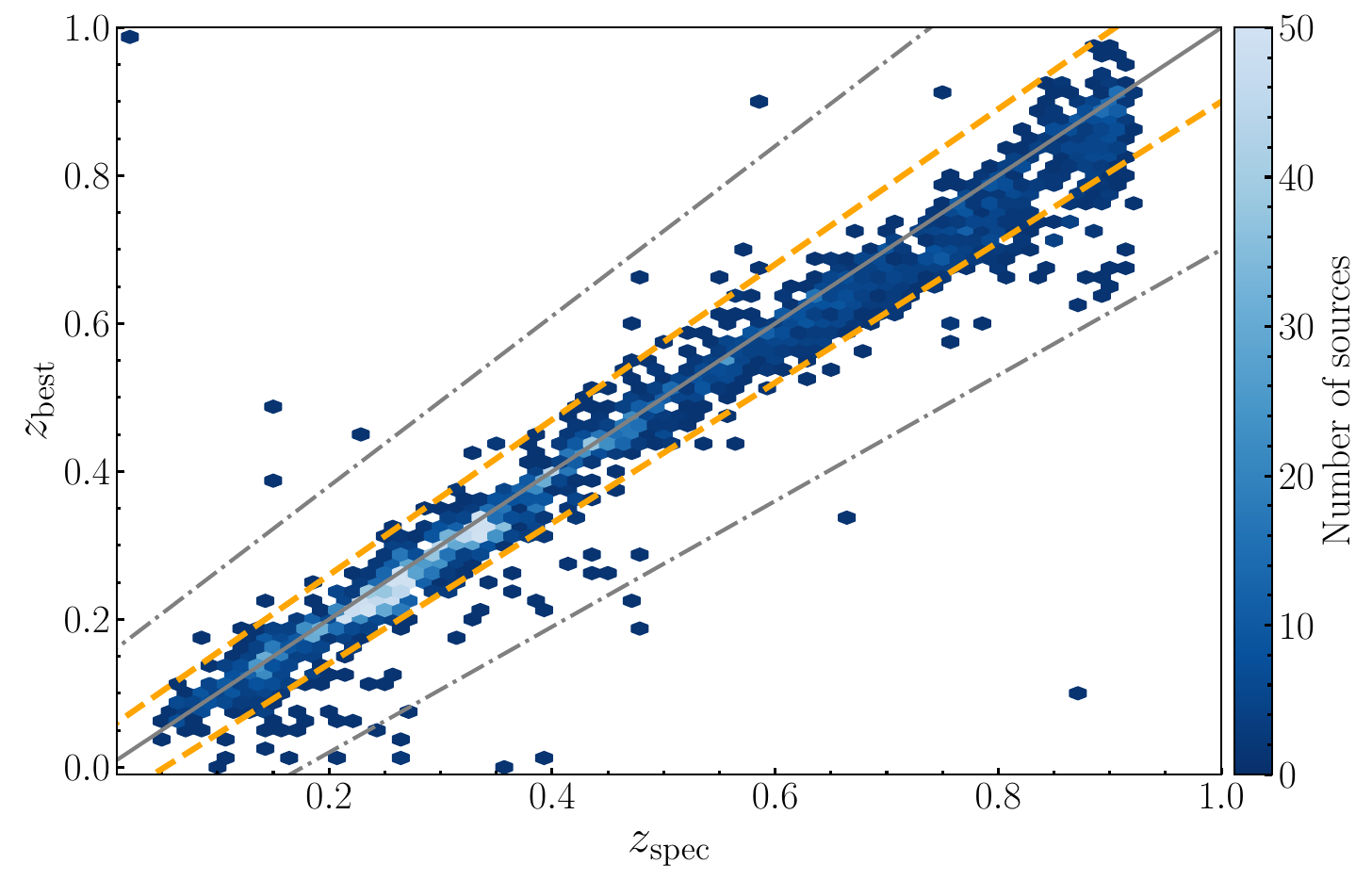}
    \caption{The relationship between the spectroscopic redshifts reported in DESI EDR ($z_{\mathrm{spec}}$), and the redshifts used by D24 ($z_{\mathrm{best}}$), which include photometric redshifts provided by \citet{duncan2021lofar} and spectroscopic redshifts from SDSS DR14. The solid line indicates equality, whereas the black dot-dashed and orange dashed lines corresponds to the $\pm0.15\times(1 + z_{\rm{spec}})$ criterion from \citet{duncan2021lofar} and the $\pm0.05\times(1 + z_{\rm{spec}})$ used in this work, respectively.}
    \label{fig:zcuts}
\end{figure}

In the absence of H$\alpha$ and [\textsc{Nii}] detections for the high-$z$ sample, we turn to the MEx diagram, which replaces the [\textsc{Nii}]$\lambda$6583/H$\alpha$ line ratio in the BPT diagram with stellar mass ($M_{*}$). As discussed by \cite{Juneau2011}, this is a suitable substitution since AGN are typically associated with higher [\textsc{NII}]$\lambda$6583/H$\alpha$ values and reside in more massive galaxies.
Here, we use the stellar mass estimates from the SED fitting catalogue from D24 for sources with acceptable SED fits (see D24 for details). However, these values were obtained by using a combination of spectroscopic redshifts obtained from the 14th data release of the Sloan Digital Sky Survey (SDSS DR14; \citealt{Abolfathi2018sdss}) and photometric redshifts from \cite{duncan2021lofar}. In the absence of spectroscopy for the majority of radio sources, \cite{duncan2021lofar} created a photometric redshift catalogue by combining template fitting and machine learning techniques. The authors noted that the outlier fraction (OLF) based on a threshold of $\delta=(|z_{\rm{phot}}-z_{\rm{spec}}|)/(1+z_{\rm{spec}})>0.15$ is only $\sim1.5-1.8\%$ for galaxies (or host-dominated sources), but for sources selected as AGN in the optical, IR and X-rays, the OLF is found to be higher ($18-20\%$). To evaluate the impact of this, we compare the spectroscopic redshifts from DESI EDR ($z_{\rm{spec}}$) and those used in D24 ($z_{\rm{best}}$), as shown in Figure \ref{fig:zcuts}. While the majority of sources are not classified as an outlier ($\sim99\%$, as indicated by the grey dot-dashed lines) according to the criterion from \cite{duncan2021lofar}, there remains a notable scatter between $z_{\rm{best}}$ and $z_{\rm{spec}}$ as expected. Therefore, we adopt a more stringent criterion of $\delta=0.05$ (as denoted by the orange dashed lines), where 120 sources that are part of the high-$z$ sample are identified as outliers and thus removed from our analysis. This is done because using significantly different redshifts would render the stellar mass estimates unreliable.

To distinguish between SFG and AGN, we use the two demarcation lines from \citeauthor{Juneau2014} (\citeyear{Juneau2014}; hereafter J14), which were calibrated based on the probability that a galaxy hosts an AGN, $P$(AGN), derived from the BPT diagram using a sample from the 7th data release of SDSS at $0.04<z<0.2$. Similarly to the BPT framework, the upper boundary denotes a region with high $P$(AGN), while the lower curve delineates the SFG region, where $P(\mathrm{AGN})<0.3$. In between these two lines lies the intermediate region where BPT composites typically reside. 
To use this diagnostic at higher redshifts, the authors also prescribe a horizontal offset based on the detection limit of emission lines. However, more recent works (e.g. \citealt{newman2014}; \citealt{coil2015}; \citealt{henry2021}; \citealt{cleri2023}) have found that this offset needs to better reflect the evolution of the mass-metallicity relation rather than the survey depth. Using a sample of X-ray confirmed AGN at $z \sim 1.6$, \cite{cleri2023} proposed an offset of $0.2(1+z)$ dex in stellar mass. However, it is unclear whether this relation works well at lower redshifts (e.g for our high-$z$ sample that spans $0.483<z<0.947$) and does not recover the J14 demarcation lines at $z\approx0$. Therefore, we modify the horizontal (i.e. stellar mass) axis of the MEx diagram (hereafter the `modified MEx' diagram or $\mathcal{M}$ex) such that: 
\begin{equation}
    \mathcal{M} = M_{*} - [1-\mathrm{exp}(-1.2z)].
\end{equation}

\noindent The choice of this relation is described in Appendix \ref{appendix:MEx}, where we have showed that this offset provides the optimal separation between SFGs and AGN classified using the BPT diagram while preventing a rapid increase with redshift, and is preferred over the J14 prescription.

To further establish a division line between low- and high-ionising sources, we make use of the EI as in section \ref{sec:bpt_class}.  However, due to our limited sample size and the requirement that all six emission lines contributing to the EI calculation be significantly detected ($>3\sigma$), we use the catalogue from Dr24. This catalogue contains 152,355 sources with emission line measurements from the 12th data release of SDSS \citep{thomas2013}, out of which 69,792 have stellar mass measurements from the MPA-JHU group (\citealt{kauffman2003stellar}; \citealt{Tremonti2004}; \citealt{brinchmann2004physical}). Of these, 1489 lie above the lower curve of the $\mathcal{M}$Ex diagram and have significant detections for all six required emission lines and stellar mass estimates. These sources are shown in the right panel of Figure \ref{fig:BPT_MEx}, where we have colour-coded them according to their EI values. We can see that there is a fairly good separation at EI = 0.95, with higher ionization sources showing higher [\textsc{Oiii}]/H$\beta$ ratios, as expected. As a result, we quantitatively identify the optimal dividing line as the one that best separates these sources based on their EI, which is found to be:

\begin{equation}
    \log_{10}([\textsc{Oiii}]/\mathrm{H}\beta) = 0.46 (\pm0.03)\log_{10}(M_{*}/M_{\odot}) - 4.6(\pm0.4).
\end{equation}

\noindent We note that this line is also consistent with the 13 sources for which we can estimate the EI in our sample. Therefore, in a similar fashion to the low-$z$ sample, we categorise all sources probabilistically into five $\mathcal{M}$Ex classes:
\texttt{$\mathcal{M}$Ex\_SFG},\texttt{ $\mathcal{M}$Ex\_CLIN}, \texttt{$\mathcal{M}$Ex\_SEYF}, \texttt{$\mathcal{M}$Ex\_LIN} and \texttt{$\mathcal{M}$Ex\_SEYF}, which have been labelled in panel (c) of Figure \ref{fig:spec_class}. 

In future work, we plan to explore alternative diagnostics to the $\mathcal{M}$Ex diagram that rely solely on spectroscopic data, such as the Kinematic-Excitation diagram (KEx; \citealt{Zhang2018A}) diagram. However, at present the KEx diagram does not clearly separate AGN into high- and low-excitation categories (see \citealt{Zhang2018A}), so further refinement will be needed.

\subsection{The full probabilistic approach}\label{sec:full_approach}

By combining the radio excess and BPT ($\mathcal{M}$Ex) diagnostics discussed in the previous sections, we classify sources in the low-$z$ (high-$z$) sample into four physical classes as follows: Sources identified as SFGs are those that are part of the \texttt{BPT\_SFG} (\texttt{$\mathcal{M}$Ex\_SFG}) class and do not exhibit a radio excess. Those positioned above the Ka03 (J14 lower) line without a radio excess are classified as RQ AGN. Sources exhibiting a radio excess and situated within the \texttt{BPT\_Seyf}/\texttt{BPT\_CSeyf} (\texttt{$\mathcal{M}$Ex\_CSeyf}/\texttt{$\mathcal{M}$Ex\_CSeyf}) region are categorised as HERGs, while those with a radio excess but not falling within \texttt{BPT\_Seyf}/\texttt{BPT\_CSeyf} (\texttt{$\mathcal{M}$Ex\_CSeyf}/\texttt{$\mathcal{M}$Ex\_CSeyf}) are classified as LERGs. 

To implement the full probabilistic method from Dr24, we generate 1,000 Monte Carlo realisations of this classification scheme, where in each iteration the stellar mass, radio and emission line fluxes are perturbed by randomly drawing from a normal distribution with a mean equal to the measured values and a standard deviation equal to their associated uncertainties\footnote{We note that the uncertainties on the photometric redshifts are not incorporated into the stellar mass uncertainties. However, in cases where stellar mass is used in the classification scheme, we only consider sources for which the photometric and spectroscopic redshifts from DESI EDR are in good agreement (i.e. $\delta < 0.05$; see Section~\ref{sec:MEx_diagnostic} for details), and as such, the impact of photometric redshift uncertainties is negligible.}. We then apply the radio excess diagnostic (in which the $A_{V}$ needed to correct $L_{\mathrm{H}\alpha}$ is calculated each time based on the particular realisations of the Balmer fluxes), and BPT/$\mathcal{M}$Ex diagnostic to each realisation and thus obtain the fraction of realizations (i.e the probability) for each source belonging to particular class. As discussed in Dr24, this approach allows us to select classification schemes that suit different scientific goals, such as maximizing the number of sources in a given class or ensuring high reliability of classes.

Since our workflow includes using Monte Carlo methods on measured line fluxes which can sometimes have low statistical significance, we need to keep track of realisations which have negative fluxes, since it may be possible to derive useful constraints on the properties of the radio source in the RE\slash BPT\slash $\mathcal{M}$Ex diagnostic diagrams, even though the logarithms of their flux ratios are undefined. There are three `poorly-behaved' cases: 
\begin{itemize}
    \item When the flux in the numerator is positive but that in the denominator is negative, we treat this as a `divide by zero' error and assign these values an unambiguously large value on the relevant axis (e.g. lying to the right of the maximum-starburst line on the BPT plot); 
    \item When the numerator flux is negative, but the denominator positive, we do the opposite and assign these realisations an unambiguously small value; 
    \item Finally, for the cases where both the numerator and denominator of a logged line ratio are negative, we are unable to assign a value and consider these realisations invalid. 
\end{itemize}

\noindent To enable the end user to identify sources for which these `edge cases' could have significant impact on the reliability of the source classifications, we calculate the fraction of realisations in which these `poorly-behaved' values occur for the RX (\texttt{RX\_warning\_frac}), BPT (\texttt{BPT\_warning\_frac}), and $\mathcal{M}$Ex (\texttt{$\mathcal{M}$Ex\_warning\_frac}) diagnostics, respectively.

In this work, we adopt the following high-reliability classification: sources in the low-$z$ sample are classified as SFG, RQ AGN, LERG, or HERG if 90 per cent of the realisations fall into a particular class using the RX and BPT diagnostics, and if both \texttt{RX\_warning\_frac} and \texttt{BPT\_warning\_frac} are below 10 per cent - that is, the extreme cases resulting from negative fluxes affect no more than 10 per cent of the realisations. Furthermore, sources that do not meet the 90 per cent threshold in the BPT scheme but do so in the $\mathcal{M}$Ex diagnostic are also considered, provided that \texttt{$\mathcal{M}$Ex\_warning\_frac} < 0.10.  Similarly, for the high-$z$ sample, a source is assigned a class if 90 per cent of its realisations fall into a particular category using the RX and $\mathcal{M}$Ex diagnostics and both \texttt{RX\_warning\_frac} and \texttt{$\mathcal{M}$Ex\_warning\_frac} are below 10 per cent.

The total number of sources falling in each class using this 90 per cent reliability scheme is detailed in Table \ref{tab:Number_sources}, along with the corresponding counts using the photometric classifications reported by B23 and D24, which we will further compare to in section \ref{sec:compare_SED}. 

\begin{table}
\centering
    \caption[The total number of a given class per given classification.]{
             The total number of star-forming galaxies, radio-quiet AGN, low- and high-excitation radio galaxies in the ELAIS-N1 deep field with DESI spectroscopy as identified by the photometric classifications from B23 and D24 (top two rows), alongside the 90 per cent confidence spectroscopic classifications from this work. The numbers in brackets for B23 and D24 indicate the number of sources in each class that are also included in the subset of our sample that meets the 90 per cent threshold (i.e. excluding the sources unclassified by our scheme).
    }
\resizebox{1.\linewidth}{!}{\centering
    \begin{tabular}{lccccc}
        \toprule
        & SFG & RQ AGN & LERG & HERG& Unclassified\\
        \midrule
        B23 & 3504 (2117) & 149 (82) &703 (133) & 43 (20) &72 (46)\\  
        D24 & 3519 (2057) & 263 (158) &538 (96) & 44 (25) &107 (62)\\ 
        This work & 1757& 488 & 129 & 24& 2073\\ 
        ~~~~- low-$z$ &1538& 295& 86 & 5 &1103\\
        ~~~~- high-$z$&219& 193& 43  &19 &970\\
        \midrule
    \end{tabular}
   }
    \label{tab:Number_sources}
\end{table}

\section{Validation}\label{sec:results}

\subsection{Comparing the BPT and $\mathcal{M}$Ex classification schemes}\label{sec:lowz vs highz scheme}
\begin{figure}
    \centering
    \includegraphics[trim={0cm 0.2cm 0cm 0.1cm},clip,width=1\columnwidth]{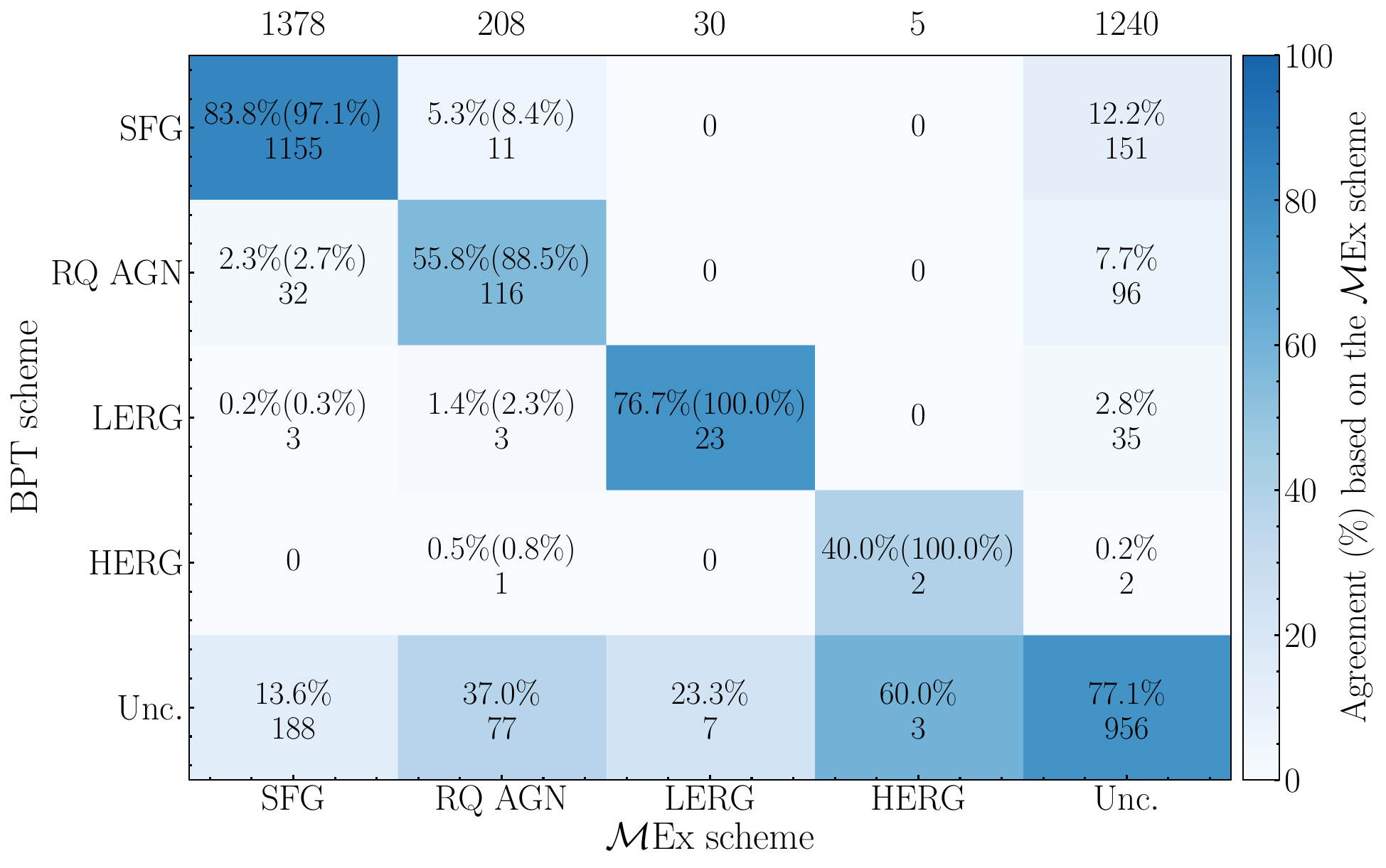}
    \caption{Confusion matrix comparing the 90 per cent spectroscopic classification of the low-$z$ sample, generated with the BPT scheme (rows) and with the $\mathcal{M}$Ex scheme (columns). Each cell represents the level of agreement based on the $\mathcal{M}$Ex scheme in percentages, also indicated by the colourbar, where we have also included the number of sources in each cell. The percentages in brackets have been calculated for the subset of sources that reach the 90 per cent threshold in both the BPT and $\mathcal{M}$Ex schemes. The total number of sources in each class per the $\mathcal{M}$Ex classification is shown at the top of each column.}
    \label{fig:conf_lowz_highz}
\end{figure}

To assess the degree of consistency between the BPT and $\mathcal{M}$Ex classification schemes, we apply both schemes to the low-$z$ sample and evaluate the results based on the 90 per cent classification threshold. Specifically, we use the probabilistic method described in section \ref{sec:full_approach}, where we use the radio excess diagnostic by incorporating the Balmer-derived $A_{V}$ and H${\alpha}$ flux, along with the BPT diagnostic for the BPT scheme. To ensure well-constrained classifications, we also require \texttt{RX\_warning\_frac} < 0.10 and \texttt{BPT\_warning\_frac} < 0.10 (see Section \ref{sec:full_approach} for details). For this test, we use the radio excess diagnostic with D23 estimates of $A_{V, \mathrm{SED}}$ from \texttt{\textsc{Prospector}} and the H$\beta$ flux (even though H$\alpha$ measurements are available for all sources), alongside the $\mathcal{M}$Ex diagnostic, again considering only sources with \texttt{RX\_warning\_frac}<0.10 and \texttt{$\mathcal{M}$Ex\_warning\_frac}<0.10. In both schemes, we further remove sources that do not satisfy the SED goodness of fit and redshift criteria. 

The overall agreement between the BPT and $\mathcal{M}$Ex classification schemes is 96 per cent considering the sources that are classified $>90$ per cent in both the BPT and $\mathcal{M}$Ex schemes; however, a more detailed evaluation across the different classes, including the unclassified sources, is presented in the confusion matrix in Figure \ref{fig:conf_lowz_highz} (see also the location of these in the BPT and $\mathcal{M}$Ex diagram in Appendix \ref{appendix:compare}). Here we present the agreement rate as the percentage of sources commonly classified according to the $\mathcal{M}$Ex scheme (as denoted by the colourbar), with the subset that meets the 90 per cent threshold for both schemes shown in brackets (i.e. excluding sources that are unclassified by the BPT and $\mathcal{M}$Ex scheme). The number of sources in each cell is also included for reference.

In the first column, we observe strong agreement between the SFG classifications, regardless of whether the unclassified sources are included or excluded. We note that the 13.6 per cent that are unclassified in the BPT scheme, but are identified in the $\mathcal{M}$Ex scheme populate the region close to the Ka03 line or the radio-excess demarcation line from section \ref{sec:radio_excess}. A similar trend is found for the RQ AGN (second column), where the BPT unclassified sources are found near the demarcation lines in the BPT or $L^{\mathrm{corr}}_{\mathrm{H}\alpha}$-$L_{\mathrm{150\mathrm{MHz}}}$ diagnostic. We also find that 2.3(2.7) percent of SFGs identified in the $\mathcal{M}$Ex scheme are classified as RQ AGN according to the BPT scheme. This likely arises from our use of the lower curve from J14 to distinguish between SF and AGN contributions. While the MEx diagnostic is based on the BPT diagram, \cite{Juneau2011} found that the MEx intermediate region (the area between the lower and upper MEx curve where composite sources lie) includes not only \texttt{BPT\_Composite} sources, but also a significant number of \texttt{BPT\_SFG} sources (albeit with slightly different demarcation lines). This is also found to be the case with the offset introduced in section \ref{sec:MEx_diagnostic} (see Appendix \ref{appendix:MEx} for a discussion). 
However, we use this demarcation line to maintain consistency with the BPT classification approach, where the composite sources are assigned to the AGN class.
In the third column, we can see that 76.7 (100) per cent of the LERG population is consistently identified in both schemes, where the 23.3 per cent unclassified in the BPT are again found near the RX demarcation line when using the Balmer-derived $A_{V}$ instead of the $A_{V, \mathrm{SED}}$ from \texttt{\textsc{Prospector}}. The HERG classifications are also found to be in good agreement (although note that this result is based on low-number statistics), where the 3 HERGs that are classified according to MEx but unclassified in the BPT are similarly located near the demarcation lines in the RX or BPT diagram. Finally, both schemes are unable to classify 77.1 per cent of sources. The differences among the remaining classified sources -- where the BPT can assign a class but the $\mathcal{M}$Ex cannot -- arise primarily from the additional diagnostic power provided by the [\textsc{Nii}]/H$\alpha$ ratio in the BPT diagram, which directly traces ionisation mechanisms, and enables a less ambiguous separation between star-forming and AGN-dominated systems than the stellar mass alone.

These results show that the $\mathcal{M}$Ex is a good substitute for the BPT at higher redshift, where H$\alpha$ and [\textsc{Nii}] fall outside the wavelength coverage of the DESI spectrograph. Furthermore, given that the sources classified by $\mathcal{M}$Ex but unclassified by BPT lie near the diagnostic boundaries, we argue that it is reasonable to apply the $\mathcal{M}$Ex in such cases to increase the classified fraction of the low-$z$ sample.

\begin{figure}
    \centering
    \includegraphics[width=1\columnwidth]{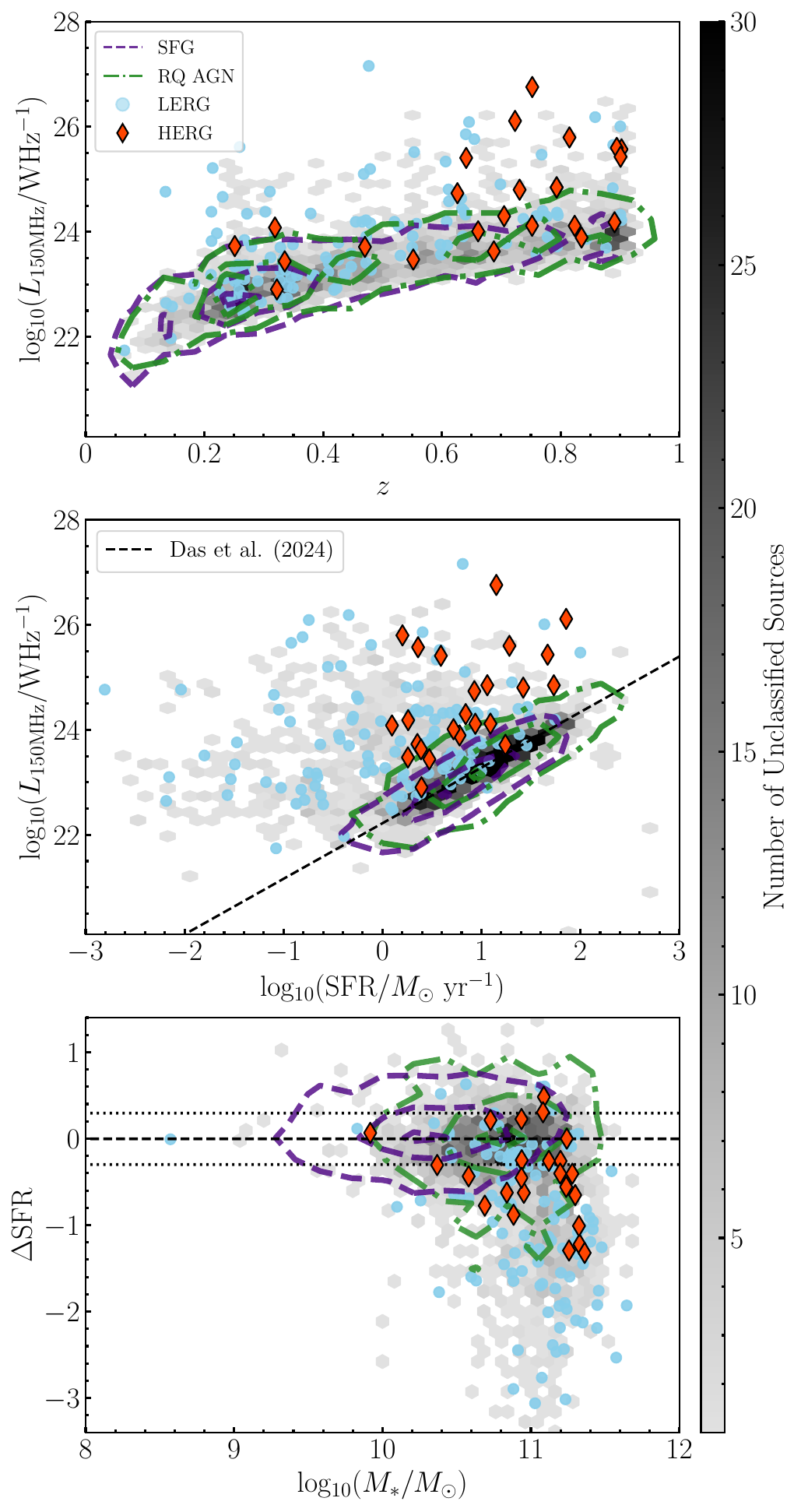}
    \caption{The physical properties of each class of radio source aligning with our expectations. \textbf{Top:} the $z-L_{150\mathrm{MHz}}$ plane for all sources in a given class with >90 per cent reliability, as indicated in the legend at the top left corner. \textbf{Middle:} the $L_{150\mathrm{MHz}}$--SFR plane, where the dashed line indicates the stellar mass independent $L_{150\mathrm{MHz}}$--SFR relation from D24. \textbf{Bottom:} the SFR relative to that expected for galaxies on the star formation rate - stellar mass relation ($\Delta$SFR), derived from using the redshift dependent relation from \citet{schreiber2015herschel}, as a function of stellar mass. The dashed line indicates the main-sequence, whereas the dotted lines denote a scatter of $\pm0.3$\,dex. The contour levels in each panel for the SFG and RQ AGN population are selected to encompass 5, 50 and 90 per cent of the corresponding sample. The grey scale and colourbar indicate the distribution of unclassified sources in each panel. The SFR and stellar mass estimates are taken from the value-added catalogue from D24.}
    \label{fig:phys_prop}
\end{figure}

\subsection{The physical properties of the sources by classification}\label{sec:properties}

To investigate the physical properties of the individual classes and thus validate our spectral classification scheme, we once again use the SED fitting catalogue from D24. This catalogue includes not only stellar mass estimates used for the high-$z$ classification, but also SFRs for all sources in our sample. To ensure reliability of results, we again consider only sources that satisfy the SED goodness of fit and redshift criteria\footnote{Note that these cuts were previously applied only to sources classified with the $\mathcal{M}$Ex scheme.} due to the use of $A_{V,\mathrm{SED}}$ and $M_{*}$ in the classification. Applying a 90 per cent confidence threshold to our spectroscopic classifications, we are able to identify a total of 1659 SFGs, 475 RQ AGN, 120 LERGs and 24 HERGs.

In the top panel of Figure \ref{fig:phys_prop}, we show the $z$ vs $L_{150\mathrm{MHz}}$ plane for each class, highlighting trends characteristic of their respective categories (e.g. \citealt{gurkan2018lofar}; \citealt{hardcastle2019}). We can see that SFGs (shown in purple contours) are more prevalent at lower redshifts ($z~<~0.75$) and moderate radio powers ($\log_{10}L_\mathrm{150MHz}~<~24$). With increasing redshift, however, the RQ AGN population (green contours) becomes more prominent, while LERGs (blue circles) and HERGs (red diamonds) are found to be associated with the highest $L_{150\mathrm{MHz}}$ at any given redshift (albeit with some overlap).

In the middle panel of Figure \ref{fig:phys_prop}, we further compare the SFR against the 150\,MHz luminosity. We can see that the SFG class lies tightly around the $L_{150\mathrm{MHz}}$-SFR relation (denoted by the dotted line which is taken from the mass independent relation in D24) as expected for galaxies whose emission is dominated by star-forming processes. Similarly to the finding by Dr24, the RQ AGN are found to largely overlap with the SFGs. However, our sample does not extend to low SFR values ($\log_{10}\mathrm{SFR} \lesssim 0$), where D24 and Dr24 observe an increased scatter above the $L_{150\mathrm{MHz}}$–SFR relation, potentially indicative of small-scale jets or highly-obscured star formation. 
The majority of LERGs and HERGs, on the other hand, lie as expected above this relation, which gives us confidence in our spectral fitting and classification schemes as the SFRs are independently estimated. We note that $\sim30$ per cent of the unclassified sources that lie above the D24 relation satisfy the 90 per cent threshold in our radio excess diagnostic, but not in the BPT or MEx classifications. These sources can therefore be considered genuine radio-excess AGN, even though they cannot be confidently assigned to the LERG or HERG classes. While this work focuses primarily on distinguishing between LERGs and HERGs, this population can be further investigated by users, as the relevant radio-excess information is provided in the accompanying catalogue.
The rest of the unclassified sources are near the boundary of our diagnostic, discussed in section \ref{sec:radio_excess}, or in the SFG region, likely a result of inconsistencies between the SFR estimates based on SED fitting and H$\alpha$.

In the bottom panel of Figure \ref{fig:phys_prop}, we also examine the difference between the estimated SFR and that expected if the sources lie on the star formation rate-stellar mass relation, also known as the `main-sequence' ($\Delta$SFR), computed using the redshift-dependent relation from \citeauthor{schreiber2015herschel} (\citeyear{schreiber2015herschel}; converted to the adopted IMF in D24), as a function of stellar mass. We can see once more that the SFG class follows the trend expected for star-forming galaxies, where the majority of sources ($\sim80$ per cent) are found to lie within the typical $\pm0.3$\,dex of the main-sequence relation (e.g. \citealt{tacchella2016confinement}). RQ AGNs occupy a similar parameter space to SFGs, with some scatter below the main sequence, as observed by D24 and Dr24. 
In contrast, $\sim$70 per cent of LERGs fall below the main sequence (i.e. $\Delta$SFR < –0.3), indicating suppressed star formation for their stellar mass, reinforcing the idea that radio-mode AGN may play a role in star formation quenching (e.g. \citealt{heckman2014coevolution}; \citealt{comerford2020}; \citealt{Magliocchetti2022}; \citealt{jin2025radioAGN}). However, the rest lie within the main sequence, corresponding to the star-forming LERG population identified in \citet{kondapally2022lofar, kondapally2024}, suggesting that not all LERGs are hosted by quiescent galaxies. These two populations will be investigated further in a future work (Arnaudova et al. in prep).
Furthermore, all three AGN classes are found to populate the plane at higher stellar masses ($\gtrsim10^{10}$\,M$_{\odot}$), which is consistent with previous findings suggesting that the most massive galaxies host a radio AGN, even if it is at relatively low radio powers (e.g. \citealt{gurkan2018lofar}; \citealt{sabater2019lotss}; \citealt{kondapally2022lofar}). 
Overall, our classification aligns well with the physical characteristics expected for each class.
 
\begin{figure*}
    \centering
    \includegraphics[width=0.95\textwidth]{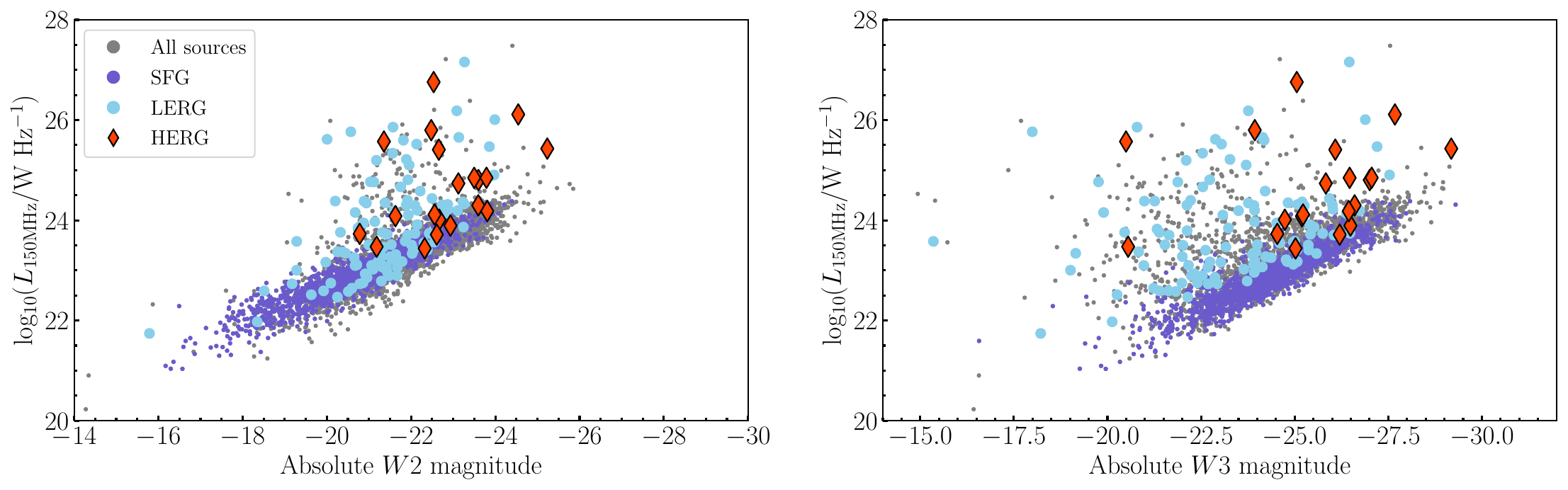}
    \caption{ The radio luminosity as a function of $W2$ (left) and $W3$ (right) absolute magnitude for sources spectroscopically classified at 90 per cent confidence. The different classes are shown in the legend in the upper left corner.}
    \label{fig:WISE_L150}
\end{figure*}

\subsection{The mid-IR/radio relation}

Further validation of our method can be accomplished by examining the distribution of SFGs and radio-excess galaxies (RXGs; LERGs and HERGs) on the mid-IR–radio luminosity plane, as was done by Dr24 (for a direct comparison with the classification from Dr24, see Appendix \ref{appendix:compare}) and further explored in \cite{hardcastle2025}. To do this, we use the AGN Host Galaxies Physical Properties value-added catalogue \citep{Siudek2024}, which provides mid-IR data at 3.4, 4.6, 12, and 22$\mu$m from the Wide-field Infrared Survey Explorer (WISE; \citealt{wright2010}). For the subsequent analysis, we note that 864 sources do not have a $W3$ detection (SNR\_$W3<3$); these sources are not excluded but their magnitudes are used as an upper limit. Although deeper IR data are available for our sample, we use the WISE bands to enable a more direct comparison with Dr24.

Figure \ref{fig:WISE_L150} displays the 150\,MHz luminosity versus $W2$ (left panel) and $W3$ (right panel) luminosities (or absolute magnitudes). The latter were calculated following the method described in \cite{hardcastle2023}, where a simple power-law extrapolation was used between $W1$ and $W2$, and $W2$ and $W3$, respectively. In the left panel, we can see that the SFGs and RXGs occupy distinct regions: SFGs follow a clear correlation, while LERGs and HERGs appear at higher $L_{150\mathrm{MHz}}$ and $W2$ values. Since the $W2$ band serves as a proxy for stellar mass, the distribution for SFGs reflects the main-sequence relation. In contrast, RXGs are found in more massive galaxies (as seen in Figure \ref{fig:phys_prop}) and exhibit an excess of radio luminosity beyond what is expected from star-forming processes, aligning with expectations and results from Dr24.
The right panel is also similar to that obtained for the wide-area LoTSS sources in Dr24; we find a tight correlation for SFGs, with RXGs residing above this correlation (albeit with some overlap). Here the $W3$ band traces the dust luminosity, such that we are observing the well-known far-infrared/radio correlation for star-forming galaxies (e.g. \citealt{Yun2001}; \citealt{smith2014}; \citealt{read2018}). While conceptually similar to the results in the previous section, these findings are derived from independent mid-infrared measurements (as opposed to model-dependent SED fitting), offering a complementary perspective based on directly observed photometric properties. As such, they provide further validation of our classification approach.

\begin{figure*}
    \centering
    \includegraphics[width=0.95\textwidth]{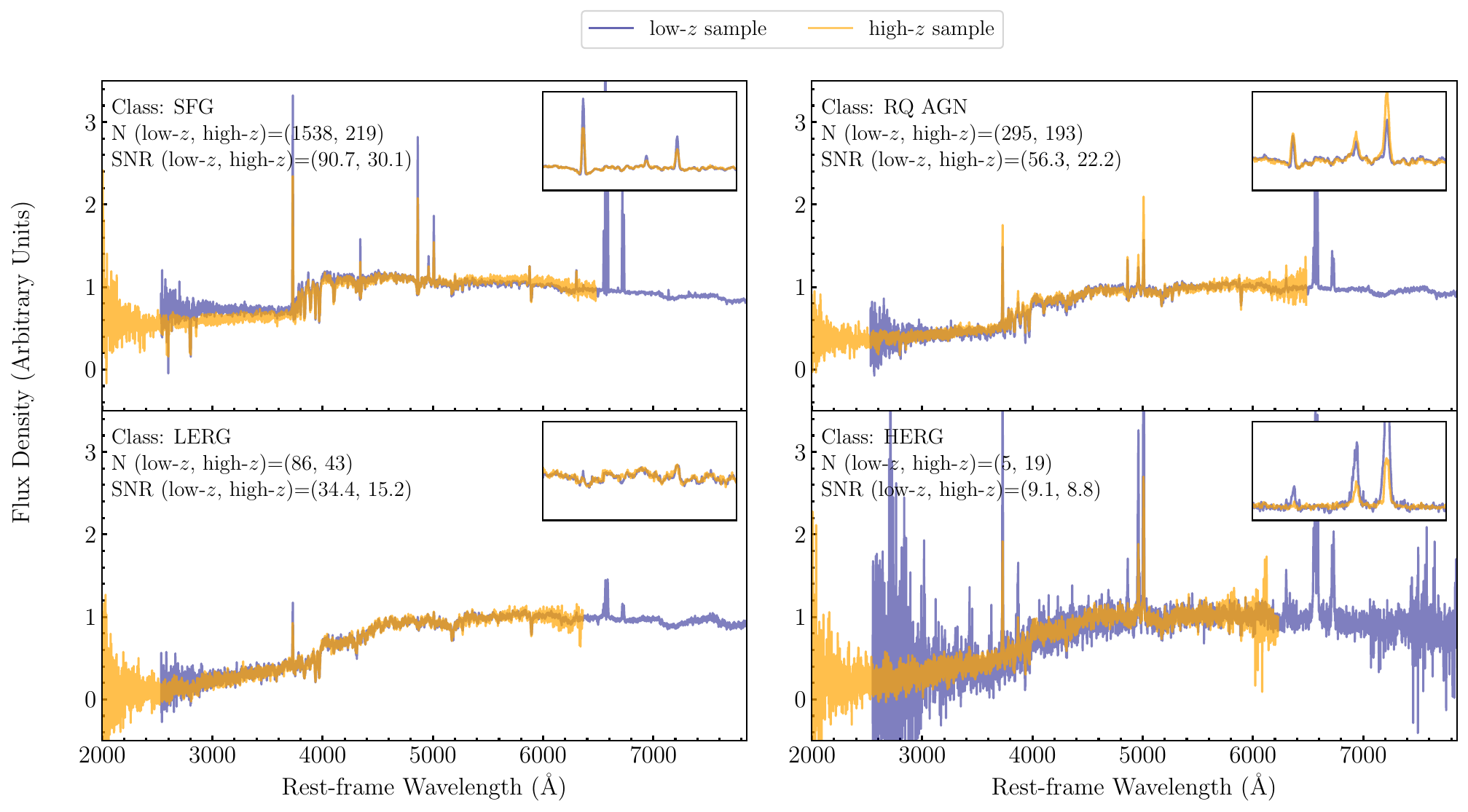}
    \caption{The composite spectrum of each class with $>90$  per cent confidence for the low-$z$ (blue) and high-$z$ (orange) sample. The class type, number of sources in each stack and its median S/N are indicated in the top left corner. The inset in each panel provides a zoomed-in view of the $4800 < \lambda < 5100$ \AA\ range, enabling a clearer comparison of the emission lines between the low-$z$ and high-$z$ stacks.}
    \label{fig:stacks}
\end{figure*}

\subsection{The average spectra of each class}\label{sec:stacks}

In this section, we further evaluate the average properties of each class by examining the composite spectrum of each population. To achieve this, we use \textsc{\texttt{SpecStacker}}: The Spectral Stacking code\footnote{https://github.com/m-arnaudova/SpecStacker}, which is fully described by \cite{arnaudova2024}. To summarise, all spectra within a given class are shifted to the rest frame and resampled onto a common wavelength grid. The spectra are then normalized by taking the median in the reddest region, corresponding to 10 per cent of the wavelength grid, where all spectra contribute and prominent emission lines are masked out. Finally, a composite spectrum is created by calculating the median flux at each rest-frame wavelength pixel. We note that while applying an inverse variance-weighted stacking method can yield higher-SNR composite spectra (e.g. \citealt{Rigby2018}), as explained in \cite{Calabro2021}, this tends to bias the result toward spectra with higher individual SNRs, which is why we have chosen to adopt median stacking. The uncertainties are estimated through bootstrapping, with an additional correction applied to account for spectral resolution changes during the de-redshifting process, as well as to address potential issues in the normalisation of each spectrum. 

Figure \ref{fig:stacks} shows the composite spectra for each class in the low-$z$ (blue) and high-$z$ (orange) samples, with an inset zoomed in on the H$\beta$ and [\textsc{Oiii}] emission lines. We can see that the composite spectra of the SFG class for both the low-$z$ and high-$z$ sample contain a bright blue continuum with stronger Balmer lines compared to forbidden lines, as expected for star-forming galaxies. The composite spectra of the RQ AGN class are also associated with a relatively blue continuum, but with stronger [\textsc{Oiii}]\,$\lambda\lambda$4959, 5007 than H$\beta$. The prominent [\textsc{Oii}]\,$\lambda\lambda$3726, 3728 emission is intriguing as it can be used as a star formation tracer, even in the presence of an AGN (e.g. \citealt{maddox2018}; \citealt{arnaudova2024}). This suggests that these sources may also have a significant SF contribution, as was demonstrated in the SFR-$M_{*}$ plane, where RQ AGN were found to occupy a similar parameter space to SFGs.  
The stacked spectra of LERGs and HERGs also exhibit features typical of these populations. The low-$z$ and high-$z$ HERGs show strong [\textsc{Oiii}] emission, typically associated with a high AGN bolometric luminosity (e.g. \citealt{best2012on}), along with additional high-ionisation lines. LERG spectra, on the other hand, are dominated by the underlying continuum emission, with very faint emission from [\textsc{Oii}], H$\alpha$, [\textsc{Nii}]\,$\lambda\lambda$6548, 6583 and [\textsc{Sii}]\,$\lambda\lambda$6716, 6732.

We note that significant selection effects arising from both the spectroscopic data (DESI) and the radio selection (LoTSS) complicate direct comparisons between the low- and high-$z$ samples. The low-$z$ sample is more representative of DESI’s BGS, while the high-$z$ sample is increasingly dominated by LRGs. The LoTSS selection adds further complexity: its sensitivity to both star-forming galaxies and AGN means that at low $z$, the radio-selected sample spans a broad range of activity types, whereas at high $z$ it becomes increasingly biased toward radio-bright AGN or massive SFGs. These combined effects mean that the stacked spectra at low and high redshift are not directly comparable and likely represent intrinsically different populations.
However, these stacks are the first high quality composite spectra representing the four classes up to $z\sim1$ and will serve as excellent templates for refining the redshift estimates of future spectroscopic surveys, such as WEAVE-LOFAR and ORCHIDSS. 

\begin{figure*}
    \centering
    \includegraphics[width=0.95\textwidth]{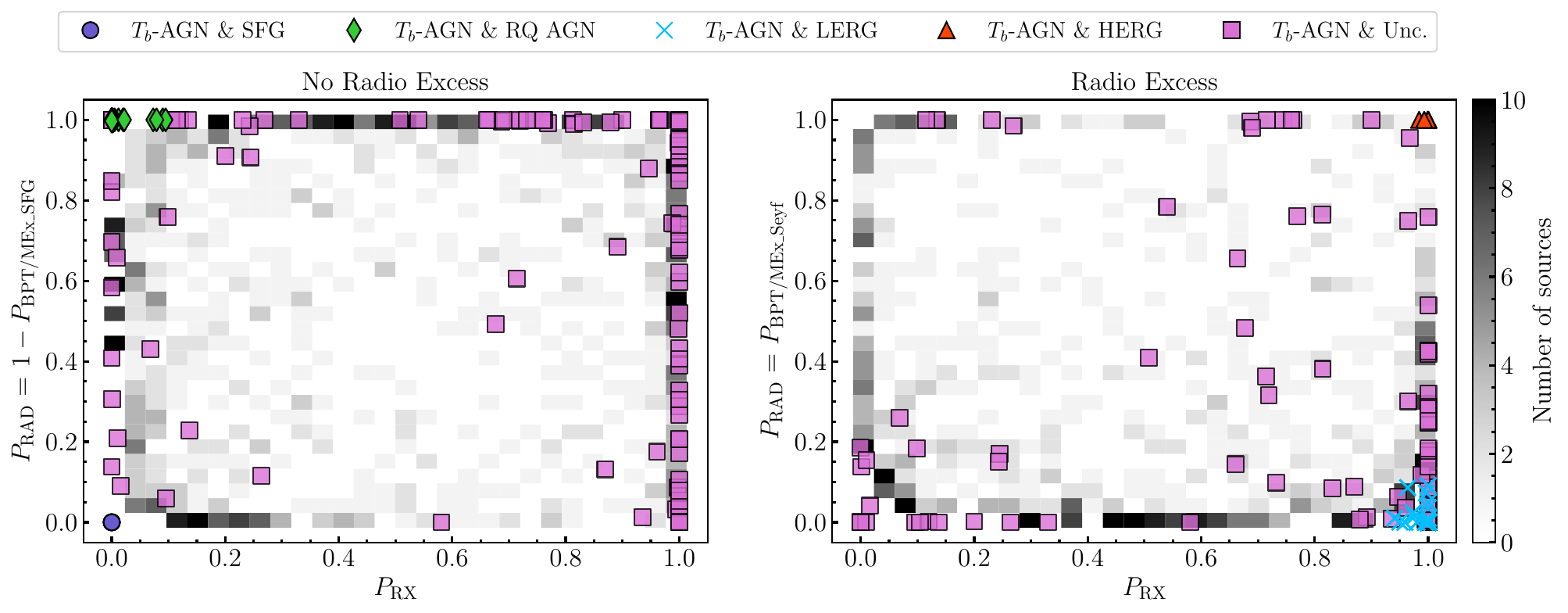}
    \caption{The probability of a source exhibiting a radio excess ($P_{\mathrm{RX}}$) versus the probability of it having a radiatively-efficient AGN ($P_{\mathrm{RAD}}$) for the brightness temperature-identified AGN ($T_{b}$-AGN). In the left panel, $P_{\mathrm{RAD}}$ is defined as the probability of a source being located outside the \texttt{BPT/$\mathcal{M}$Ex\_SFG} region, which is used for sources without a radio excess. While in the right panel, $P_{\mathrm{RAD}}$ is defined as the probability of a source being located inside the \texttt{BPT/$\mathcal{M}$Ex\_Seyf+CSeyf} region, as used for radio-excess sources, in order to select only HERGs and not LERGs. The $T_{b}$-AGN are further categorized based on our spectroscopic classification at $>90$ per cent confidence, as indicated by the legend at the top. The colourbar indicates the remainder of sources (that is not $T_{b}$-AGN) for both panels.}
    \label{fig:tb_comparison}
\end{figure*}

\subsection{The brightness temperature diagnostic}\label{sec:tb}

Finally, an interesting comparison can be made with the brightness temperature method used to identify AGN activity (\citealt{morabito2022identifying, morabito2025}) using sub-arcsecond imaging at 144\,MHz with the international LOFAR stations, now available for a subset of sources in ELAIS-N1 \citep{dejong2024}. This relies on the fact that star formation can only produce radio emission up to a certain maximum surface brightness (\citealt{condon1992radio}), and therefore has a corresponding upper limit on the brightness temperature ($T_b$), above which the emission is likely dominated by AGN activity. Using the same conservative assumptions as in \cite{morabito2022identifying}, \cite{morabito2025} calculated the $T_{b}$ for 5,245 sources in ELAIS-N1, out of which 3,503 are in our sample and 175 are identified as $T_{b}$-AGN. Comparing this with our 90 per cent spectroscopic classification, we find that these $T_{b}$-AGN are also classified as 4 SFGs, 23 RQ AGN, 25 LERGs, and 8 HERGs.

To determine the nature of the $T_{b}$-AGN unclassified in our sample (``$T_{b}$-AGN \& Unc.''), we examine their distribution in terms of the probability of a source having a radio excess ($P_{\mathrm{RX}}$) versus the probability of being a radiatively-efficient AGN ($P_{\mathrm{RAD}}$; see Figure \ref{fig:tb_comparison}). We define $P_{\mathrm{RX}}$ as the probability of a source having a $\log L_{150\mathrm{MHz}}-\log L_{\mathrm{H\alpha}}^{\mathrm{corr}}>14.86$ (see section \ref{sec:radio_excess}). Due to our classification scheme, we define $P_{\mathrm{RAD}}$ as the probability of a source being located outside the \texttt{BPT/$\mathcal{M}$Ex\_SFG} region ($1~-~P_{\mathrm{BPT/\mathcal{M}Ex\_SFG}}$) for sources without a radio excess (left panel) and as the probability of a source falling in the \texttt{BPT/$\mathcal{M}$Ex\_Seyf+CSeyf} region 
($P_{\mathrm{BPT/\mathcal{M}Ex\_Seyf}}$) for sources with a radio excess (right panel). We further present the $T_{b}$-AGN classified in our sample as a reference; ``$T_{b}$-AGN \& SFG'' and ``$T_{b}$-AGN \& RQ AGN'' sources by definition have $P_{\mathrm{RX}}$<0.10, and \texttt{BPT/$\mathcal{M}$Ex\_SFG}<0.10 and \texttt{BPT/$\mathcal{M}$Ex\_SFG}>0.90, respectively. Similarly, ``$T_{b}$-AGN \& LERG'' and ``$T_{b}$-AGN \& HERG'' sources are located at $P_{\mathrm{RX}}$>0.90, and are associated with $P_{\mathrm{BPT/\mathcal{M}Ex\_Seyf}}$<0.10 and $P_{\mathrm{BPT/\mathcal{M}Ex\_Seyf}}$>0.90, respectively.

The majority of ``$T_{b}$-AGN \& Unc.'' sources are found to exhibit either high 
$P_{\mathrm{RX}}$ values (similar to those of ``$T_{b}$-AGN \& LERG'' and ``$T_{b}$-AGN \& HERG'') or a high probability of falling outside the \texttt{BPT/$\mathcal{M}$Ex\_SFG} or inside the 
\texttt{BPT/$\mathcal{M}$Ex\_Seyf+CSeyf} region (as seen in ``$T_{b}$-AGN \& RQ AGN'' or ``$T_{b}$-AGN \& HERGs''). However, these probabilities fall just below our 90 per cent classification threshold, which is why these sources remain unclassified. However, there are also ``$T_{b}$-AGN \& Unc.'' sources with low $P_{\mathrm{RE}}$ but high $P_{\mathrm{BPT/MEx\_SFG}}$ and low $P_{\mathrm{BPT/\mathcal{M}Ex\_Seyf}}$ values. These sources along with the presence of ``$T_{b}$-AGN \& SFGs'' are likely due to a compact, sub-dominant AGN that is detected by LOFAR’s long baselines but remain undetected in optical spectroscopy. Examination of the 0.3\arcsec resolution radio images reveals that all four sources exhibit extended, not jet-like emission, supporting the interpretation that they are likely composite systems hosting both star formation and AGN activity. 
A similar scenario is also found for the ``$T_{b}$-AGN \& RQ AGN'' sources, where we find that 10 out of 23 exhibit a flux deficit at 0.3\arcsec compared to the total flux in the 6\arcsec\ LoTSS maps. This suggests that some of the emission is extended and thus might be associated with star formation, potentially reinforcing the idea that AGN as well as star formation contribute to 150\,MHz emission in RQ AGN (see \citealt{panessa2019} for a review).

While this comparison overall supports our classification approach, it also highlights the necessity of high-resolution radio imaging to uncover AGN populations that are otherwise missed by spectroscopic diagnostics.

\begin{figure*}
    \centering
    \includegraphics[width=1\textwidth]{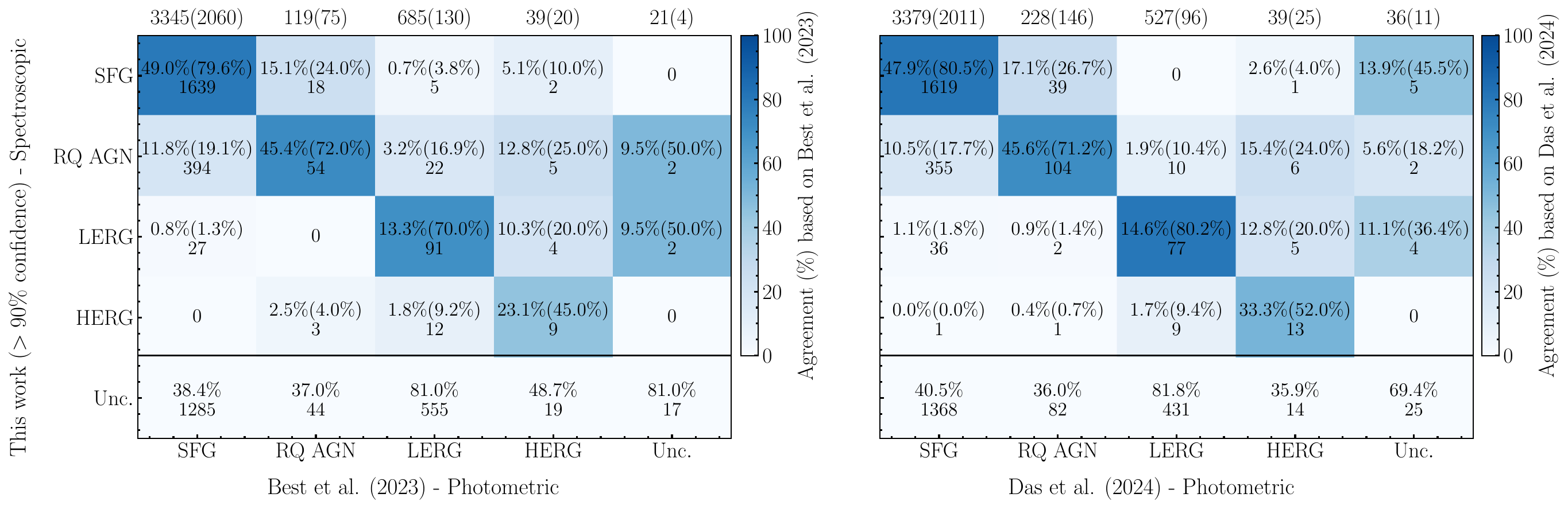}
    \caption{Confusion matrices comparing the 90 per cent spectroscopic classification (rows) with the photometric classifications by B23 (columns in the left panel), and D24 (columns in the right). Each cell represents the level of agreement based on the photometric classifications in percentages, where we have also included the number of sources in each cell. The percentages in brackets are calculated for all sources that reach our 90 per cent threshold, which is also indicated by the colour bar. The total number of sources in each class per photometric classification (and those that are all classified in our 90 per cent classification) is shown at the top. }
    \label{fig:conf_90}
\end{figure*}

\section{Comparison with photometric classifications}\label{sec:compare_SED}

\begin{figure*}
    \centering
    \includegraphics[width=0.95\textwidth]{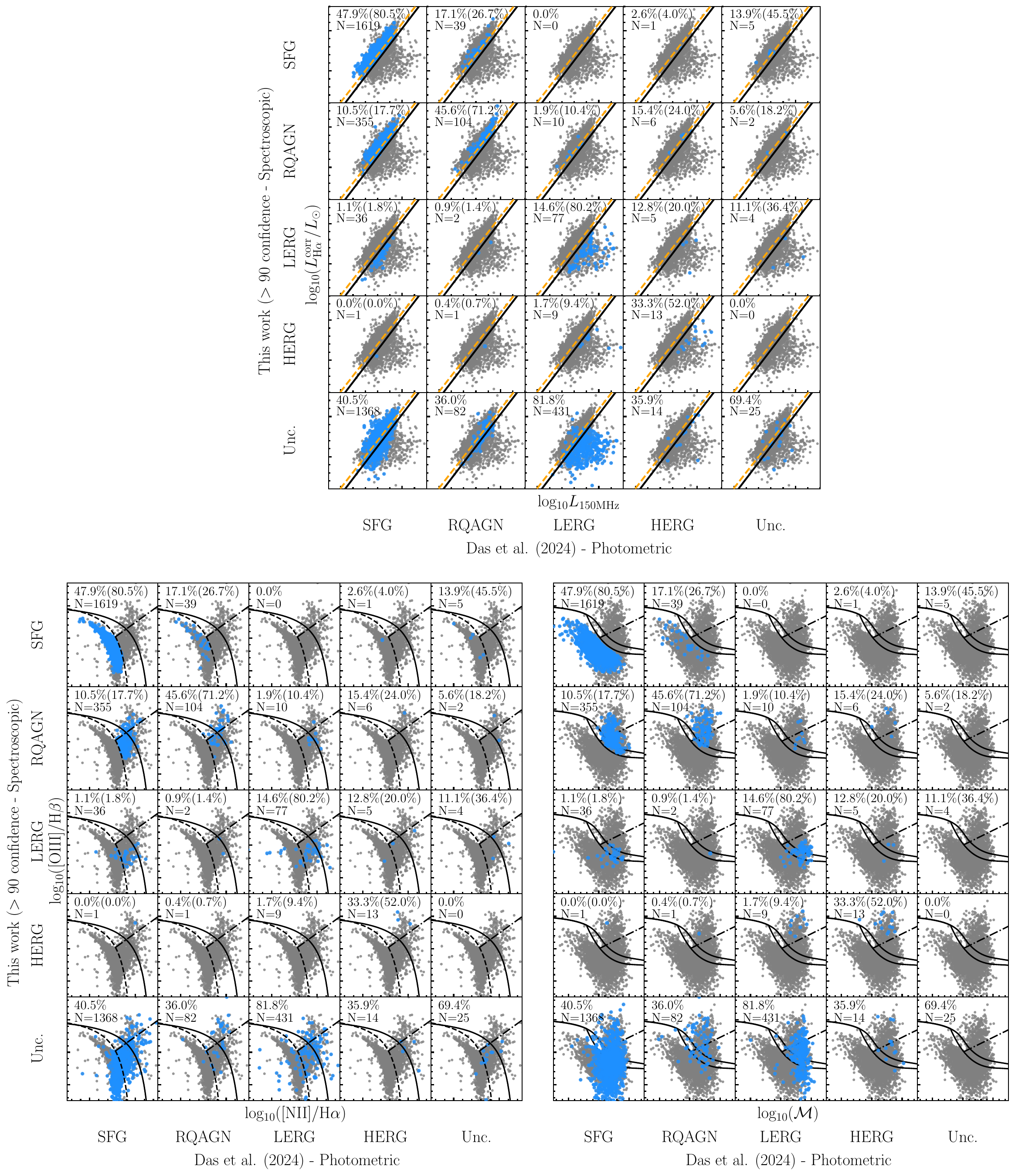}
    \caption{The location of the sources in the radio-excess diagram (top), the BPT diagram (bottom left) and the MEx diagram (bottom right) for each cell of the confusion matrix, comparing the 90 per cent classification with D24 (blue points). The grey dots in each diagram represent the total number of sources in the relevant samples. The level of agreement compared to D24, as well as the number of sources in each cell are given in the top left corner. The demarcation lines, labels and ranges are the same as in Figure \ref{fig:spec_class}, where the translated radio-excess definition from D24 (dashed orange line) is also included in the top panel. }
    \label{fig:conf_diag}
\end{figure*}

\subsection{Overall results}
To evaluate the performance of using photometric SED fitting to classify radio sources, we compare the results of our highest purity classification (with $>90$ per cent reliability) with those derived in ELAIS-N1 using panchromatic photometry with state-of-the-art models from B23 and D24. Since our sample is a subset of those included in B23 and D24, we are able to make a direct comparison. 
We make this comparison initially by calculating the change in the percentage of sources classified into each class, both overall, and (in parenthesis) if we consider only the subset of sources that exceed our 90 per cent classification threshold, using the numbers from Table \ref{tab:Number_sources}. It is clear that there are significant differences between the spectroscopic and photometric classifications of the same sources; compared to B23, we observe $\sim50$ (17) per cent fewer SFGs, 227 (495) per cent more RQ AGN, and 82 per cent fewer  LERGs. However, the number of LERGs becomes comparable when considering only those exceeding our high-reliability threshold. Similarly, when compared to D24, there are about 50 (15) per cent fewer SFGs, 86 (209) per cent more RQ AGN, 76 per cent fewer LERGs, but 34 per cent more when removing the spectroscopically unclassified sources. For the HERG class, although affected by small-number statistics, we find $\sim44$ per cent fewer HERGs relative to both B23 and D24. However, when applying the 90 per cent reliability cut, our counts are 20 per cent higher than B23 and similar to D24.

The overall agreement with B23 and D24 is $\sim 42$ per cent comparing all sources, and $\sim77$ per cent when considering only those that meet out 90 per cent criterion (i.e. not including the sources in our unclassified branch). But, to investigate this further and make a more detailed comparison, we construct confusion matrices for our 90 per cent sample with respect to B23 and D24, as shown in Figure \ref{fig:conf_90}, where we again consider only the sources that satisfy the redshift criteria. The results are presented as percentages, showing the level of agreement when considering the photometric classifications as the true values, by including all sources and those that only reach our 90 per cent threshold. The number of sources in each cell is also included for reference. Considering the percentages associated with the whole sample, we can see that a large number of sources classified in B23 and D24 are unclassified according to our classification scheme (bottom row). This is a result of our high reliability threshold ($>90$ per cent), which ensures strong confidence in the classification, unlike B23 and D24, which aimed to provide a best-estimate classification for as many sources as possible irrespective of the confidence in the results. Therefore, we discuss the remainder of this analysis by disregarding the sources in our unclassified branch (i.e. we consider the results in brackets). Focusing on the first two columns, we find 79.6 and 80.5 per cent agreement for the SFGs, and 72 and 71.2 per cent for RQ AGN, respectively in B23 and D24. There appears to be a large discrepancy between the SFG and RQ AGN classes, where 19.1 and 17.7 per cent of the photometrically classified SFGs in B23 and D24, respectively, fall into our RQ AGN class, and 24 and 26.7 per cent vice versa. Note that 20 per cent in the first column corresponds to about 400 sources in both classifications, which is associated with the relatively large number of RQ AGN in our spectroscopic classification.  
Moving to the third column, we find 70 and 80.2 per cent agreement with the LERGs identified in B23 and D24, where 16.9 and 10.4 per cent of the photometrically-classified LERGs fall into our RQ AGN class. For the sources classified as HERGs in B23 and D24, we find 45 and 52 per cent agreement, with 25 and 24 per cent classified as RQ AGN, and 20 per cent for both works classified as LERGs in our scheme (although note the low number of HERGs). 
The reasons behind these differences are investigated in the following section.

\begin{figure*}
    \centering
    \includegraphics[trim={0cm 2.4cm 0cm 0cm},clip,width=1\textwidth]{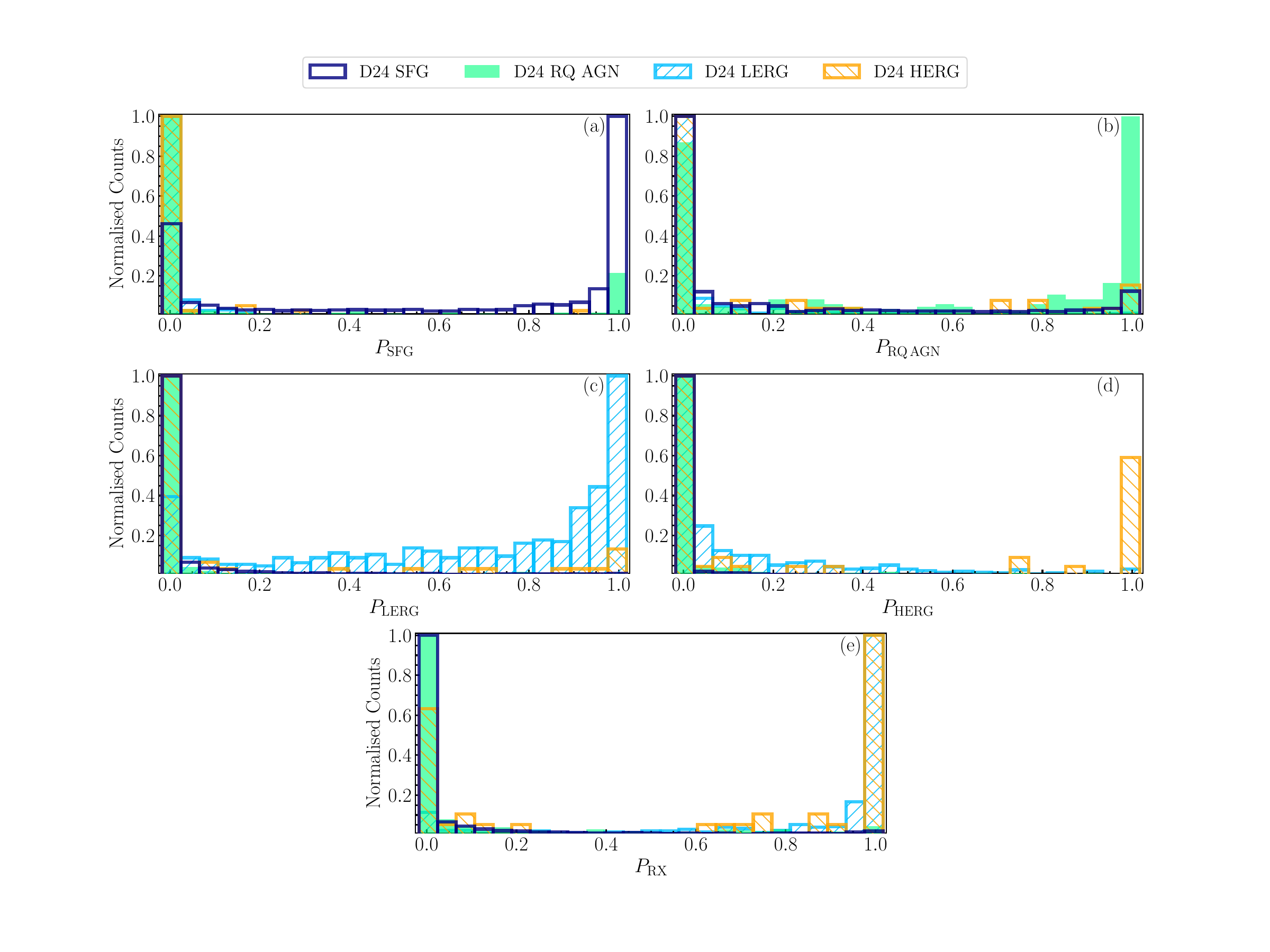}
    \caption{The probability distribution of SFGs (panel a), RQ AGN (panel b), LERGs (panel c), HERGs (panel d), and radio-excess sources (panel e), as defined by our probabilistic spectroscopic method based on the RX and BPT diagnostics for sources at $0<z<0.483$, and on the RX and $\mathcal{M}$Ex diagnostics for sources at $0.483<z<0.947$. Each panel displays the corresponding distribution for sources photometrically classified by D24, as indicated in the legend at the top.}
    \label{fig:prob_hist}
\end{figure*}

\subsection{The photometric classifications in our classification scheme}\label{sec:phot_diag}

To understand the differences between the spectroscopic and photometric classifications, we investigate the location of the sources in our diagnostic diagrams. This is shown in Figure \ref{fig:conf_diag}, where we have overlaid all sources in the radio excess (top panel), as well as the BPT (bottom left) and $\mathcal{M}$Ex (bottom right) diagrams, depending on which diagnostic is applicable to each source, for each cell of the confusion matrix from Figure \ref{fig:conf_90}. Here, we focus only on D24 classification since the overall results are comparable with B23.

In the top panel, we have included the radio excess definition from D24 (dashed orange line), which is based on a 3$\sigma$ offset from the SFR-$L_{150\mathrm{MHz}}$ mass independent relation, where we have used the $L_{\mathrm{H}\alpha}$-SFR relation from \cite{kennicutt1998} to overlay it in our diagram. As can be seen in the figure, this definition matches roughly with our demarcation line (black solid line). Therefore, we expect to have comparable results for identifying radio excess sources. While this is found to be the case for the majority of sources, there is a subsample (e.g. third and fifth row of the first column) for which this is not the case. This could be due to the fact that the SFR estimates from \texttt{\textsc{Prospector}} probe longer timescales than those traced by H$\alpha$, and therefore may not directly reflect the same SF activity (see comparison in Appendix~\ref{sec:SFR}). This results in the misclassification of LERGs as SFGs, and similarly HERGs as RQ AGN, although in both cases the numbers are small.

Moving to the bottom panels, we can start to better understand the large discrepancies with the RQ AGN class. In both the BPT (left) and $\mathcal{M}$Ex diagram (right) we find higher ionisation sources which are classified as SFG or LERGs in the photometric classifications. This may arise because photometric classifications, although using information across the electromagnetic spectrum, lack the fine resolution needed to identify a radiatively-efficient AGN, especially when the galaxy's overall SED may be dominated by SF activity. In contrast, the higher resolution of spectroscopy allows for a more direct probe of the ionization state of the gas, providing a clearer signature of an AGN activity. This may have an effect not only on the separation of RQ AGN and SFGs (e.g second row, first column), but also on LERGs and HERGs, as can be seen in the fourth row of the third column of the bottom right panel (see also section \ref{sec:accretion_rates} for a discussion of how photometric classification can lead to incorrect interpretations, particularly in distinguishing between the accretion properties of LERGs and HERGs). In addition, the inability of SED fitting to identify high-ionisation lines, combined with the differences between the SFR used could explain why some RQ AGN are photometrically classified as LERGs.
However, we also identify sources that populate the \texttt{BPT\_SFG} and \texttt{$\mathcal{M}$Ex\_SFG} region, which are classified as RQ AGN in photometry, suggesting that these may be obscured AGN. Such sources can be identified with SED fitting through access to mid-IR information, which is sensitive to the reprocessed emission from the AGN's torus (e.g. see \citealt{williams2018lofar}; B23 and D24 for details).

\subsection{The spectroscopic probabilities of the photometric classifications}

In the previous sections, we examined the photometric classifications based on our high-purity spectroscopic classification (with $>90$ per cent confidence). We now consider the distribution of classification probabilities for sources photometrically classified in each class.

Figure \ref{fig:prob_hist} shows the distribution of probabilities derived from our probabilistic spectroscopic method for the photometric classes in D24. In panel (a), we present the probability of a source being spectroscopically classified as an SFG ($P_{\mathrm{SFG}}$) across the four classes in D24. We observe that D24 SFGs (dark blue) correctly exhibit high probabilities, with low values of $P_{\mathrm{SFG}}$ for LERGs (light blue), HERGs (red), and some RQ AGN (green) classified in D24. However, there are also RQ AGN associated with a high probability of them being classified as an SFG ($P_{\mathrm{SFG}}$>0.9). These sources likely indicate the mid-IR AGN that our spectroscopic method cannot identify, which we previously discussed in section \ref{sec:phot_diag}.
Panel (b) displays the spectroscopic probability distribution for RQ AGN ($P_{\mathrm{RQAGN}}$), where we can see that not only RQ AGN exhibit high $P_{\mathrm{RQAGN}}$, but also some D24 SFGs and D24 HERGs. These D24 SFGs could be associated with radiatively-efficient AGN that show strong forbidden lines, which cannot be easily picked up by SED fitting methods, while the presence of D24 HERGs at high $P_{\mathrm{RQAGN}}$ likely point to differences between the photometrically-derived SFR and those inferred from H$\alpha$, affecting the radio-excess diagnostic.

In panel (c), we show the spectroscopic probability distribution that a source is a LERG ($P_{\mathrm{LERG}}$), where it is evident that the D24 LERGs are generally associated with high $P_{\mathrm{LERG}}$. However, unlike the previous two cases, their distribution is more spread out, indicating greater uncertainty within the LERG population; this is probably due to the typically weaker emission lines in LERGs, leading to larger uncertainties in their location in the emission line diagnostics. The D24 SFGs and D24 RQ AGN are associated with low $P_{\mathrm{LERG}}$, but while the majority of HERGs are also found at low $P_{\mathrm{LERG}}$, a fraction of them lie at $P_{\mathrm{LERG}}$ > 0.8, suggesting that some HERGs may be misclassified as LERGs. The probability distribution of HERGs presented in panel (d) is less ambiguous, as generally only D24 HERGs are associated with high $P_{\mathrm{HERG}}$ values. However, we can still see that the majority of D24 HERGs are actually associated with $P_{\mathrm{HERG}}$ = 0. These sources are likely associated with strong forbidden lines that SED fitting cannot identify, or arise from inconsistencies between the SFR estimates based on SED fitting and H$\alpha$.

Finally, panel (e) shows the probability distribution for radio-excess sources ($P_{\mathrm{RX}}$). The D24 LERGs and D24 HERGs are correctly found to be the only populations associated with high values of $P_{\mathrm{RX}}$, although there are still some HERGs with $P_{\mathrm{RX}}$ = 0. The presence of HERGs at low $P_{\mathrm{RX}}$ values is again likely due to the differences in the SFRs.

These comparisons show that photometric methods alone cannot accurately classify the radio source population. While spectroscopy enables the identification of additional AGN through high-ionisation emission lines, our method has been unable to identify mid-IR AGN, as well as compact radio AGN detectable only through sub-arcsecond radio imaging (see section~\ref{sec:tb}). Ultimately, a comprehensive census of the radio source population will require a combination of optical spectroscopy and multi-wavelength SED fitting (Das et al. \textit{in prep.}), alongside high-resolution radio imaging.

\subsection{Implications of using photometric classifications}\label{sec:accretion_rates}

In the previous sections, we saw that photometric classifications have difficulties in identifying radiatively-efficient AGN, which resulted in the misidentification of RQ AGN as SFG and LERGs as HERGs. We are interested in seeing whether this can explain recent results using photometric classifications showing that the Eddington-scaled accretion rates ($\lambda_{\mathrm{Edd}}$) between LERGs and HERGs are no longer distinct at higher redshift and at lower radio powers (\citealt{whittam2022mightee}). 

To do this, we define $\lambda_{\mathrm{Edd}}$ as the ratio between the total energetic output of the black hole, which includes the bolometric radiative ($L_{\mathrm{bol}}$) and jet mechanical luminosity ($L_{\mathrm{\mathrm{mech}}}$), and the Eddington luminosity ($L_{\mathrm{Edd}}$). 
For the bolometric luminosity, we consider the $L_{\mathrm{bol}}=3500L_{\mathrm{[OIII]}}$ relation from \cite{heckman2004}, where $L_{\mathrm{[OIII]}}$ is the [\textsc{Oiii}] luminosity. For the 74 LERGs where [\textsc{Oiii}]$\lambda$5007 is not significantly detected (i.e. SNR$<3$), we impose a 3$\sigma$ upper limit. To estimate $L_{\mathrm{mech}}$, we use the relationship from \cite{cavagnolo2010}, $L_{\mathrm{mech}}=7.3\times10^{36}(L_{1.4\mathrm{GHz}}/10^{24}\mathrm{WHz}^{-1})^{0.7}$W, where we assume a spectral index of $\alpha=-0.7$ to convert the $L_{150\mathrm{MHz}}$ to the 1.4\,GHz radio luminosity ($L_{1.4\mathrm{GHz}}$). Finally, the Eddington luminosity is calculated as $L_{\mathrm{Edd}} = 1.3\times10^{31}M_{\mathrm{BH}}/M_{\odot}$W, where $M_{\mathrm{BH}}$ is the black hole mass. For consistency with \cite{whittam2022mightee}, we use the stellar mass - black hole relation from \cite{haring2004}, for which we use $M_{*}$ from D24. 

\begin{figure}
    \centering
    \includegraphics[width=1\columnwidth]{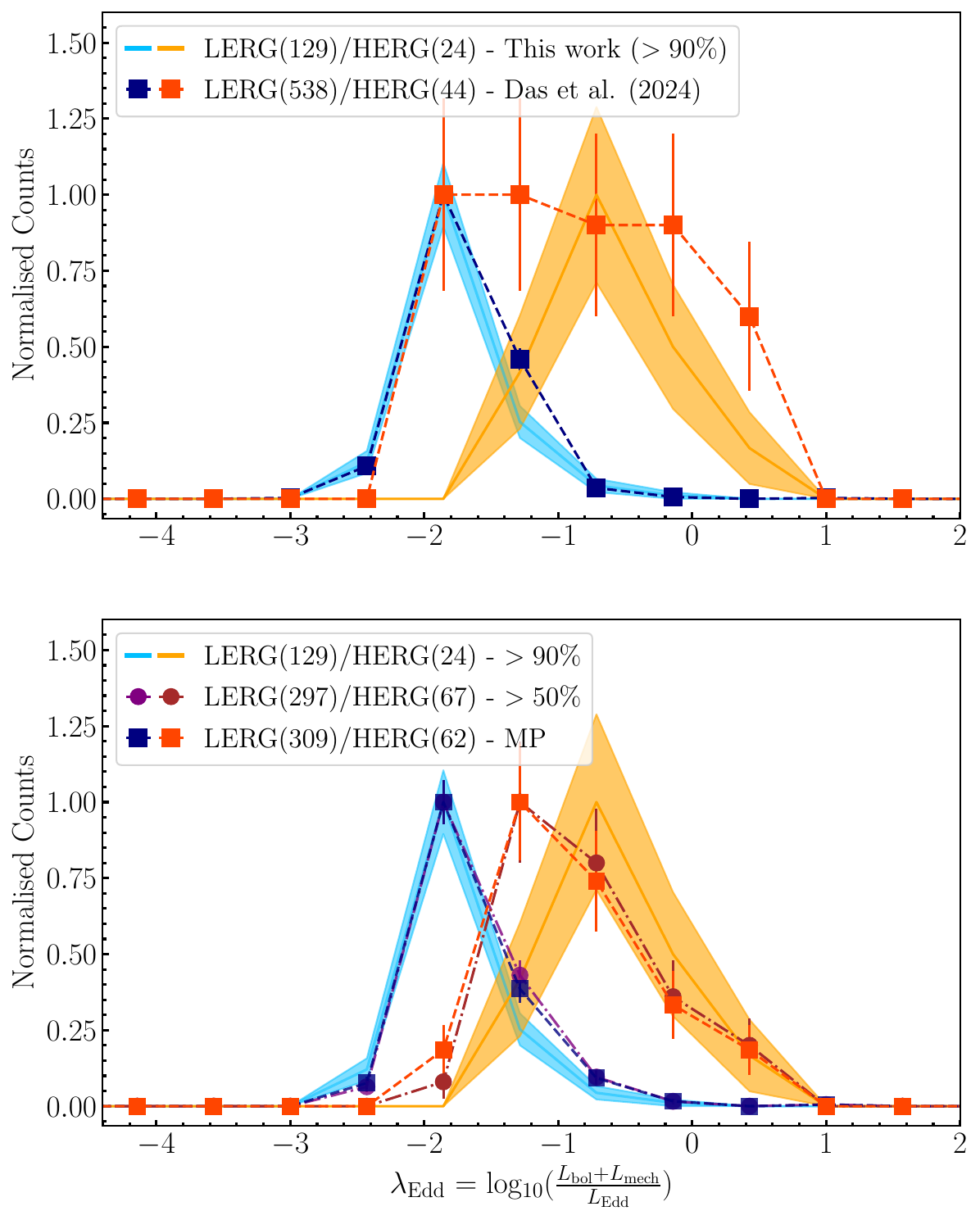}
    \caption{The distribution of Eddington-scaled accretion rates for LERGs and HERGs. The top panel compares the results obtained with our spectroscopic classification at $>90$ per cent confidence with the classifications from D24, as indicated by the legend in the top left corner. Similarly, the bottom panel compares the accretion rate distributions for LERGs and HERGs classified at $>90\%$, $>50\%$ threshold and using the most probable classification, as indicated by the legend. The number of sources in each class is included in the brackets, while the errorbars represent Poisson uncertainties. }
    \label{fig:accretion rates}
\end{figure}

The Eddington-scaled accretion rate distributions for the LERGs and HERGs classified in our spectroscopic samples, as well as the corresponding values for sources photometrically classified by D24 are presented in the top panel of Figure \ref{fig:accretion rates}. We can see that the spectroscopically-classified LERGs and HERGs are clearly separated, where the HERG distribution peaks at $\lambda_{\mathrm{Edd}}\sim0.1$, while the LERG distribution is shifted towards lower values, as expected (e.g. \citealt{best2012on}; \citealt{mingo2014}). Considering the LERGs and HERGs classified according to D24, however, we can see that the distributions are less distinct. The D24 LERG distribution is similar to LERGs identified in this work, but the D24 HERGs spreads towards lower $\lambda_{\mathrm{Edd}}$ values. This could potentially explain the results seen in \cite{whittam2022mightee}, although we note that our sample spans a significantly smaller range in redshifts (only up to $z\sim1$) than was probed in that work. 

In the bottom panel of Figure \ref{fig:accretion rates}, we also investigate the accretion rate distribution of LERGs and HERGs when applying a less stringent threshold to our classifications, by lowering the requirement from 90 to 50 per cent, as well as considering the most probable class (MP classification\footnote{Note that the number of HERGs identified using the MP classification is lower than in the $50\%$ classification, as the latter combines both BPT and MEx diagnostics for the low-$z$ sample, while the MP classification relies solely on the BPT diagram at $0<z<0.483$.}) in a manner similar in intent to the photometric classifications. Compared to the 90 per cent classification, the HERG distribution in both the 50 per cent and MP classifications is shifted towards lower values of $\lambda_{\mathrm{Edd}}$, while the LERG distribution remains largely unchanged. Nevertheless, LERGs and HERGs continue to exhibit distinct accretion rate distributions. Similar results are obtained by considering the MP classifications for secure radio-excess sources (i.e. those with RE>0.90), however, the current sample size is too small to allow for a meaningful statistical analysis.

These results underscore the importance of robust, probabilistic spectroscopic classifications in reliably separating AGN accretion modes. While photometric classifications and lower-threshold probabilistic approaches offer broader completeness, they risk introducing contamination between populations, particularly by misclassifying HERGs with lower accretion rates. Our analysis highlights that even under more relaxed criteria, the underlying distinction between LERGs and HERGs persists, but is more clearly resolved when high-confidence spectroscopic diagnostics are used.  This work will be extended in Arnaudova et al. (\textit{in prep.}), where this probabilistic spectroscopic framework will be applied to the LoTSS wide-area sample, which now consists of $\sim350,000$ radio sources with spectroscopic data up to $z=1$ from DESI-DR1, enabling a more statistically powerful investigation of the AGN accretion properties.

\section{Summary}\label{sec:ch4_summary}

In this work, we have combined the first data release of the LOFAR Two-Metre Sky Survey Deep Fields (LoTSS) with the early data release of the Dark Energy Spectroscopic Instrument (DESI) survey to spectroscopically classify sources as star-forming galaxies (SFGs), radio-quiet AGN (RQ AGN), low-excitation radio galaxies (LERGs), and high-excitation radio galaxies (HERGs) in the ELAIS-N1 field. To achieve this, we have performed spectral fitting on 5,109 sources using the stellar population library from \cite{bruzual2003stellar}, combined with one or two Gaussian components per emission line, to extract the necessary emission line information for the classification scheme. Expanding on the probabilistic method introduced by Dr24, we have used a combination of a radio excess and the BPT diagnostic, along with our newly developed modified MEx ($\mathcal{M}$Ex) diagnostic to classify 4,471 sources, by generating multiple Monte Carlo realizations. Due to the probabilistic nature of our approach, we are able to produce high purity (with $>90$ per cent confidence) and more complete classifications (assigning all sources to their most likely class) out to $z=1$, nearly doubling the redshift range of Dr24.

Using the 90 per cent reliability sample, we identified 1,757 SFGs, 488 RQ AGN, 129 LERGs and 24 HERGs. These classes are found to populate the expected regions of the $z$ vs $L_{150\mathrm{MHz}}$, SFR vs $L_{150\mathrm{MHz}}$, $M_{*}$ vs SFR and the mid-IR/radio luminosity plane, produce composite spectra with characteristic features, and comparable results with the brightness temperature method, thus validating our classification method. Comparing these results to the photometric classifications from B23 and D24 for the same sample, we find up to $\sim77$ per cent agreement, though with some notable discrepancies. For some sources, differences in the timescales probed by H$\alpha$ and photometric SFR estimates can lead to variations in the identification of LERGs as SFGs and HERGs as RQ AGN, or vice versa. However, this affects only a small subsample. A greater discrepancy involves the identification of radiatively-efficient AGN, where we find that the higher-resolution nature of spectroscopy allows us to identify such objects more efficiently, particularly within the RQ AGN class, where we find $\times5$ and  $\times2.1$ more sources than B23 and D24, respectively. Additionally, we find that this issue may impact the separation between LERGs and HERGs, causing their accretion properties to appear less distinct (as observed by \citealt{whittam2022mightee}).

However, a potential limitation of our spectroscopic approach is the difficulty to detect obscured AGN, which can be revealed through SED fitting, as well as compact AGN that are primarily detected with high-resolution radio imaging. This will be addressed in future works (e.g., Das et al. \textit{in prep.}), where SED fitting combining photometry and spectroscopy will provide a more comprehensive classification of the radio source population. Ongoing and future high-resolution imaging with \texttt{LOFAR2.0} will also enhance our ability to identify compact AGN missed by other methods. Furthermore, with the launch of the WEAVE-LOFAR survey (\citealt{smith2016weave}), we will obtain over a million spectra of sources identified in the LOFAR surveys, including not only complete spectroscopy of sources in the LoTSS Deep Fields but also large samples in the wide-field LoTSS areas. This dramatic increase in sample size will allow us to probe the interplay between star formation and AGN activity, as well as the different accretion modes, thereby providing us with a deeper and more complete understanding of the low-frequency radio population. Additionally, investigating whether photometric classifications can be improved based on the spectroscopic results presented here will be an important avenue for future work, particularly in the era of the SKA, where a vast number of radio sources will remain without spectroscopic observations.

\bibliographystyle{mnras}
\bibliography{ref} 

\begin{thebibliography}{}
\makeatletter
\relax
\def\mn@urlcharsother{\let\do\@makeother \do\$\do\&\do\#\do\^\do\_\do\%\do\~}
\def\mn@doi{\begingroup\mn@urlcharsother \@ifnextchar [ {\mn@doi@} {\mn@doi@[]}}
\def\mn@doi@[#1]#2{\def\@tempa{#1}\ifx\@tempa\@empty \href {http://dx.doi.org/#2} {doi:#2}\else \href {http://dx.doi.org/#2} {#1}\fi \endgroup}
\def\mn@eprint#1#2{\mn@eprint@#1:#2::\@nil}
\def\mn@eprint@arXiv#1{\href {http://arxiv.org/abs/#1} {{\tt arXiv:#1}}}
\def\mn@eprint@dblp#1{\href {http://dblp.uni-trier.de/rec/bibtex/#1.xml} {dblp:#1}}
\def\mn@eprint@#1:#2:#3:#4\@nil{\def\@tempa {#1}\def\@tempb {#2}\def\@tempc {#3}\ifx \@tempc \@empty \let \@tempc \@tempb \let \@tempb \@tempa \fi \ifx \@tempb \@empty \def\@tempb {arXiv}\fi \@ifundefined {mn@eprint@\@tempb}{\@tempb:\@tempc}{\expandafter \expandafter \csname mn@eprint@\@tempb\endcsname \expandafter{\@tempc}}}

\bibitem[\protect\citeauthoryear{{Abolfathi} et~al.,}{{Abolfathi} et~al.}{2018}]{Abolfathi2018sdss}
{Abolfathi} B.,  et~al., 2018, \mn@doi [\apjs] {10.3847/1538-4365/aa9e8a}, \href {https://ui.adsabs.harvard.edu/abs/2018ApJS..235...42A} {235, 42}

\bibitem[\protect\citeauthoryear{{Arnaudova} et~al.,}{{Arnaudova} et~al.}{2024a}]{arnaudova2024}
{Arnaudova} M.~I.,  et~al., 2024a, \mn@doi [\mnras] {10.1093/mnras/stae233}, \href {https://ui.adsabs.harvard.edu/abs/2024MNRAS.528.4547A} {528, 4547}

\bibitem[\protect\citeauthoryear{{Arnaudova} et~al.,}{{Arnaudova} et~al.}{2024b}]{arnaudova2024b}
{Arnaudova} M.~I.,  et~al., 2024b, \mn@doi [\mnras] {10.1093/mnras/stae2235}, \href {https://ui.adsabs.harvard.edu/abs/2024MNRAS.535.2269A} {535, 2269}

\bibitem[\protect\citeauthoryear{{Baldwin}, {Phillips}  \& {Terlevich}}{{Baldwin} et~al.}{1981}]{baldwin1981}
{Baldwin} J.~A.,  {Phillips} M.~M.,   {Terlevich} R.,  1981, \mn@doi [\pasp] {10.1086/130766}, \href {https://ui.adsabs.harvard.edu/abs/1981PASP...93....5B} {93, 5}

\bibitem[\protect\citeauthoryear{{Balokovi{\'c}}, {Smol{\v{c}}i{\'c}}, {Ivezi{\'c}}, {Zamorani}, {Schinnerer}  \& {Kelly}}{{Balokovi{\'c}} et~al.}{2012}]{balokovic2012disclosing}
{Balokovi{\'c}} M.,  {Smol{\v{c}}i{\'c}} V.,  {Ivezi{\'c}} {\v{Z}}.,  {Zamorani} G.,  {Schinnerer} E.,   {Kelly} B.~C.,  2012, \mn@doi [\apj] {10.1088/0004-637X/759/1/30}, \href {https://ui.adsabs.harvard.edu/abs/2012ApJ...759...30B} {759, 30}

\bibitem[\protect\citeauthoryear{{Begelman}, {Blandford}  \& {Rees}}{{Begelman} et~al.}{1984}]{begelman1984theory}
{Begelman} M.~C.,  {Blandford} R.~D.,   {Rees} M.~J.,  1984, \mn@doi [Reviews of Modern Physics] {10.1103/RevModPhys.56.255}, \href {https://ui.adsabs.harvard.edu/abs/1984RvMP...56..255B} {56, 255}

\bibitem[\protect\citeauthoryear{{Bernardi}, {Sheth}, {Tundo}  \& {Hyde}}{{Bernardi} et~al.}{2007}]{bernardi2007selection}
{Bernardi} M.,  {Sheth} R.~K.,  {Tundo} E.,   {Hyde} J.~B.,  2007, \mn@doi [\apj] {10.1086/512719}, \href {https://ui.adsabs.harvard.edu/abs/2007ApJ...660..267B} {660, 267}

\bibitem[\protect\citeauthoryear{{Best} \& {Heckman}}{{Best} \& {Heckman}}{2012}]{best2012on}
{Best} P.~N.,  {Heckman} T.~M.,  2012, \mn@doi [\mnras] {10.1111/j.1365-2966.2012.20414.x}, \href {https://ui.adsabs.harvard.edu/abs/2012MNRAS.421.1569B} {421, 1569}

\bibitem[\protect\citeauthoryear{{Best}, {Kauffmann}, {Heckman}, {Brinchmann}, {Charlot}, {Ivezi{\'c}}  \& {White}}{{Best} et~al.}{2005}]{best2005}
{Best} P.~N.,  {Kauffmann} G.,  {Heckman} T.~M.,  {Brinchmann} J.,  {Charlot} S.,  {Ivezi{\'c}} {\v{Z}}.,   {White} S.~D.~M.,  2005, \mn@doi [\mnras] {10.1111/j.1365-2966.2005.09192.x}, \href {https://ui.adsabs.harvard.edu/abs/2005MNRAS.362...25B} {362, 25}

\bibitem[\protect\citeauthoryear{{Best} et~al.,}{{Best} et~al.}{2023}]{best2023lofar}
{Best} P.~N.,  et~al., 2023, \mn@doi [\mnras] {10.1093/mnras/stad1308}, \href {https://ui.adsabs.harvard.edu/abs/2023MNRAS.523.1729B} {523, 1729}

\bibitem[\protect\citeauthoryear{{Boyle} \& {Terlevich}}{{Boyle} \& {Terlevich}}{1998}]{boyle1998cosmological}
{Boyle} B.~J.,  {Terlevich} R.~J.,  1998, \mn@doi [\mnras] {10.1046/j.1365-8711.1998.01264.x}, \href {https://ui.adsabs.harvard.edu/abs/1998MNRAS.293L..49B} {293, L49}

\bibitem[\protect\citeauthoryear{{Brinchmann}, {Charlot}, {White}, {Tremonti}, {Kauffmann}, {Heckman}  \& {Brinkmann}}{{Brinchmann} et~al.}{2004}]{brinchmann2004physical}
{Brinchmann} J.,  {Charlot} S.,  {White} S.~D.~M.,  {Tremonti} C.,  {Kauffmann} G.,  {Heckman} T.,   {Brinkmann} J.,  2004, \mn@doi [\mnras] {10.1111/j.1365-2966.2004.07881.x}, \href {https://ui.adsabs.harvard.edu/abs/2004MNRAS.351.1151B} {351, 1151}

\bibitem[\protect\citeauthoryear{{Bruzual} \& {Charlot}}{{Bruzual} \& {Charlot}}{2003}]{bruzual2003stellar}
{Bruzual} G.,  {Charlot} S.,  2003, \mn@doi [\mnras] {10.1046/j.1365-8711.2003.06897.x}, \href {https://ui.adsabs.harvard.edu/abs/2003MNRAS.344.1000B} {344, 1000}

\bibitem[\protect\citeauthoryear{{Buttiglione}, {Capetti}, {Celotti}, {Axon}, {Chiaberge}, {Macchetto}  \& {Sparks}}{{Buttiglione} et~al.}{2010}]{Buttiglione2010optical}
{Buttiglione} S.,  {Capetti} A.,  {Celotti} A.,  {Axon} D.~J.,  {Chiaberge} M.,  {Macchetto} F.~D.,   {Sparks} W.~B.,  2010, \mn@doi [\aap] {10.1051/0004-6361/200913290}, \href {https://ui.adsabs.harvard.edu/abs/2010A&A...509A...6B} {509, A6}

\bibitem[\protect\citeauthoryear{{Calabr{\`o}} et~al.,}{{Calabr{\`o}} et~al.}{2021}]{Calabro2021}
{Calabr{\`o}} A.,  et~al., 2021, \mn@doi [\aap] {10.1051/0004-6361/202039244}, \href {https://ui.adsabs.harvard.edu/abs/2021A&A...646A..39C} {646, A39}

\bibitem[\protect\citeauthoryear{{Calzetti}, {Armus}, {Bohlin}, {Kinney}, {Koornneef}  \& {Storchi-Bergmann}}{{Calzetti} et~al.}{2000}]{calzetti2000dust}
{Calzetti} D.,  {Armus} L.,  {Bohlin} R.~C.,  {Kinney} A.~L.,  {Koornneef} J.,   {Storchi-Bergmann} T.,  2000, \mn@doi [\apj] {10.1086/308692}, \href {https://ui.adsabs.harvard.edu/abs/2000ApJ...533..682C} {533, 682}

\bibitem[\protect\citeauthoryear{{Carnall}}{{Carnall}}{2017}]{carnall2017spectres}
{Carnall} A.~C.,  2017, \mn@doi [arXiv e-prints] {10.48550/arXiv.1705.05165}, \href {https://ui.adsabs.harvard.edu/abs/2017arXiv170505165C} {p. arXiv:1705.05165}

\bibitem[\protect\citeauthoryear{{Cavagnolo}, {McNamara}, {Nulsen}, {Carilli}, {Jones}  \& {B{\^\i}rzan}}{{Cavagnolo} et~al.}{2010}]{cavagnolo2010}
{Cavagnolo} K.~W.,  {McNamara} B.~R.,  {Nulsen} P.~E.~J.,  {Carilli} C.~L.,  {Jones} C.,   {B{\^\i}rzan} L.,  2010, \mn@doi [\apj] {10.1088/0004-637X/720/2/1066}, \href {https://ui.adsabs.harvard.edu/abs/2010ApJ...720.1066C} {720, 1066}

\bibitem[\protect\citeauthoryear{{Chabrier}}{{Chabrier}}{2003}]{chabrier2003}
{Chabrier} G.,  2003, \mn@doi [\pasp] {10.1086/376392}, \href {https://ui.adsabs.harvard.edu/abs/2003PASP..115..763C} {115, 763}

\bibitem[\protect\citeauthoryear{{Cid Fernandes}, {Stasi{\'n}ska}, {Schlickmann}, {Mateus}, {Vale Asari}, {Schoenell}  \& {Sodr{\'e}}}{{Cid Fernandes} et~al.}{2010}]{cidfernandes2010alternative}
{Cid Fernandes} R.,  {Stasi{\'n}ska} G.,  {Schlickmann} M.~S.,  {Mateus} A.,  {Vale Asari} N.,  {Schoenell} W.,   {Sodr{\'e}} L.,  2010, \mn@doi [\mnras] {10.1111/j.1365-2966.2009.16185.x}, \href {https://ui.adsabs.harvard.edu/abs/2010MNRAS.403.1036C} {403, 1036}

\bibitem[\protect\citeauthoryear{{Cirasuolo}, {Magliocchetti}, {Celotti}  \& {Danese}}{{Cirasuolo} et~al.}{2003a}]{cirasuolo2003radio}
{Cirasuolo} M.,  {Magliocchetti} M.,  {Celotti} A.,   {Danese} L.,  2003a, \mn@doi [\mnras] {10.1046/j.1365-8711.2003.06485.x}, \href {https://ui.adsabs.harvard.edu/abs/2003MNRAS.341..993C} {341, 993}

\bibitem[\protect\citeauthoryear{{Cirasuolo}, {Celotti}, {Magliocchetti}  \& {Danese}}{{Cirasuolo} et~al.}{2003b}]{cirasuolo2003there}
{Cirasuolo} M.,  {Celotti} A.,  {Magliocchetti} M.,   {Danese} L.,  2003b, \mn@doi [\mnras] {10.1046/j.1365-2966.2003.07105.x}, \href {https://ui.adsabs.harvard.edu/abs/2003MNRAS.346..447C} {346, 447}

\bibitem[\protect\citeauthoryear{{Cleri} et~al.,}{{Cleri} et~al.}{2023}]{cleri2023}
{Cleri} N.~J.,  et~al., 2023, \mn@doi [\apj] {10.3847/1538-4357/acc1e6}, \href {https://ui.adsabs.harvard.edu/abs/2023ApJ...948..112C} {948, 112}

\bibitem[\protect\citeauthoryear{{Cochrane} et~al.,}{{Cochrane} et~al.}{2023}]{cocrane2023lofar}
{Cochrane} R.~K.,  et~al., 2023, \mn@doi [\mnras] {10.1093/mnras/stad1602}, \href {https://ui.adsabs.harvard.edu/abs/2023MNRAS.523.6082C} {523, 6082}

\bibitem[\protect\citeauthoryear{{Coil} et~al.,}{{Coil} et~al.}{2015}]{coil2015}
{Coil} A.~L.,  et~al., 2015, \mn@doi [\apj] {10.1088/0004-637X/801/1/35}, \href {https://ui.adsabs.harvard.edu/abs/2015ApJ...801...35C} {801, 35}

\bibitem[\protect\citeauthoryear{{Comerford} et~al.,}{{Comerford} et~al.}{2020}]{comerford2020}
{Comerford} J.~M.,  et~al., 2020, \mn@doi [\apj] {10.3847/1538-4357/abb2ae}, \href {https://ui.adsabs.harvard.edu/abs/2020ApJ...901..159C} {901, 159}

\bibitem[\protect\citeauthoryear{{Condon}}{{Condon}}{1992}]{condon1992radio}
{Condon} J.~J.,  1992, \mn@doi [\araa] {10.1146/annurev.aa.30.090192.003043}, \href {https://ui.adsabs.harvard.edu/abs/1992ARA&A..30..575C} {30, 575}

\bibitem[\protect\citeauthoryear{DESI~Collaboration Aghamousa et~al.,}{DESI~Collaboration et~al.}{2016a}]{desi2016desiI}
DESI~Collaboration Aghamousa A.,  et~al., 2016a, \mn@doi [arXiv e-prints] {10.48550/arXiv.1611.00036}, \href {https://ui.adsabs.harvard.edu/abs/2016arXiv161100036D} {p. arXiv:1611.00036}

\bibitem[\protect\citeauthoryear{DESI~Collaboration Aghamousa et~al.,}{DESI~Collaboration et~al.}{2016b}]{desi2016desiII}
DESI~Collaboration Aghamousa A.,  et~al., 2016b, \mn@doi [arXiv e-prints] {10.48550/arXiv.1611.00037}, \href {https://ui.adsabs.harvard.edu/abs/2016arXiv161100037D} {p. arXiv:1611.00037}

\bibitem[\protect\citeauthoryear{{DESI Collaboration} et~al.,}{{DESI Collaboration} et~al.}{2023}]{desi2023early}
{DESI Collaboration} et~al., 2023, \mn@doi [arXiv e-prints] {10.48550/arXiv.2306.06308}, \href {https://ui.adsabs.harvard.edu/abs/2023arXiv230606308D} {p. arXiv:2306.06308}

\bibitem[\protect\citeauthoryear{{Das} et~al.,}{{Das} et~al.}{2024}]{das2024lofar}
{Das} S.,  et~al., 2024, \mn@doi [\mnras] {10.1093/mnras/stae1204}, \href {https://ui.adsabs.harvard.edu/abs/2024MNRAS.531..977D} {531, 977}

\bibitem[\protect\citeauthoryear{{Dom{\'\i}nguez} et~al.,}{{Dom{\'\i}nguez} et~al.}{2013}]{dominguez2013dust}
{Dom{\'\i}nguez} A.,  et~al., 2013, \mn@doi [\apj] {10.1088/0004-637X/763/2/145}, \href {https://ui.adsabs.harvard.edu/abs/2013ApJ...763..145D} {763, 145}

\bibitem[\protect\citeauthoryear{{Drake} et~al.,}{{Drake} et~al.}{2024}]{drake2024lofar}
{Drake} A.~B.,  et~al., 2024, \mn@doi [\mnras] {10.1093/mnras/stae2117}, \href {https://ui.adsabs.harvard.edu/abs/2024MNRAS.534.1107D} {534, 1107}

\bibitem[\protect\citeauthoryear{{Duncan} et~al.,}{{Duncan} et~al.}{2021}]{duncan2021lofar}
{Duncan} K.~J.,  et~al., 2021, \mn@doi [\aap] {10.1051/0004-6361/202038809}, \href {https://ui.adsabs.harvard.edu/abs/2021A&A...648A...4D} {648, A4}

\bibitem[\protect\citeauthoryear{{Duncan} et~al.,}{{Duncan} et~al.}{2023}]{duncan2023}
{Duncan} K.,  et~al., 2023, \mn@doi [The Messenger] {10.18727/0722-6691/5306}, \href {https://ui.adsabs.harvard.edu/abs/2023Msngr.190...25D} {190, 25}

\bibitem[\protect\citeauthoryear{{Escott} et~al.,}{{Escott} et~al.}{2025}]{Escott2025}
{Escott} E.~L.,  et~al., 2025, \mn@doi [\mnras] {10.1093/mnras/stae2645}, \href {https://ui.adsabs.harvard.edu/abs/2025MNRAS.536.1166E} {536, 1166}

\bibitem[\protect\citeauthoryear{{Ferrarese} \& {Ford}}{{Ferrarese} \& {Ford}}{2005}]{ferrarese2005supermassive}
{Ferrarese} L.,  {Ford} H.,  2005, \mn@doi [\ssr] {10.1007/s11214-005-3947-6}, \href {https://ui.adsabs.harvard.edu/abs/2005SSRv..116..523F} {116, 523}

\bibitem[\protect\citeauthoryear{{Ferrarese} \& {Merritt}}{{Ferrarese} \& {Merritt}}{2000}]{Ferrarese2000}
{Ferrarese} L.,  {Merritt} D.,  2000, \mn@doi [\apjl] {10.1086/312838}, \href {https://ui.adsabs.harvard.edu/abs/2000ApJ...539L...9F} {539, L9}

\bibitem[\protect\citeauthoryear{{Fitzpatrick}}{{Fitzpatrick}}{1999}]{fitzpatrick1999correcting}
{Fitzpatrick} E.~L.,  1999, \mn@doi [\pasp] {10.1086/316293}, \href {https://ui.adsabs.harvard.edu/abs/1999PASP..111...63F} {111, 63}

\bibitem[\protect\citeauthoryear{{Gebhardt} et~al.,}{{Gebhardt} et~al.}{2000}]{Gebhardt2000}
{Gebhardt} K.,  et~al., 2000, \mn@doi [\apjl] {10.1086/312840}, \href {https://ui.adsabs.harvard.edu/abs/2000ApJ...539L..13G} {539, L13}

\bibitem[\protect\citeauthoryear{{Girdhar} et~al.,}{{Girdhar} et~al.}{2022}]{girdhar2022quasar}
{Girdhar} A.,  et~al., 2022, \mn@doi [\mnras] {10.1093/mnras/stac073}, \href {https://ui.adsabs.harvard.edu/abs/2022MNRAS.512.1608G} {512, 1608}

\bibitem[\protect\citeauthoryear{{Graham}}{{Graham}}{2016}]{graham2016galaxy}
{Graham} A.~W.,  2016, in {Laurikainen} E.,  {Peletier} R.,   {Gadotti} D.,  eds,  Astrophysics and Space Science Library Vol. 418, Galactic Bulges. p.~263 (\mn@eprint {arXiv} {1501.02937}), \mn@doi{10.1007/978-3-319-19378-6_11}

\bibitem[\protect\citeauthoryear{{G{\"u}rkan} et~al.,}{{G{\"u}rkan} et~al.}{2018}]{gurkan2018lofar}
{G{\"u}rkan} G.,  et~al., 2018, \mn@doi [\mnras] {10.1093/mnras/sty016}, \href {https://ui.adsabs.harvard.edu/abs/2018MNRAS.475.3010G} {475, 3010}

\bibitem[\protect\citeauthoryear{{G{\"u}rkan} et~al.,}{{G{\"u}rkan} et~al.}{2019}]{gurkan2019lotss}
{G{\"u}rkan} G.,  et~al., 2019, \mn@doi [\aap] {10.1051/0004-6361/201833892}, \href {https://ui.adsabs.harvard.edu/abs/2019A&A...622A..11G} {622, A11}

\bibitem[\protect\citeauthoryear{{Guy} et~al.,}{{Guy} et~al.}{2023}]{guy2023spec}
{Guy} J.,  et~al., 2023, \mn@doi [\aj] {10.3847/1538-3881/acb212}, \href {https://ui.adsabs.harvard.edu/abs/2023AJ....165..144G} {165, 144}

\bibitem[\protect\citeauthoryear{{Hardcastle} \& {Croston}}{{Hardcastle} \& {Croston}}{2020}]{hardcastle2020radio}
{Hardcastle} M.~J.,  {Croston} J.~H.,  2020, \mn@doi [\nar] {10.1016/j.newar.2020.101539}, \href {https://ui.adsabs.harvard.edu/abs/2020NewAR..8801539H} {88, 101539}

\bibitem[\protect\citeauthoryear{{Hardcastle}, {Evans}  \& {Croston}}{{Hardcastle} et~al.}{2007}]{hardcastle2007}
{Hardcastle} M.~J.,  {Evans} D.~A.,   {Croston} J.~H.,  2007, \mn@doi [\mnras] {10.1111/j.1365-2966.2007.11572.x}, \href {https://ui.adsabs.harvard.edu/abs/2007MNRAS.376.1849H} {376, 1849}

\bibitem[\protect\citeauthoryear{{Hardcastle} et~al.,}{{Hardcastle} et~al.}{2019}]{hardcastle2019}
{Hardcastle} M.~J.,  et~al., 2019, \mn@doi [\aap] {10.1051/0004-6361/201833893}, \href {https://ui.adsabs.harvard.edu/abs/2019A&A...622A..12H} {622, A12}

\bibitem[\protect\citeauthoryear{{Hardcastle} et~al.,}{{Hardcastle} et~al.}{2023}]{hardcastle2023}
{Hardcastle} M.~J.,  et~al., 2023, \mn@doi [\aap] {10.1051/0004-6361/202347333}, \href {https://ui.adsabs.harvard.edu/abs/2023A&A...678A.151H} {678, A151}

\bibitem[\protect\citeauthoryear{{Hardcastle} et~al.,}{{Hardcastle} et~al.}{2025}]{hardcastle2025}
{Hardcastle} M.~J.,  et~al., 2025, \mn@doi [\mnras] {10.1093/mnras/staf622}, \href {https://ui.adsabs.harvard.edu/abs/2025MNRAS.539.1856H} {539, 1856}

\bibitem[\protect\citeauthoryear{{H{\"a}ring} \& {Rix}}{{H{\"a}ring} \& {Rix}}{2004}]{haring2004}
{H{\"a}ring} N.,  {Rix} H.-W.,  2004, \mn@doi [\apjl] {10.1086/383567}, \href {https://ui.adsabs.harvard.edu/abs/2004ApJ...604L..89H} {604, L89}

\bibitem[\protect\citeauthoryear{{Harrison}}{{Harrison}}{2017}]{harrison2017impact}
{Harrison} C.~M.,  2017, \mn@doi [Nature Astronomy] {10.1038/s41550-017-0165}, \href {https://ui.adsabs.harvard.edu/abs/2017NatAs...1E.165H} {1, 0165}

\bibitem[\protect\citeauthoryear{{Heckman} \& {Best}}{{Heckman} \& {Best}}{2014}]{heckman2014coevolution}
{Heckman} T.~M.,  {Best} P.~N.,  2014, \mn@doi [\araa] {10.1146/annurev-astro-081913-035722}, \href {https://ui.adsabs.harvard.edu/abs/2014ARA&A..52..589H} {52, 589}

\bibitem[\protect\citeauthoryear{{Heckman}, {Kauffmann}, {Brinchmann}, {Charlot}, {Tremonti}  \& {White}}{{Heckman} et~al.}{2004}]{heckman2004}
{Heckman} T.~M.,  {Kauffmann} G.,  {Brinchmann} J.,  {Charlot} S.,  {Tremonti} C.,   {White} S. D.~M.,  2004, \mn@doi [\apj] {10.1086/422872}, \href {https://ui.adsabs.harvard.edu/abs/2004ApJ...613..109H} {613, 109}

\bibitem[\protect\citeauthoryear{{Henry} et~al.,}{{Henry} et~al.}{2021}]{henry2021}
{Henry} A.,  et~al., 2021, \mn@doi [\apj] {10.3847/1538-4357/ac1105}, \href {https://ui.adsabs.harvard.edu/abs/2021ApJ...919..143H} {919, 143}

\bibitem[\protect\citeauthoryear{{Hickox} \& {Alexander}}{{Hickox} \& {Alexander}}{2018}]{hickox2018obscured}
{Hickox} R.~C.,  {Alexander} D.~M.,  2018, \mn@doi [\araa] {10.1146/annurev-astro-081817-051803}, \href {https://ui.adsabs.harvard.edu/abs/2018ARA&A..56..625H} {56, 625}

\bibitem[\protect\citeauthoryear{{Ho}}{{Ho}}{2008}]{ho2008nuclear}
{Ho} L.~C.,  2008, \mn@doi [\araa] {10.1146/annurev.astro.45.051806.110546}, \href {https://ui.adsabs.harvard.edu/abs/2008ARA&A..46..475H} {46, 475}

\bibitem[\protect\citeauthoryear{{Hopkins}, {Richards}  \& {Hernquist}}{{Hopkins} et~al.}{2007}]{hopkins2007observational}
{Hopkins} P.~F.,  {Richards} G.~T.,   {Hernquist} L.,  2007, \mn@doi [\apj] {10.1086/509629}, \href {https://ui.adsabs.harvard.edu/abs/2007ApJ...654..731H} {654, 731}

\bibitem[\protect\citeauthoryear{{Ivezi{\'c}} et~al.,}{{Ivezi{\'c}} et~al.}{2002}]{ivezic2002optical}
{Ivezi{\'c}} {\v{Z}}.,  et~al., 2002, \mn@doi [\aj] {10.1086/344069}, \href {https://ui.adsabs.harvard.edu/abs/2002AJ....124.2364I} {124, 2364}

\bibitem[\protect\citeauthoryear{{Jarvis} et~al.,}{{Jarvis} et~al.}{2019}]{jarvis2019prevalence}
{Jarvis} M.~E.,  et~al., 2019, \mn@doi [\mnras] {10.1093/mnras/stz556}, \href {https://ui.adsabs.harvard.edu/abs/2019MNRAS.485.2710J} {485, 2710}

\bibitem[\protect\citeauthoryear{{Jin}, {Kauffmann}, {Best}, {Shenoy}  \& {Ma{\l}ek}}{{Jin} et~al.}{2025}]{jin2025radioAGN}
{Jin} G.,  {Kauffmann} G.,  {Best} P.~N.,  {Shenoy} S.,   {Ma{\l}ek} K.,  2025, \mn@doi [\aap] {10.1051/0004-6361/202451974}, \href {https://ui.adsabs.harvard.edu/abs/2025A&A...694A.309J} {694, A309}

\bibitem[\protect\citeauthoryear{{Jonas} \& {MeerKAT Team}}{{Jonas} \& {MeerKAT Team}}{2016}]{jonas2016meerkat}
{Jonas} J.,  {MeerKAT Team} 2016, in MeerKAT Science: On the Pathway to the SKA. p.~1, \mn@doi{10.22323/1.277.0001}

\bibitem[\protect\citeauthoryear{{Juneau}, {Dickinson}, {Alexander}  \& {Salim}}{{Juneau} et~al.}{2011}]{Juneau2011}
{Juneau} S.,  {Dickinson} M.,  {Alexander} D.~M.,   {Salim} S.,  2011, \mn@doi [\apj] {10.1088/0004-637X/736/2/104}, \href {https://ui.adsabs.harvard.edu/abs/2011ApJ...736..104J} {736, 104}

\bibitem[\protect\citeauthoryear{{Juneau} et~al.,}{{Juneau} et~al.}{2014}]{Juneau2014}
{Juneau} S.,  et~al., 2014, \mn@doi [\apj] {10.1088/0004-637X/788/1/88}, \href {https://ui.adsabs.harvard.edu/abs/2014ApJ...788...88J} {788, 88}

\bibitem[\protect\citeauthoryear{{Kalfountzou}, {Jarvis}, {Bonfield}  \& {Hardcastle}}{{Kalfountzou} et~al.}{2012}]{kalfountzou2012star}
{Kalfountzou} E.,  {Jarvis} M.~J.,  {Bonfield} D.~G.,   {Hardcastle} M.~J.,  2012, \mn@doi [\mnras] {10.1111/j.1365-2966.2012.22093.x}, \href {https://ui.adsabs.harvard.edu/abs/2012MNRAS.427.2401K} {427, 2401}

\bibitem[\protect\citeauthoryear{{Kauffmann} et~al.,}{{Kauffmann} et~al.}{2003a}]{kauffman2003stellar}
{Kauffmann} G.,  et~al., 2003a, \mn@doi [\mnras] {10.1046/j.1365-8711.2003.06291.x}, \href {https://ui.adsabs.harvard.edu/abs/2003MNRAS.341...33K} {341, 33}

\bibitem[\protect\citeauthoryear{{Kauffmann} et~al.,}{{Kauffmann} et~al.}{2003b}]{kauffmann2003host}
{Kauffmann} G.,  et~al., 2003b, \mn@doi [\mnras] {10.1111/j.1365-2966.2003.07154.x}, \href {https://ui.adsabs.harvard.edu/abs/2003MNRAS.346.1055K} {346, 1055}

\bibitem[\protect\citeauthoryear{{Kellermann}, {Sramek}, {Schmidt}, {Shaffer}  \& {Green}}{{Kellermann} et~al.}{1989}]{kellermann1989vla}
{Kellermann} K.~I.,  {Sramek} R.,  {Schmidt} M.,  {Shaffer} D.~B.,   {Green} R.,  1989, \mn@doi [\aj] {10.1086/115207}, \href {https://ui.adsabs.harvard.edu/abs/1989AJ.....98.1195K} {98, 1195}

\bibitem[\protect\citeauthoryear{{Kennicutt}}{{Kennicutt}}{1998}]{kennicutt1998}
{Kennicutt} Robert~C. J.,  1998, \mn@doi [\araa] {10.1146/annurev.astro.36.1.189}, \href {https://ui.adsabs.harvard.edu/abs/1998ARA&A..36..189K} {36, 189}

\bibitem[\protect\citeauthoryear{{Kennicutt} \& {Evans}}{{Kennicutt} \& {Evans}}{2012}]{kennicutt2012}
{Kennicutt} R.~C.,  {Evans} N.~J.,  2012, \mn@doi [\araa] {10.1146/annurev-astro-081811-125610}, \href {https://ui.adsabs.harvard.edu/abs/2012ARA&A..50..531K} {50, 531}

\bibitem[\protect\citeauthoryear{{Kewley}, {Dopita}, {Sutherland}, {Heisler}  \& {Trevena}}{{Kewley} et~al.}{2001}]{kewley2001theoretical}
{Kewley} L.~J.,  {Dopita} M.~A.,  {Sutherland} R.~S.,  {Heisler} C.~A.,   {Trevena} J.,  2001, \mn@doi [\apj] {10.1086/321545}, \href {https://ui.adsabs.harvard.edu/abs/2001ApJ...556..121K} {556, 121}

\bibitem[\protect\citeauthoryear{{Kewley}, {Groves}, {Kauffmann}  \& {Heckman}}{{Kewley} et~al.}{2006}]{kewley2006host}
{Kewley} L.~J.,  {Groves} B.,  {Kauffmann} G.,   {Heckman} T.,  2006, \mn@doi [\mnras] {10.1111/j.1365-2966.2006.10859.x}, \href {https://ui.adsabs.harvard.edu/abs/2006MNRAS.372..961K} {372, 961}

\bibitem[\protect\citeauthoryear{{Kewley}, {Nicholls}  \& {Sutherland}}{{Kewley} et~al.}{2019}]{kewley2019understanding}
{Kewley} L.~J.,  {Nicholls} D.~C.,   {Sutherland} R.~S.,  2019, \mn@doi [\araa] {10.1146/annurev-astro-081817-051832}, \href {https://ui.adsabs.harvard.edu/abs/2019ARA&A..57..511K} {57, 511}

\bibitem[\protect\citeauthoryear{{Kondapally} et~al.,}{{Kondapally} et~al.}{2021}]{kondapally2021lofar}
{Kondapally} R.,  et~al., 2021, \mn@doi [\aap] {10.1051/0004-6361/202038813}, \href {https://ui.adsabs.harvard.edu/abs/2021A&A...648A...3K} {648, A3}

\bibitem[\protect\citeauthoryear{{Kondapally} et~al.,}{{Kondapally} et~al.}{2022}]{kondapally2022lofar}
{Kondapally} R.,  et~al., 2022, \mn@doi [\mnras] {10.1093/mnras/stac1128}, \href {https://ui.adsabs.harvard.edu/abs/2022MNRAS.513.3742K} {513, 3742}

\bibitem[\protect\citeauthoryear{{Kondapally} et~al.,}{{Kondapally} et~al.}{2024}]{kondapally2024}
{Kondapally} R.,  et~al., 2024, \mn@doi [arXiv e-prints] {10.48550/arXiv.2411.08104}, \href {https://ui.adsabs.harvard.edu/abs/2024arXiv241108104K} {p. arXiv:2411.08104}

\bibitem[\protect\citeauthoryear{{Macfarlane} et~al.,}{{Macfarlane} et~al.}{2021}]{macfarlane2021radio}
{Macfarlane} C.,  et~al., 2021, \mn@doi [\mnras] {10.1093/mnras/stab1998}, \href {https://ui.adsabs.harvard.edu/abs/2021MNRAS.506.5888M} {506, 5888}

\bibitem[\protect\citeauthoryear{{Madau} \& {Dickinson}}{{Madau} \& {Dickinson}}{2014}]{madau2014cosmic}
{Madau} P.,  {Dickinson} M.,  2014, \mn@doi [\araa] {10.1146/annurev-astro-081811-125615}, \href {https://ui.adsabs.harvard.edu/abs/2014ARA&A..52..415M} {52, 415}

\bibitem[\protect\citeauthoryear{{Maddox}}{{Maddox}}{2018}]{maddox2018}
{Maddox} N.,  2018, \mn@doi [\mnras] {10.1093/mnras/sty2201}, \href {https://ui.adsabs.harvard.edu/abs/2018MNRAS.480.5203M} {480, 5203}

\bibitem[\protect\citeauthoryear{{Magliocchetti}}{{Magliocchetti}}{2022}]{Magliocchetti2022}
{Magliocchetti} M.,  2022, \mn@doi [\aapr] {10.1007/s00159-022-00142-1}, \href {https://ui.adsabs.harvard.edu/abs/2022A&ARv..30....6M} {30, 6}

\bibitem[\protect\citeauthoryear{{Maiolino} et~al.,}{{Maiolino} et~al.}{2017}]{maiolino2017star}
{Maiolino} R.,  et~al., 2017, \mn@doi [\nat] {10.1038/nature21677}, \href {https://ui.adsabs.harvard.edu/abs/2017Natur.544..202M} {544, 202}

\bibitem[\protect\citeauthoryear{{Mingo}, {Hardcastle}, {Croston}, {Dicken}, {Evans}, {Morganti}  \& {Tadhunter}}{{Mingo} et~al.}{2014}]{mingo2014}
{Mingo} B.,  {Hardcastle} M.~J.,  {Croston} J.~H.,  {Dicken} D.,  {Evans} D.~A.,  {Morganti} R.,   {Tadhunter} C.,  2014, \mn@doi [\mnras] {10.1093/mnras/stu263}, \href {https://ui.adsabs.harvard.edu/abs/2014MNRAS.440..269M} {440, 269}

\bibitem[\protect\citeauthoryear{{Molyneux}, {Harrison}  \& {Jarvis}}{{Molyneux} et~al.}{2019}]{molyneux2019extreme}
{Molyneux} S.~J.,  {Harrison} C.~M.,   {Jarvis} M.~E.,  2019, \mn@doi [\aap] {10.1051/0004-6361/201936408}, \href {https://ui.adsabs.harvard.edu/abs/2019A&A...631A.132M} {631, A132}

\bibitem[\protect\citeauthoryear{{Morabito} et~al.,}{{Morabito} et~al.}{2022}]{morabito2022identifying}
{Morabito} L.~K.,  et~al., 2022, \mn@doi [\mnras] {10.1093/mnras/stac2129}, \href {https://ui.adsabs.harvard.edu/abs/2022MNRAS.515.5758M} {515, 5758}

\bibitem[\protect\citeauthoryear{{Morabito} et~al.,}{{Morabito} et~al.}{2025}]{morabito2025}
{Morabito} L.~K.,  et~al., 2025, \mn@doi [\mnras] {10.1093/mnrasl/slae104}, \href {https://ui.adsabs.harvard.edu/abs/2025MNRAS.536L..32M} {536, L32}

\bibitem[\protect\citeauthoryear{{Moustakas}, {Buhler}, {Scholte}, {Dey}  \& {Khederlarian}}{{Moustakas} et~al.}{2023}]{Moustakas2023fastspecfit}
{Moustakas} J.,  {Buhler} J.,  {Scholte} D.,  {Dey} B.,   {Khederlarian} A.,  2023, {FastSpecFit: Fast spectral synthesis and emission-line fitting of DESI spectra}, Astrophysics Source Code Library, record ascl:2308.005 (\mn@eprint {ascl} {2308.005})

\bibitem[\protect\citeauthoryear{{Myers} et~al.,}{{Myers} et~al.}{2023}]{Myers2023target}
{Myers} A.~D.,  et~al., 2023, \mn@doi [\aj] {10.3847/1538-3881/aca5f9}, \href {https://ui.adsabs.harvard.edu/abs/2023AJ....165...50M} {165, 50}

\bibitem[\protect\citeauthoryear{{Narayan} \& {Yi}}{{Narayan} \& {Yi}}{1994}]{Narayan1994}
{Narayan} R.,  {Yi} I.,  1994, \mn@doi [\apjl] {10.1086/187381}, \href {https://ui.adsabs.harvard.edu/abs/1994ApJ...428L..13N} {428, L13}

\bibitem[\protect\citeauthoryear{{Narayan} \& {Yi}}{{Narayan} \& {Yi}}{1995}]{Narayan1995}
{Narayan} R.,  {Yi} I.,  1995, \mn@doi [\apj] {10.1086/176343}, \href {https://ui.adsabs.harvard.edu/abs/1995ApJ...452..710N} {452, 710}

\bibitem[\protect\citeauthoryear{{Newman} et~al.,}{{Newman} et~al.}{2014}]{newman2014}
{Newman} S.~F.,  et~al., 2014, \mn@doi [\apj] {10.1088/0004-637X/781/1/21}, \href {https://ui.adsabs.harvard.edu/abs/2014ApJ...781...21N} {781, 21}

\bibitem[\protect\citeauthoryear{{Novak} et~al.,}{{Novak} et~al.}{2017}]{novak2017}
{Novak} M.,  et~al., 2017, \mn@doi [\aap] {10.1051/0004-6361/201629436}, \href {https://ui.adsabs.harvard.edu/abs/2017A&A...602A...5N} {602, A5}

\bibitem[\protect\citeauthoryear{{Osterbrock} \& {Ferland}}{{Osterbrock} \& {Ferland}}{2006}]{Osterbrock2006}
{Osterbrock} D.~E.,  {Ferland} G.~J.,  2006, {Astrophysics of gaseous nebulae and active galactic nuclei}

\bibitem[\protect\citeauthoryear{{Panessa}, {Baldi}, {Laor}, {Padovani}, {Behar}  \& {McHardy}}{{Panessa} et~al.}{2019}]{panessa2019}
{Panessa} F.,  {Baldi} R.~D.,  {Laor} A.,  {Padovani} P.,  {Behar} E.,   {McHardy} I.,  2019, \mn@doi [Nature Astronomy] {10.1038/s41550-019-0765-4}, \href {https://ui.adsabs.harvard.edu/abs/2019NatAs...3..387P} {3, 387}

\bibitem[\protect\citeauthoryear{{Papovich} et~al.,}{{Papovich} et~al.}{2022}]{Papovich2022}
{Papovich} C.,  et~al., 2022, \mn@doi [\apj] {10.3847/1538-4357/ac8058}, \href {https://ui.adsabs.harvard.edu/abs/2022ApJ...937...22P} {937, 22}

\bibitem[\protect\citeauthoryear{{Pirie} et~al.,}{{Pirie} et~al.}{2025}]{pirie2025}
{Pirie} C.~A.,  et~al., 2025, \mn@doi [\mnras] {10.1093/mnras/staf1006}, \href {https://ui.adsabs.harvard.edu/abs/2025MNRAS.541.1348P} {541, 1348}

\bibitem[\protect\citeauthoryear{{Pirzkal} et~al.,}{{Pirzkal} et~al.}{2024}]{pirzkal2024}
{Pirzkal} N.,  et~al., 2024, \mn@doi [\apj] {10.3847/1538-4357/ad429c}, \href {https://ui.adsabs.harvard.edu/abs/2024ApJ...969...90P} {969, 90}

\bibitem[\protect\citeauthoryear{{Read} et~al.,}{{Read} et~al.}{2018}]{read2018}
{Read} S.~C.,  et~al., 2018, \mn@doi [\mnras] {10.1093/mnras/sty2198}, \href {https://ui.adsabs.harvard.edu/abs/2018MNRAS.480.5625R} {480, 5625}

\bibitem[\protect\citeauthoryear{{Rigby} et~al.,}{{Rigby} et~al.}{2018}]{Rigby2018}
{Rigby} J.~R.,  et~al., 2018, \mn@doi [\apj] {10.3847/1538-4357/aaa2fc}, \href {https://ui.adsabs.harvard.edu/abs/2018ApJ...853...87R} {853, 87}

\bibitem[\protect\citeauthoryear{{Sabater} et~al.,}{{Sabater} et~al.}{2019}]{sabater2019lotss}
{Sabater} J.,  et~al., 2019, \mn@doi [\aap] {10.1051/0004-6361/201833883}, \href {https://ui.adsabs.harvard.edu/abs/2019A&A...622A..17S} {622, A17}

\bibitem[\protect\citeauthoryear{{Sabater} et~al.,}{{Sabater} et~al.}{2021}]{sabater2021lofar}
{Sabater} J.,  et~al., 2021, \mn@doi [\aap] {10.1051/0004-6361/202038828}, \href {https://ui.adsabs.harvard.edu/abs/2021A&A...648A...2S} {648, A2}

\bibitem[\protect\citeauthoryear{{Schlegel}, {Finkbeiner}  \& {Davis}}{{Schlegel} et~al.}{1998}]{schlegel1998maps}
{Schlegel} D.~J.,  {Finkbeiner} D.~P.,   {Davis} M.,  1998, \mn@doi [\apj] {10.1086/305772}, \href {https://ui.adsabs.harvard.edu/abs/1998ApJ...500..525S} {500, 525}

\bibitem[\protect\citeauthoryear{{Schreiber} et~al.,}{{Schreiber} et~al.}{2015}]{schreiber2015herschel}
{Schreiber} C.,  et~al., 2015, \mn@doi [\aap] {10.1051/0004-6361/201425017}, \href {https://ui.adsabs.harvard.edu/abs/2015A&A...575A..74S} {575, A74}

\bibitem[\protect\citeauthoryear{{Shimwell} et~al.,}{{Shimwell} et~al.}{2022}]{shimwell2022lofar}
{Shimwell} T.~W.,  et~al., 2022, \mn@doi [\aap] {10.1051/0004-6361/202142484}, \href {https://ui.adsabs.harvard.edu/abs/2022A&A...659A...1S} {659, A1}

\bibitem[\protect\citeauthoryear{{Shimwell} et~al.,}{{Shimwell} et~al.}{2025}]{shimwell2025}
{Shimwell} T.~W.,  et~al., 2025, \mn@doi [arXiv e-prints] {10.48550/arXiv.2501.04093}, \href {https://ui.adsabs.harvard.edu/abs/2025arXiv250104093S} {p. arXiv:2501.04093}

\bibitem[\protect\citeauthoryear{{Silk}}{{Silk}}{2013}]{silk2013unleashing}
{Silk} J.,  2013, \mn@doi [\apj] {10.1088/0004-637X/772/2/112}, \href {https://ui.adsabs.harvard.edu/abs/2013ApJ...772..112S} {772, 112}

\bibitem[\protect\citeauthoryear{{Siudek} et~al.,}{{Siudek} et~al.}{2024}]{Siudek2024}
{Siudek} M.,  et~al., 2024, \mn@doi [\aap] {10.1051/0004-6361/202451761}, \href {https://ui.adsabs.harvard.edu/abs/2024A&A...691A.308S} {691, A308}

\bibitem[\protect\citeauthoryear{{Smith} et~al.,}{{Smith} et~al.}{2011}]{smith2011herschel}
{Smith} D.~J.~B.,  et~al., 2011, \mn@doi [\mnras] {10.1111/j.1365-2966.2011.18827.x}, \href {https://ui.adsabs.harvard.edu/abs/2011MNRAS.416..857S} {416, 857}

\bibitem[\protect\citeauthoryear{{Smith} et~al.,}{{Smith} et~al.}{2014}]{smith2014}
{Smith} D.~J.~B.,  et~al., 2014, \mn@doi [\mnras] {10.1093/mnras/stu1830}, \href {https://ui.adsabs.harvard.edu/abs/2014MNRAS.445.2232S} {445, 2232}

\bibitem[\protect\citeauthoryear{{Smith} et~al.,}{{Smith} et~al.}{2016}]{smith2016weave}
{Smith} D.~J.~B.,  et~al., 2016, in {Reyl{\'e}} C.,  {Richard} J.,  {Cambr{\'e}sy} L.,  {Deleuil} M.,  {P{\'e}contal} E.,  {Tresse} L.,   {Vauglin} I.,  eds, SF2A-2016: Proceedings of the Annual meeting of the French Society of Astronomy and Astrophysics. pp 271--280 (\mn@eprint {arXiv} {1611.02706}), \mn@doi{10.48550/arXiv.1611.02706}

\bibitem[\protect\citeauthoryear{{Smith} et~al.,}{{Smith} et~al.}{2021}]{smith2021lofar}
{Smith} D.~J.~B.,  et~al., 2021, \mn@doi [\aap] {10.1051/0004-6361/202039343}, \href {https://ui.adsabs.harvard.edu/abs/2021A&A...648A...6S} {648, A6}

\bibitem[\protect\citeauthoryear{{Smol{\v{c}}i{\'c}} et~al.,}{{Smol{\v{c}}i{\'c}} et~al.}{2017}]{smolcic2017vla}
{Smol{\v{c}}i{\'c}} V.,  et~al., 2017, \mn@doi [\aap] {10.1051/0004-6361/201630223}, \href {https://ui.adsabs.harvard.edu/abs/2017A&A...602A...2S} {602, A2}

\bibitem[\protect\citeauthoryear{{Stasi{\'n}ska}, {Cid Fernandes}, {Mateus}, {Sodr{\'e}}  \& {Asari}}{{Stasi{\'n}ska} et~al.}{2006}]{stasinska2006semi}
{Stasi{\'n}ska} G.,  {Cid Fernandes} R.,  {Mateus} A.,  {Sodr{\'e}} L.,   {Asari} N.~V.,  2006, \mn@doi [\mnras] {10.1111/j.1365-2966.2006.10732.x}, \href {https://ui.adsabs.harvard.edu/abs/2006MNRAS.371..972S} {371, 972}

\bibitem[\protect\citeauthoryear{{Sutherland} \& {Saunders}}{{Sutherland} \& {Saunders}}{1992}]{sutherland1992likelihood}
{Sutherland} W.,  {Saunders} W.,  1992, \mn@doi [\mnras] {10.1093/mnras/259.3.413}, \href {https://ui.adsabs.harvard.edu/abs/1992MNRAS.259..413S} {259, 413}

\bibitem[\protect\citeauthoryear{{Tacchella}, {Dekel}, {Carollo}, {Ceverino}, {DeGraf}, {Lapiner}, {Mandelker}  \& {Primack Joel}}{{Tacchella} et~al.}{2016}]{tacchella2016confinement}
{Tacchella} S.,  {Dekel} A.,  {Carollo} C.~M.,  {Ceverino} D.,  {DeGraf} C.,  {Lapiner} S.,  {Mandelker} N.,   {Primack Joel} R.,  2016, \mn@doi [\mnras] {10.1093/mnras/stw131}, \href {https://ui.adsabs.harvard.edu/abs/2016MNRAS.457.2790T} {457, 2790}

\bibitem[\protect\citeauthoryear{{Tasse} et~al.,}{{Tasse} et~al.}{2021}]{tasse2021lofar}
{Tasse} C.,  et~al., 2021, \mn@doi [\aap] {10.1051/0004-6361/202038804}, \href {https://ui.adsabs.harvard.edu/abs/2021A&A...648A...1T} {648, A1}

\bibitem[\protect\citeauthoryear{{Thomas} et~al.,}{{Thomas} et~al.}{2013}]{thomas2013}
{Thomas} D.,  et~al., 2013, \mn@doi [\mnras] {10.1093/mnras/stt261}, \href {https://ui.adsabs.harvard.edu/abs/2013MNRAS.431.1383T} {431, 1383}

\bibitem[\protect\citeauthoryear{{Tremonti} et~al.,}{{Tremonti} et~al.}{2004}]{Tremonti2004}
{Tremonti} C.~A.,  et~al., 2004, \mn@doi [\apj] {10.1086/423264}, \href {https://ui.adsabs.harvard.edu/abs/2004ApJ...613..898T} {613, 898}

\bibitem[\protect\citeauthoryear{{Whittam} et~al.,}{{Whittam} et~al.}{2022}]{whittam2022mightee}
{Whittam} I.~H.,  et~al., 2022, \mn@doi [\mnras] {10.1093/mnras/stac2140}, \href {https://ui.adsabs.harvard.edu/abs/2022MNRAS.516..245W} {516, 245}

\bibitem[\protect\citeauthoryear{{Williams} et~al.,}{{Williams} et~al.}{2018}]{williams2018lofar}
{Williams} W.~L.,  et~al., 2018, \mn@doi [\mnras] {10.1093/mnras/sty026}, \href {https://ui.adsabs.harvard.edu/abs/2018MNRAS.475.3429W} {475, 3429}

\bibitem[\protect\citeauthoryear{{Williams} et~al.,}{{Williams} et~al.}{2019}]{williams2019}
{Williams} W.~L.,  et~al., 2019, \mn@doi [\aap] {10.1051/0004-6361/201833564}, \href {https://ui.adsabs.harvard.edu/abs/2019A&A...622A...2W} {622, A2}

\bibitem[\protect\citeauthoryear{{Wright} et~al.,}{{Wright} et~al.}{2010}]{wright2010}
{Wright} E.~L.,  et~al., 2010, \mn@doi [\aj] {10.1088/0004-6256/140/6/1868}, \href {https://ui.adsabs.harvard.edu/abs/2010AJ....140.1868W} {140, 1868}

\bibitem[\protect\citeauthoryear{{Yue} et~al.,}{{Yue} et~al.}{2024}]{yue2024}
{Yue} B.~H.,  et~al., 2024, \mn@doi [\mnras] {10.1093/mnras/stae725}, \href {https://ui.adsabs.harvard.edu/abs/2024MNRAS.529.3939Y} {529, 3939}

\bibitem[\protect\citeauthoryear{{Yue} et~al.,}{{Yue} et~al.}{2025}]{yue2025}
{Yue} B.~H.,  et~al., 2025, \mn@doi [\mnras] {10.1093/mnras/staf077}, \href {https://ui.adsabs.harvard.edu/abs/2025MNRAS.537..858Y} {537, 858}

\bibitem[\protect\citeauthoryear{{Yun}, {Reddy}  \& {Condon}}{{Yun} et~al.}{2001}]{Yun2001}
{Yun} M.~S.,  {Reddy} N.~A.,   {Condon} J.~J.,  2001, \mn@doi [\apj] {10.1086/323145}, \href {https://ui.adsabs.harvard.edu/abs/2001ApJ...554..803Y} {554, 803}

\bibitem[\protect\citeauthoryear{{Zavala} et~al.,}{{Zavala} et~al.}{2021}]{Zavala2021evolution}
{Zavala} J.~A.,  et~al., 2021, \mn@doi [\apj] {10.3847/1538-4357/abdb27}, \href {https://ui.adsabs.harvard.edu/abs/2021ApJ...909..165Z} {909, 165}

\bibitem[\protect\citeauthoryear{{Zhang} \& {Hao}}{{Zhang} \& {Hao}}{2018}]{Zhang2018A}
{Zhang} K.,  {Hao} L.,  2018, \mn@doi [\apj] {10.3847/1538-4357/aab207}, \href {https://ui.adsabs.harvard.edu/abs/2018ApJ...856..171Z} {856, 171}

\bibitem[\protect\citeauthoryear{{Zou} et~al.,}{{Zou} et~al.}{2024}]{Zou2024large}
{Zou} H.,  et~al., 2024, \mn@doi [\apj] {10.3847/1538-4357/ad1409}, \href {https://ui.adsabs.harvard.edu/abs/2024ApJ...961..173Z} {961, 173}

\bibitem[\protect\citeauthoryear{{de Jong} et~al.,}{{de Jong} et~al.}{2024}]{dejong2024}
{de Jong} J.~M.~G.~H.~J.,  et~al., 2024, \mn@doi [\aap] {10.1051/0004-6361/202450595}, \href {https://ui.adsabs.harvard.edu/abs/2024A&A...689A..80D} {689, A80}

\bibitem[\protect\citeauthoryear{{van Haarlem} et~al.,}{{van Haarlem} et~al.}{2013}]{vanhaarlem2013lofar}
{van Haarlem} M.~P.,  et~al., 2013, \mn@doi [\aap] {10.1051/0004-6361/201220873}, \href {https://ui.adsabs.harvard.edu/abs/2013A&A...556A...2V} {556, A2}

\makeatother
\end{thebibliography}

\section*{Acknowledgments}

The authors sincerely thank the anonymous reviewer for a report which significantly improved the quality of this paper.
MIA acknowledges support from the UK Science and Technology Facilities Council (STFC) studentship under the grant ST/V506709/1.
MIA, DJBS and MJH acknowledge support from the STFC under grant ST/V000624/1. DJBS and LRH acknowledge support from STFC under grant ST/Y001028/1.
MIA and PNB acknowledge support from the STFC under grant ST/Y000951/1.
PNB is grateful for support from the UK STFC via grant ST/V000594/1.
SD acknowledges support from the STFC studentship via grant ST/W507490/1.
SS acknowledges support from the STFC via
studentship grant number ST/X508408/1.
KJD acknowledges support from the STFC through an Ernest Rutherford Fellowship (grant number ST/W003120/1).
RK is grateful for support from the UK STFC via grant ST/V000594/1.
LKM is grateful for support from a UKRI FLF [MR/Y020405/1].

LOFAR is the Low Frequency Array designed and constructed by ASTRON. It has observing, data processing, and data storage facilities in several countries, which are owned by various parties (each with their own funding sources), and that are collectively operated by the ILT foundation under a joint scientific policy. The ILT resources have benefited from the following recent major funding sources: CNRS-INSU, Observatoire de Paris and Université d’Orléans, France; BMBF, MIWFNRW, MPG, Germany; Science Foundation Ireland (SFI), Department of Business, Enterprise and Innovation (DBEI), Ireland; NWO,
The Netherlands; The Science and Technology Facilities Council,
UK; Ministry of Science and Higher Education, Poland; The Istituto
Nazionale di Astrofisica (INAF), Italy.
This research made use of the Dutch national e-infrastructure with
support of the SURF Cooperative (e-infra 180169) and the LOFAR
e-infra group. The Jülich LOFAR Long Term Archive and the German LOFAR network are both coordinated and operated by the Jülich
Supercomputing Centre (JSC), and computing resources on the supercomputer JUWELS at JSC were provided by the Gauss Centre for
Supercomputing e.V. (grant CHTB00) through the John von Neumann Institute for Computing (NIC).
This research made use of the University of Hertfordshire high performance computing facility and the LOFAR-UK computing facility located at the University of Hertfordshire and supported by
STFC [ST/V002414/1], and of the Italian LOFAR IT computing infrastructure supported and operated by INAF, and by the Physics
Department of Turin University (under an agreement with Consorzio
Interuniversitario per la Fisica Spaziale) at the C3S Supercomputing
Centre, Italy.

\section*{Data Availability}
The radio and spectroscopic data used in this work are publicly available and can be found at the LOFAR Surveys (\href{www.lofar-surveys.org}{www.lofar-surveys.org}) and the Dark Energy Spectroscopic Instrument website (\href{www.sdss.org}{https://data.desi.lbl.gov}.), respectively. The output catalogue, including emission line flux estimates and the probabilities used for the classifications, is also made available at \href{https://lofar-surveys.org/deepfields_public_en1.html}{lofar-surveys.org/deepfields\_public\_en1.html} as part of this paper.



\appendix

\section{Comparison of H$\alpha$-Derived and Photometric Star Formation Rates}\label{sec:SFR}

In this section, we compare our star formation rate (SFR) estimates based on H$\alpha$ luminosity to the SFRs derived from \texttt{\textsc{Prospector}} in D24. This comparison serves both to assess the validity of our aperture corrections based on the total-to-fibre $r$-band flux ratio and to highlight differences in the radio-excess diagnostic between our work and D24, which will be relevant for the discussion in section~\ref{sec:compare_SED}. To do this, we use the H$\beta$ and H$\alpha$ fluxes from our spectral fitting, corrected for aperture losses, and compute the dust-corrected H$\alpha$ luminosity following the procedure described in section~\ref{sec:radio_excess}. Specifically, we apply $A_V$ derived from the Balmer decrement where available, and use the \texttt{Prospector}-based $A_V$ in cases where H$\alpha$ is not detected ($z>0.483$). We then convert the dust-corrected H$\alpha$ luminosities into SFRs (converted to the adopted IMF in D24) using the relation from \citet{kennicutt1998}. 

\begin{figure*}
    \centering
    \includegraphics[width=0.55\textwidth]{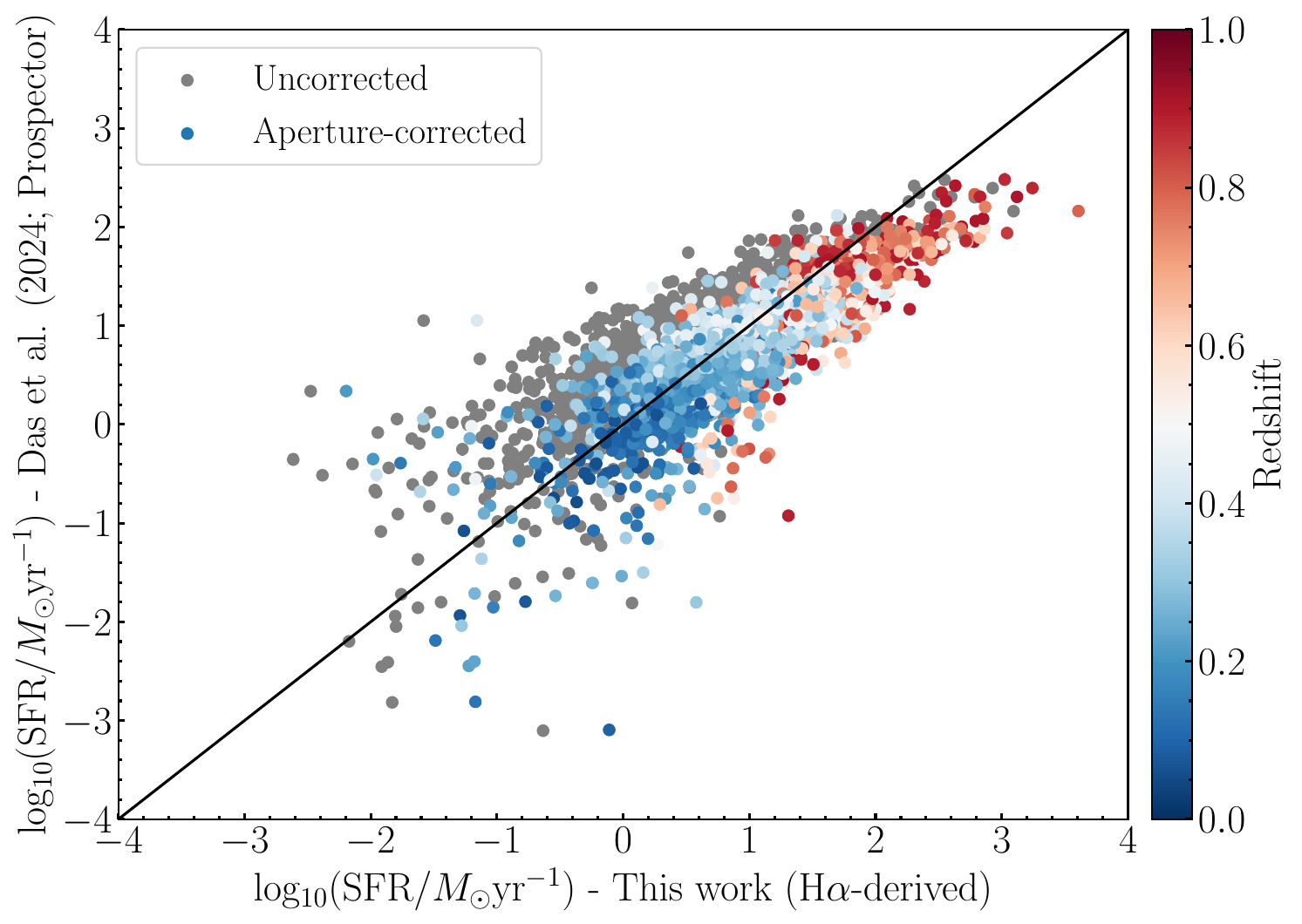}
    \caption{The H$\alpha$-derived SFR for all sources in our sample with a 3$\sigma$ detection in H$\beta$ and H$\alpha$ compared to the SFR from \texttt{\textsc{Prospector}} with (colour-coded) and without an aperture correction (grey). The solid line indicates equality, while the colourbar denotes the spectroscopic redshift for each source.}
    \label{fig:SFR_Ha_Prospector}
\end{figure*}

The comparison is shown in Figure~\ref{fig:SFR_Ha_Prospector}, where we present only sources with a 3$\sigma$ detection in both H$\beta$ and H$\alpha$, and which also satisfy the SED goodness-of-fit and redshift criteria described in section \ref{sec:MEx_diagnostic}. The coloured data points represent aperture-corrected SFR estimates, colour-coded by redshift to illustrate the consistency between the two methods of estimating H$\alpha$ luminosity, while the grey points show the uncorrected SFRs. It is clear that applying our total-to-fibre flux correction improves the agreement between the H$\alpha$-derived SFRs and those from \texttt{\textsc{Prospector}}. There is still, however, a noticeable scatter, with the H$\alpha$-based SFRs appearing systematically higher by $\sim$0.2dex. 
As noted in previous studies (e.g. \citealt{gurkan2018lofar} and references therein), this offset may arise from several factors, including differences in dust attenuation corrections, aperture effects, or the timescales over which star formation rates are measured. For instance, \texttt{\textsc{Prospector}} averages the star formation rate over the past 100\,Myr, whereas H$\alpha$ emission traces more recent star formation  ($\sim$10\,Myr; \citealt{kennicutt2012}). Therefore, the larger offset seen at high SFRs could be due to the fact that these sources are bursty systems that have a higher SFR in the last 10\,Myrs than the last 100\,Myrs (e.g. \citealt{pirie2025}). This could potentially lead to inconsistencies between the radio-excess sources classified by photometric classifications and this work, but as discussed in section \ref{sec:compare_SED} the differences are small.

\section{Emission line flux comparison with DESI EDR value added catalogues}\label{appendix:fastspec}

\begin{figure*}
    \centering
    \includegraphics[trim={1.5cm 0.5cm 5cm 1.5cm},clip,width=1.0\textwidth]{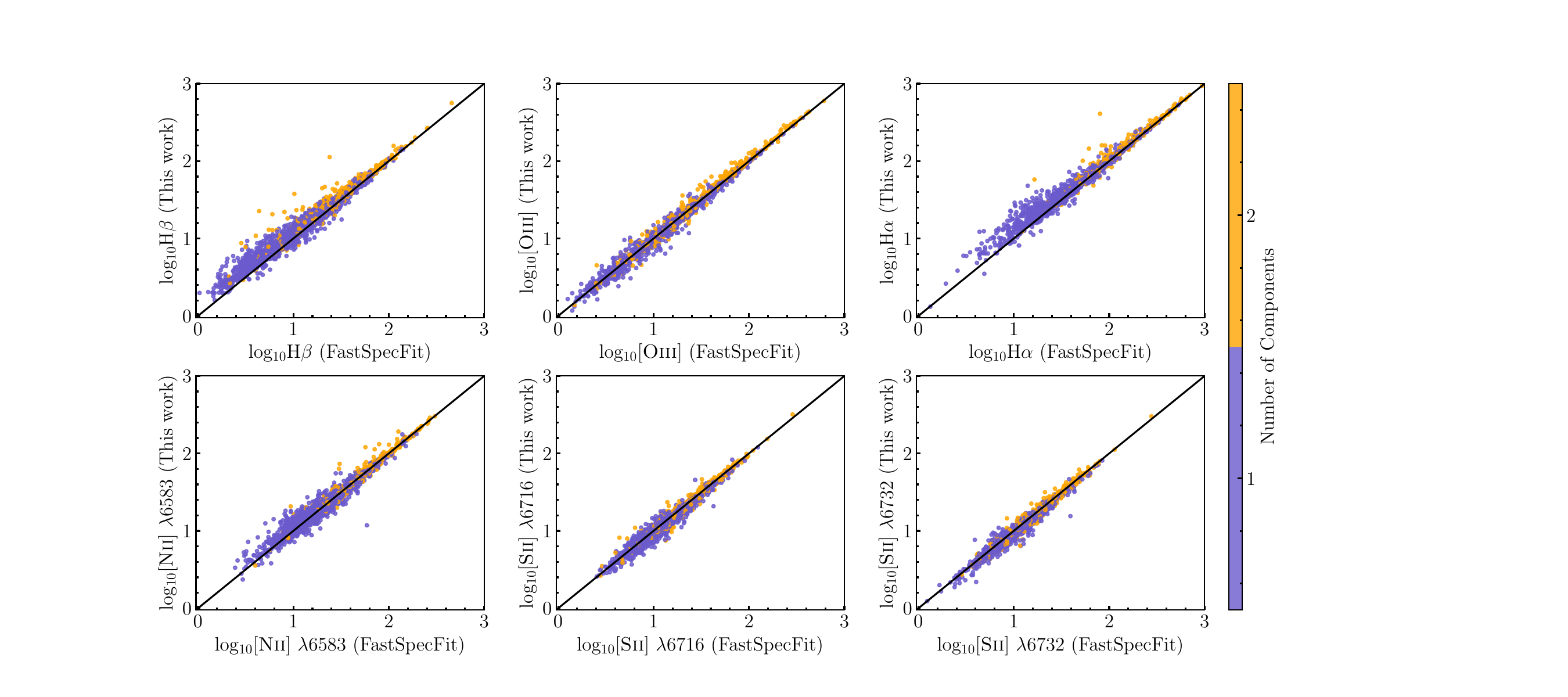}
    \caption{Emission line fluxes obtained with our fitting code, compared to the measurements provided by \texttt{\textsc{FastSpecFit}} for all sources with a 3$\sigma$ detection in each line considered. The solid line indicates equality, while the colourbar denotes the number of Gaussian components used in our fit.}
    \label{fig:fastspec}
\end{figure*}

In this section, we investigate how our aperture-uncorrected emission line fluxes, obtained with our new spectral fitting code described in section \ref{sec:line_fitting}, compare with those from the DESI EDR value added catalogue, \textsc{\texttt{FastSpecFit}} (\citealt{Moustakas2023fastspecfit}; Moustakas et al. \textit{in prep.}). \textsc{\texttt{FastSpecFit}} is a stellar continuum and emission-
line fitting code that uses physically motivated stellar
continuum and emission-line templates to simultaneously model DESI spectroscopy and broadband photometry, making it a relevant product for comparison. 

Figure \ref{fig:fastspec} shows the emission line fluxes for H$\beta$, [\textsc{Oiii}]\,$\lambda$5007, H$\alpha$, [\textsc{Nii}]\,$\lambda$6583, and [\textsc{Sii}]\,$\lambda\lambda$6717,6731 for all sources with a $>3\sigma$ detection, colour-coded based on the number of Gaussian components used in our fitting technique. Overall, we observe good agreement across all emission lines, where we find that two-Gaussian components are predominantly needed at higher line flux values. This is expected, as higher flux values typically correspond to higher signal-to-noise ratios, allowing for a more reliable detection of multiple kinematic components. We do see a slight median offset for H$\beta$ ($\sim0.03$dex) and H$\alpha$ ($\sim0.02$dex) at low line fluxes, which is likely due to the different stellar population models used in both tools. Nonetheless, the overall agreement between our fluxes and those from \texttt{FastSpecFit} suggests that both fitting methods produce consistent results, highlighting the reliability of our fitting technique and its potential for use in future spectral analyses.

\section{Updated MEx source classification criteria}\label{appendix:MEx}

To ensure that the MEx diagram effectively distinguishes between SFGs and AGN in our high-$z$ sample, we use the BPT diagnostic diagram to modify the $x$-axis in a redshift dependent way (hereafter the modified MEx diagram or $\mathcal{M}$Ex). Using the Ka03 line, we classify all sources in our low-$z$ sample with a $>3\sigma$ detection in H$\beta$, [\textsc{Oiii}]$\lambda$5007, H$\alpha$ and [\textsc{Nii}]$\lambda$6583 as BPT\_SFG or BPT\_AGN and use this classification to assess the performance of various offsets as a function of $z$. We test linear, logarithmic, and exponential relations, where the $x$-axis of the MEx diagram becomes $\mathcal{M} = M_{*} - \text{offset}$, and use the J14 lower curve at $z\sim0$ to define the $\mathcal{M}$Ex\_SFG and $\mathcal{M}$Ex\_AGN regions. For each case, we determine the optimal relation that minimizes the contamination of BPT\_SFG sources falling in the $\mathcal{M}$Ex\_AGN region and similarly BPT\_AGN sources in the $\mathcal{M}$Ex\_SFG region. To evaluate the performance of each relation, including the line detection limit prescription from J14, we investigate the fraction of SFGs/AGN that are classified as BPT\_SFG/BPT\_AGN and fall within the $\mathcal{M}$Ex\_SFG/$\mathcal{M}$Ex\_AGN regions. We find that an offset of $1-\mathrm{exp}(-1.2z)$ strikes a good balance between maintaining purity and avoiding a rapid increase at higher redshifts, with a fraction of correctly identified SFGs and AGN reaching up to 91 and 81 per cent. In addition, this is found to perform better than the J14 line detection limit prescription, where the fraction of \texttt{BPT\_SFG} located in the \texttt{$\mathcal{M}$Ex\_SFG} region is 92 per cent, but the \texttt{BPT\_AGN} within the \texttt{$\mathcal{M}$Ex\_AGN} region is $\sim70$ per cent.

\begin{figure*}
    \centering
    \includegraphics[width=1\textwidth]{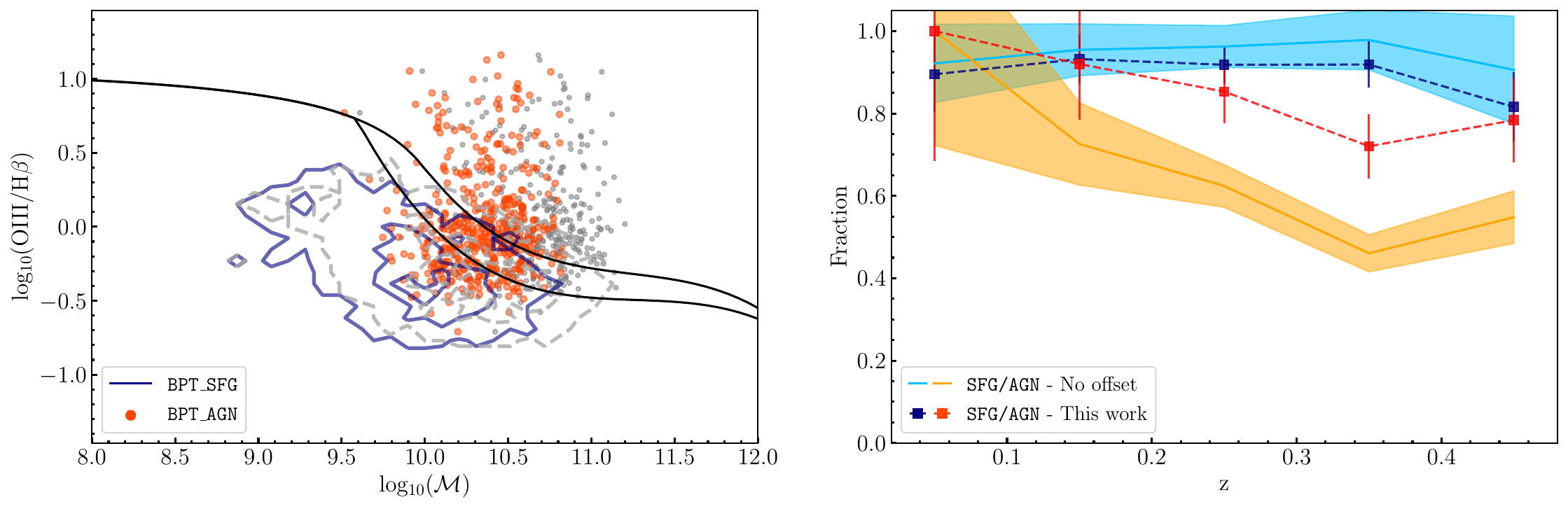}

    \caption{Left panel: The distribution of \texttt{BPT\_SFG} (blue contours) and \texttt{BPT\_AGN} (red dots) sources on the modified MEx ($\mathcal{M}$Ex) diagram. The grey contours and points represent the distribution of the same populations but without an offset, corresponding to the original MEx diagram. The contour levels are selected to encompass 5 per cent, 50 per cent, and 90 per cent of the corresponding sample. The black solid lines represent the demarcation lines from J14 at $z=0$. Right panel: The fraction of sources classified as \texttt{BPT\_SFG} and \texttt{BPT\_AGN} that also fall within the \texttt{$\mathcal{M}$Ex\_SFG} and \texttt{$\mathcal{M}$Ex\_AGN} regions, which are defined with and without an offset as indicated by the legend. The errorbars are associated with Poisson uncertainties.}

    \label{fig:MEx_mod}
\end{figure*}

To visualize the changes made, we present the distribution of the \texttt{BPT\_SFG} and \texttt{BPT\_AGN} classes in the MEx diagram in the left panel of Figure \ref{fig:MEx_mod}, both with and without the $1-\mathrm{exp}(-1.2z)$ offset applied - effectively comparing the modified and original MEx classifications. When the offset is applied, the overlap region between the two classes largely occurs within the region delineated by the J14 lower and upper curves (the \texttt{MEx\_Comp} region), rather than the \texttt{MEx\_AGN} region, which is the case when no offset is present. The \texttt{MEx\_Comp} was originally intended to correspond to the \texttt{BPT\_Comp} region (recall that we assign this region to the AGN class), which also contained a fraction of sources from the \texttt{BPT\_SFG} region (see \citealt{Juneau2011}). This gives us confidence in our results as it demonstrates that the offset provides a reasonable adjustment while maintaining the overall classification trends. The right panel in Figure \ref{fig:MEx_mod} further shows the fraction of SFG/AGN with (dotted lines) and without the $1-\mathrm{exp}(-1.2z)$ offset (solid lines) in five redshift bins. Without the offset, only 40–60 percent of \texttt{BPT\_AGN} sources fall within the \texttt{MEx\_AGN} region in the $0.2 < z < 0.5$ range. In contrast, applying the $1-\mathrm{exp}(-1.2z)$ offset recovers 80–90 percent of the \texttt{BPT\_AGN} whilst giving consistent results for SFGs, demonstrating its effectiveness in improving consistency between the two diagnostic diagrams. Overall, we conclude that the proposed offset provides a reasonable improvement in the separation between SFG and radiatively-efficient AGN when using the MEx diagram.

\section{Additional comparison with our probabilistic spectroscopic method}\label{appendix:compare}
\subsection{The BPT and $\mathcal{M}$Ex diagnostics}

\begin{figure*}
    \centering
    \includegraphics[width=1\textwidth]{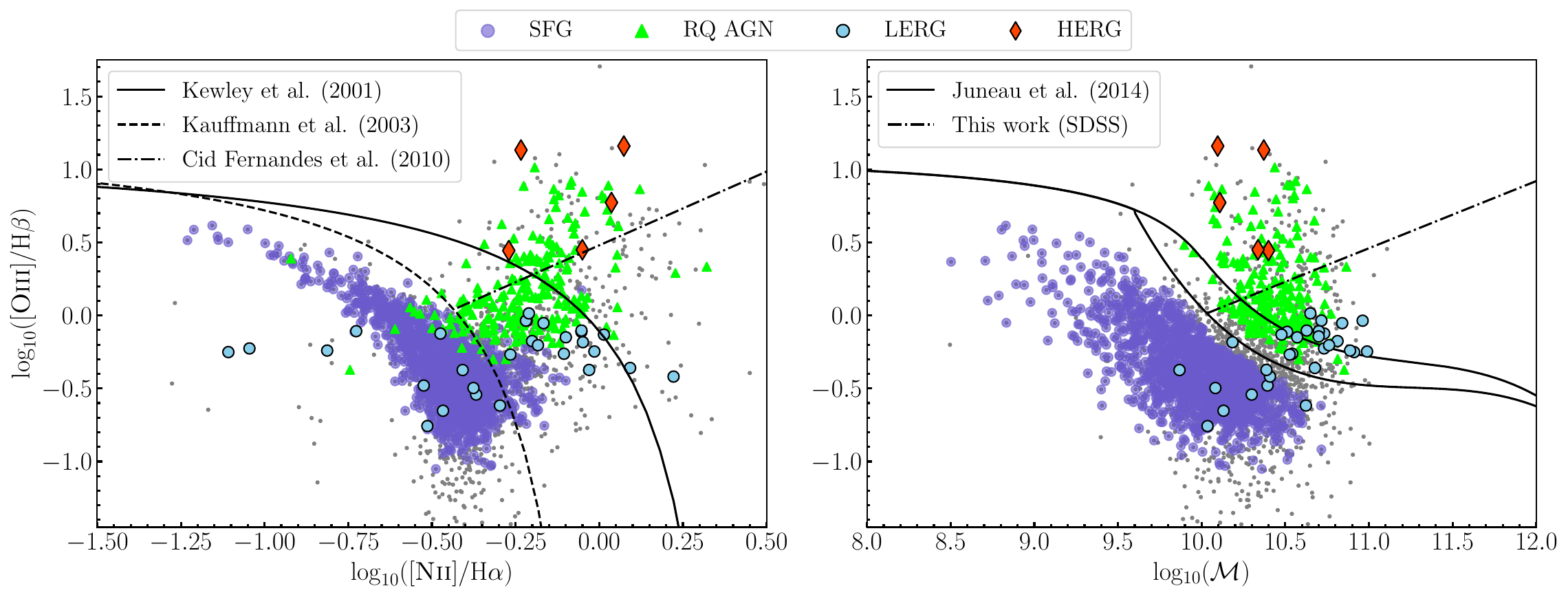}
    \caption{The distribution of SFGs (dark blue), RQ AGN (green), LERGs (light blue) and HERGs (red) in the BPT (left panel) and $\mathcal{M}$Ex diagram (right panel) for sources in the low-$z$ sample classified at $90$ per cent confidence according to $\mathcal{M}$Ex classification. The demarcation lines indicated in the legend define the regions discussed in section \ref{sec:bpt_class} and \ref{sec:MEx_diagnostic}.}
    \label{fig:BPT_MEx_test}
\end{figure*}

In this section, we provide an additional comparison between the BPT and $\mathcal{M}$Ex classifications discussed in section \ref{sec:class_scheme}, where we have applied both methods to our low-$z$ sample. 
In Figure \ref{fig:BPT_MEx_test}, we show the location of sources on the BPT (left panel) and the modified MEx diagram (right panel) colour-coded by their classification using the $\mathcal{M}$ex method. We can see an overall good agreement between the two diagrams, where the sources in each class, classified according to the $\mathcal{M}$Ex diagram, populate the expected regions of the BPT; the majority of SFGs/RQ AGN are found below/above the Ka03 line, while LERGs/HERGs populate the region below/above the C10 line. 
There are, however, some sources that occupy the region near the division lines; approximately $10\%$ of RQ AGN appear within the \texttt{BPT\_SFG} region but remain near the Ka03 line, while about $6\%$ of SFGs fall within the \texttt{BPT\_Comp} region. As discussed in section \ref{sec:lowz vs highz scheme} and Appendix \ref{appendix:MEx}, such discrepancies are to be expected given how the MEx diagram and subsequently the $\mathcal{M}$Ex were constructed. Nonetheless, the $\mathcal{M}$Ex remains a reasonable alternative to the BPT diagram at higher $z$, where [\textsc{Nii}] and H$\alpha$ are not available.

\subsection{Comparison with SDSS classifications}

\begin{figure*}
    \centering
    \includegraphics[width=0.55\textwidth]{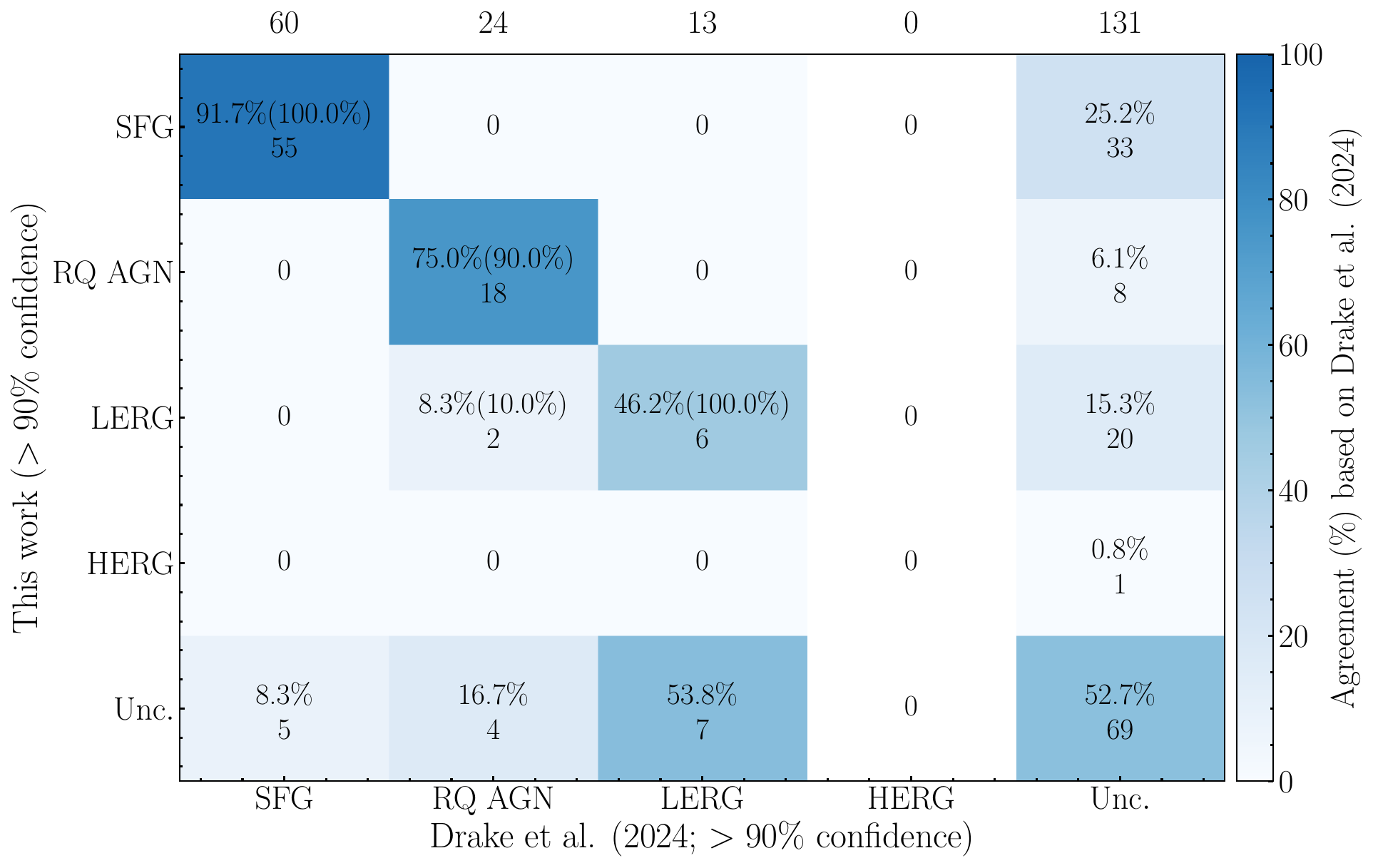}
    \caption{Confusion matrix comparing the 90 per cent spectroscopic classification of this work (rows) with the 90 per cent classification of Dr24 (columns) for the common sample of sources. Each cell represents the level of agreement
    based on Dr24 in percentages, also indicated by the colourbar, where we have also included the number of sources in each cell. The percentages in brackets are calculated for the subset of sources that reach the 90 per cent threshold in both works. The total number of sources in each class for the Dr24 classification is shown at the top.}
    \label{fig:conf_mia_abd}
\end{figure*}

To provide more confidence in our spectroscopic classifications, we further compare our results with the probabilistic spectroscopic classifications from Dr24 for the sources in common between DESI EDR and SDSS. This sample was obtained by using a positional cross-match with a maximum search radius of 1 arcsecond between our catalogue and the one provided by Dr24\footnote{Available from \url{https://lofar-surveys.org/dr2_release.html}}, which gives us a total of 228 sources. 
In Figure \ref{fig:conf_mia_abd}, we present a confusion matrix of our $>90$ per cent spectroscopic classification with the $>90$ per cent spectroscopic classification from Dr24. Each confusion matrix cell presents the agreement rate as the percentage of sources commonly classified according to Dr24, with the subset that meets the 90 per cent threshold for our scheme shown in brackets and denoted by the colourbar. Irrespective of the small sample size, we find excellent agreement when considering both the total sources and the ones that we confidently classify at 90 per cent (the values in brackets). Interestingly, we classify more sources than Dr24, which could be attributed to several factors, such as higher S/N in our detected emission lines and differences in fibre aperture sizes, since Dr24 uses SDSS data and does not apply any aperture corrections, potentially leading to different flux measurements. In fact, this prompted us to adopt a different radio excess demarcation line, as discussed in section \ref{sec:radio_excess}. Additionally, we are using the C10 line to separate LINERs and Seyferts, rather than the Ka03 line, which could also influence the classification results. Finally, we are using a new spectral fitting method, which fits multiple Gaussian components to the emission lines and uses the MCMC chains to obtain robust uncertainties that may differ from those used in Dr24.

\begin{figure*}
    \centering
    \includegraphics[width=1\textwidth]{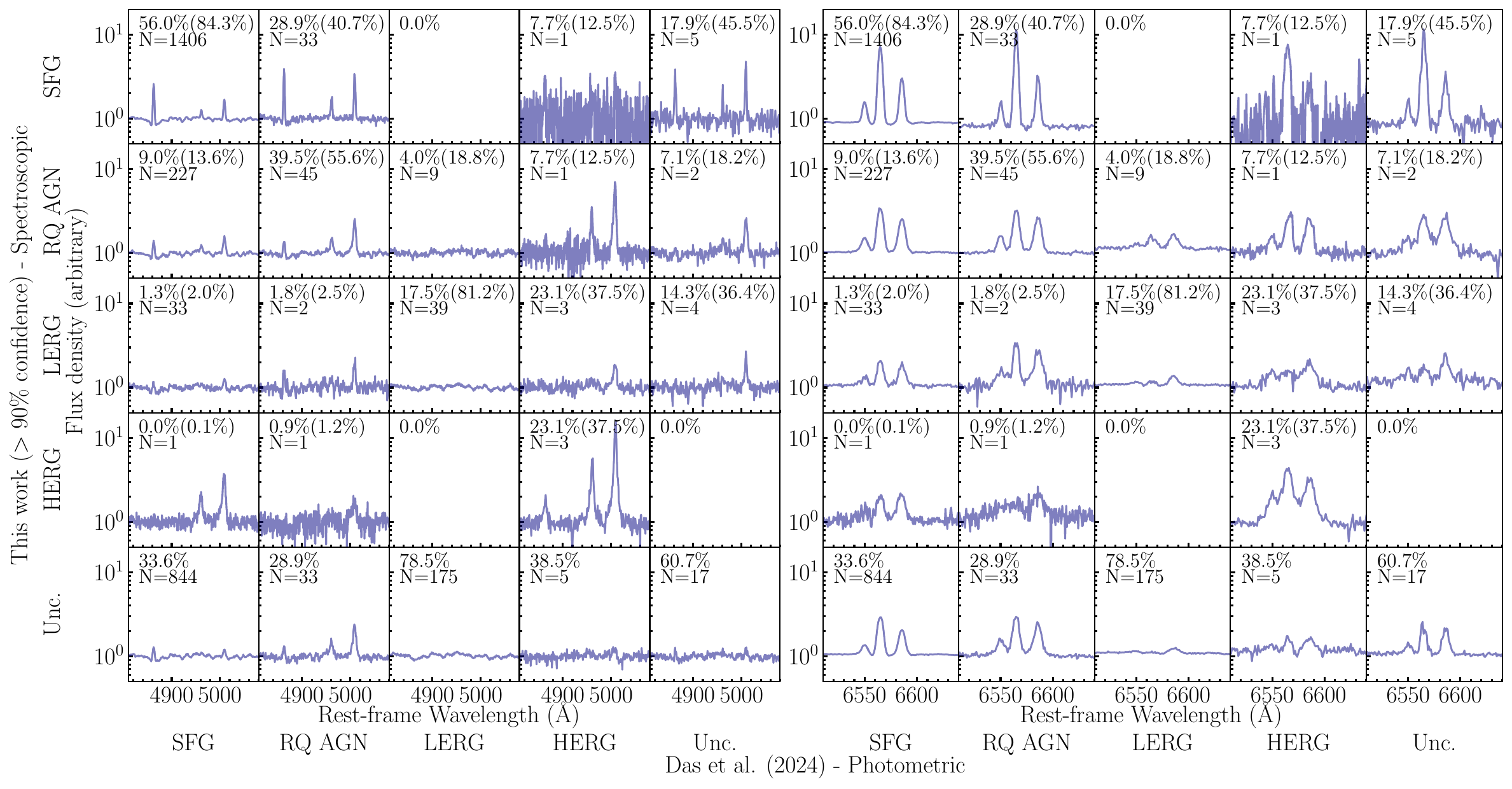}
    \caption{The composite spectra of sources, centred on the H$\beta$ and \textsc{[Oiii]} (left panel) and H$\alpha$ and \textsc{[Nii]} (right) emission lines, in each cell of the confusion matrix comparing 90 per cent spectroscopic classification with D24 for the low-$z$ sample. The level of agreement compared to D24 for the whole sample and those that are classified with 90 per cent confidence in this work, as well as the number of sources in each cell are given in the top left corner.}
    \label{fig:conf_stack_Ha_Hb}
\end{figure*}
\subsection{The composite spectra of the different classifications}

In this section, we present the composite spectra (see Figure \ref{fig:conf_stack_Ha_Hb}), centred on the H$\beta$ - [\textsc{Oiii}] and H$\alpha$ - [\textsc{Nii}] complex, from the confusion matrix comparing our $>90$ per cent classifications in the low-$z$ sample with the results from D24 (see also panel b in Figure \ref{fig:conf_diag}). These spectra were obtained with \textsc{\texttt{SpecStacker}} in a similar way as in section \ref{sec:stacks}, with the number of spectra that go in each stack presented in the top left corner of each cell. The agreement rate is also shown in the top left corner, which again is defined as the percentage of sources commonly classified according to D24, with and without the spectroscopic unclassified branch. 

We can see that the composite spectra along the diagonal and across the four top rows follow the typical characteristics of the four classes, as discussed in section \ref{sec:stacks}. 
In the second row and first column, we can see that the composite spectrum exhibits stronger [\textsc{Oiii}] and [\textsc{Nii}] emission compared to H$\beta$ and H$\alpha$, respectively, than what we observe for SFGs (as shown in the first row), again highlighting that the D24 SED fitting has incorrectly assigned these sources to the SFG class. Similarly, we can see that the spectrum in the third row and first column lacks prominent H$\beta$ and [\textsc{Oiii}] emission and is associated with faint H$\alpha$ and [\textsc{Nii}] emission, which is more characteristic of LERGs rather than SFGs, contrary to what the photometric classification from D24 suggests. This type of comparison further highlights the necessity of using spectroscopy, alongside other methods, to reliably classifying the faint radio-source population.


\bsp	
\label{lastpage}
\end{document}